\documentclass[prb,preprint]{revtex4}

\pdfoutput=1

\usepackage{graphicx}
\usepackage{latexsym}
\usepackage{amssymb}
\usepackage{amsmath} 
\usepackage{extarrows} 
\usepackage{MnSymbol} 
\usepackage{xcolor} 

\begin{document}


\title{Understanding the normal state of cuprate superconductors}

\medskip 

\date{November 11, 2026} \bigskip

\author{Manfred Bucher \\}
\affiliation{\text{\textnormal{Physics Department, California State University,}} \textnormal{Fresno,}
\textnormal{Fresno, California 93740-8031} \\}

\begin{abstract}
Doped holes in the $CuO_2$ planes reside pairwise at $O$ atoms. Their lack of net charge {\it demottes} the flanking $Cu  3d$-$O  2p$ holes of half-filling. Below the doping-dependent pseudogap (PG) temperature, $T^*(p)$, the latter are trapped between {\it motted} $Cu^{2+}$ ions. At $T > T^*$, the remaining $Cu  3d$-$O  2p$ holes are {\it thermally} demotted, initiating the {\it strange metal} (SM) phase.
 The PG energy at the antinodal points in the Brillouin zone decreases with increasing doping, as less $Cu 3d$-$O  2p$ holes need to be thermally demotted. In the SM phase, the $Cu  3d$-$O  2p$ holes hop, with Planckian scattering rate, by leaps of  lattice-constant length, $a_0$. This gives rise to a resistivity $\rho \propto T$. In the PG phase, $T < T^*$, a hole pair hops twice---zig-zag from $O$ atom to $O^{2-}$ ion---to advance by $a_0$, possibly causing $\rho \propto T^2$. $O$ atoms in the $CuO_2$ plane, from $Ae = Sr, Ba$ doping of {\it 214} cuprates or from oxygen enrichment (OE), arrange to a {\it charge order} (CO).  In the {\it 214} systems, an $O$ superlattice (SL) forms with incommensurability $q_{co}(p) \propto \sqrt{p}$. It saturates at $\hat{p}$ when a secondary SL develops with $Q_{co}(p) \propto \sqrt{p - \hat{p}}$.   
At doping $p^*$---a putative quantum critical point---$q_{co}(\hat{p}) = q_{co}(p^*) = \sqrt{2} Q_{co}(p^*)$ holds, causing a sudden collapse of $T^*(p)$. In OE cuprates, the CO dilutes the host lattice's $CuO_6$ octahedra of diagonal tilt, causing an incommensurability $q_{CuO_6}(p) = 0.35 - 0.71 p$ that is instrumental in the formation of Fermi pockets. Oxygen atoms are singlets, $O^{\uparrow \downarrow}$, in  OE cuprates but in a triplet state, $O^{\upuparrows}$, in {\it 214} systems. The latter's magnetic moment gives rise to a narrow 3D AFM doping range and to magnetic order accompanying CO (stripes). Charge-density fluctuations occur when double holes hop back and forth between $O$ atoms of the SL and neighboring $O^{2-}$ ions. Quantum oscillations in two OE cuprates result from Fermi surface reconstruction.

\end{abstract}

\maketitle

\section{INTRODUCTION}
Research on superconductivity in high-$T_c$ superconductors has been accompanied by research on the compounds' {\it normal state} from where superconductivity develops upon sufficient cooling (Bose-Einstein condensation of holes or electrons, formation of Cooper pairs). Thus, one expects similarities of the normal and superconducting states as well as differences. Many features of the normal state are not well understood: 

What causes the pseudogap (PG) and the strange-metal (SM) phase?*

What causes Fermi arcs?

What makes  doped holes or doped electrons reside in the $CuO_2$ plane?

Why does 3D antiferromagnetism collapse in lanthanum cuprates ({\it 214}) at  N\'{e}el doping \indent $p_N = 0.02$ but at $p_N = 0.08 \pm 0.03$ in oxygen enriched cuprates (OEC)?

Why is the resistivity $\rho \propto T$ in the SM phase but $\rho \propto T^2$ in the PG  phase?

Why is long-range charge order accompanied by spin order  in {\it 214} but not in   OEC?

Why does its incommensurability $q_{co}(p)$  increase with $p$ in {\it 214} but {\it decrease} in OEC ?

Why is in {\it 214} the incommensurability $q_{co}(p)$ constant above a ``watershed'' doping $\hat{p}$?

Why does the pseudogap temperature $T^*(p)$ decrease linearly with doping but discon-\indent tinuously collapse to $T^*(p^*) = 0$ at a  ``quantum  critical''  doping $p^*$?

Why is $x^* = 0.18$  in $La_{2-x}Sr_xCuO_4$, $p^* = 0.19$ in $YBa_2Cu_3O_{6+y}$ and in OEC, but $x^* = 0.23$ \indent in co-doped  $La_{2-z-x}Ln_zSr_xCuO_4, \;Ln = Nd, Eu$?

Why is short-range charge order confined to doping $0.08 < p < 0.16$  and peak at  $ \simeq 0.12$?

How are charge density fluctuations related to static charge order?

Why are quantum oscillations found in $YBa_2Cu_3O_{6+y}$ and $HgBa_2O_{4+\delta}$ but not in {\it 214}? \\

How can one begin to unravel those questions?
Concerning the {\it 214} lanthum cuprates, consensus holds on: (1) A pristine $La_2CuO_4$ crystal is a Mott insulator. 
(2) $CuO_6$ octahedra have a slight diagonal tilt to match the $CuO_2$ plane (where the basal $O^{2-}$ ions reside) with the sandwiching LaO layers (where the apical $O^{2-}$ ions reside). (3) When $La_2CuO_4$ is doped with alkaline-earth, $Ae = Sr, Ba$, beyond the N\'{e}el value $x_N$, conductivity is carried by holes.

The present approach starts with a consideration of the changes in the $CuO_2$ plane caused \underline{by one or two doped holes.}

\noindent *\footnotesize 
Non-scientific terms such as ``mysterious" and ``enigmatic'' are employed for them in the relevant literature. \normalsize 

\pagebreak

TABLE OF CONTENTS

I. Introduction 

II. Pristine and hole-doped lanthanum cuprates

III. Pseudogap and Fermi arcs

IV. Pseudogap temperature

V. Charge transport in the $CuO_2$ plane

VI. Charge transport in electron-doped cuprates

VII. Resistivity

VIII. Other hole-doped superconducting cuprates

IX. Collapse of 3D antiferromagnetism

X. Tilt of $CuO_6$ octahedra

XI. Charge order in {\it 214} lanthanum cuprates

XII. Quantum critical point of hole-doped {\it 214} cuprates

XIII. Selection of data

XIV. Collapse of the pseudogap temperature line $T^*(x)$

XV. Low-temperature crystal phases of $p$-doped {\it 214} cuprates

XVI. Short-range charge-order dome

XVII. Charge order in oxygen-doped/enriched cuprates

XVIII. Quantum critical point of oxygen-doped/enriched cuprates

XIX. Charge density fluctuations

XX. Fermi surface reconstruction and quantum oscillations

XXI. Fermi arcs revisited


XXII. Summary

Appendix A: Contraction of planar lattice constants

Appendix B: Superconducting dome

Appendix C: Commensurate superlattices


\pagebreak 

\includegraphics[width=5.7in]{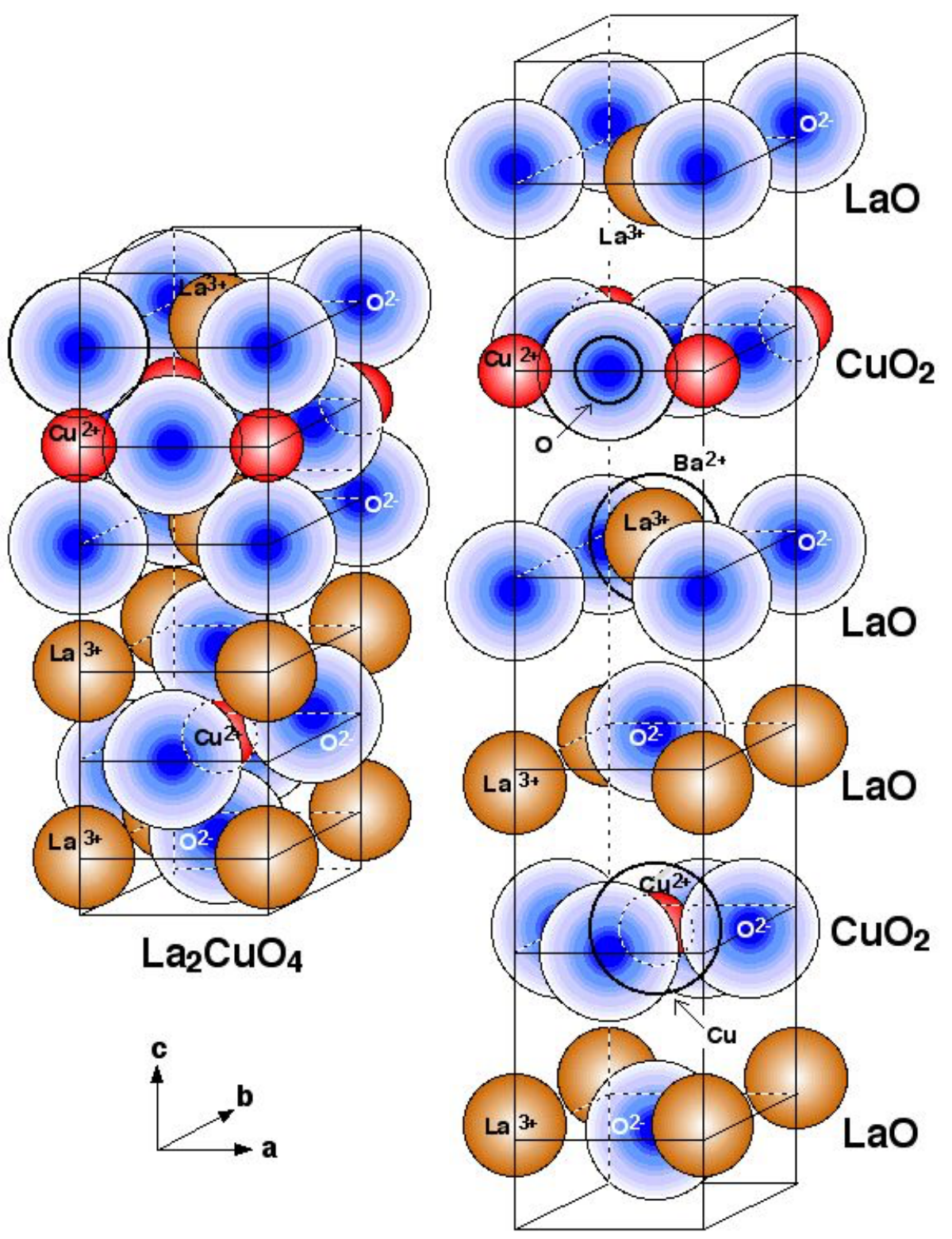}  

\noindent \footnotesize FIG. 1. Two unit cells of $La_2CuO_{4}$, staggered by half the planar lattice constants, ($\frac{a_0}{2},\frac{b_0}{2}$), in hard-sphere ion display (left), and vertically exploded (right) to better show the ion planes as noted on the side. When doped, $La^{3+}$ ions are substituted by larger-sized $Sr^{2+}$ or $Ba^{2+}$, as illustrated on the right (third plane from the top). Size and position of a lattice-defect $O$ atom that hosts two doped holes is also shown in the second plane. Likewise, a lattice-defect $Cu$ atom that hosts two doped electrons is shown in the second plane from the bottom. A $CuO_6$ octahedron can be seen in the bottom unit cell. \normalsize

\section{PRISTINE AND HOLE-DOPED LANTHANUM CUPRATES}

Pristine $La_2CuO_4$ has a lamellar crystal structure. A central $CuO_2$ plane, sandwiched between two $LaO$ layers, repeats periodically (see Fig. 1). When doped with alkaline-earth, $Ae = Sr, Ba$, substitution of $La \rightarrow La^{3+} + 3 e^-$ by $Ae \rightarrow Ae^{2+} + 2 e^-$ causes a lack of electrons in the $LaO$ layers. The lack is remedied by a transfer of electrons from the $CuO_2$ plane, leaving holes behind---the ``doped holes.'' Why are those electrons taken from the center $CuO_2$ plane instead of from within each $LaO$ layer? The reason lies with the marginal stability of $O^{2-}$ ions: Being unstable as a {\it free} closed-shell ion, $O^{2-}$ is stabilized by the crystal lattice. Both the number and charge of cations per $O^{2-}$ ion serve as a rough indication of the latter's stability in its respective environment: In each $LaO$ layer there are as many $La^{3+}$ ions as $O^{2-}$ ions, providing for good $O^{2-}$ stability, accompanied by a substantial binding energy against $O^{2-} \rightarrow O^- + e^-$ or  $O^{2-} \rightarrow O + 2e^-$ dissociation. Conversely, in the $CuO_2$ plane there are only half as many $Cu^{2+}$ ions as $O^{2-}$  ions. This renders less  stability to $O^{2-}$ ions in the $CuO_2$ plane, making them more prone to $O^{2-}$ dissociation. 

In diffraction experiments, deviations from the perfect $CuO_2$ plane, say by holes hosted at $O^-$ ions, or by double holes hosted at $O$ atoms, are sources of secondary spherical wavelets, according to Huygens principle. The doping dependence of charge-density waves and spin-density waves (``stripes'') is only compatible with double holes at $O$ atoms but \emph{not} with single holes at $O^-$ ions (see Ref. 1). This motivates

    \textbf{Assumption 1:}  \textbf{Doped holes in the} $\mathbf{CuO_2}$ \textbf{plane attach \emph{pairwise} to} $\mathbf{O^{2-}}$ \textbf{ions to form neutral}  $\mathbf{O}$ \textbf{ atoms}, $\mathbf{O^{2-} + 2e^+ \rightarrow O}$. 

\noindent The assumption is supported by the observation that $O^-$ is rarely stable in crystalline lattices.

How does doping of $LaCuO_4$ with $Ae = Sr, Ba$ affect the planar lattice constants, $a_0(x)$ and $b_0(x)$, of $La_{2-x}Ae_xCuO_4$? Although the radius $r(Ae^{2+})$ is larger than $r(La^{3+})$---see Table I below---the planar lattice constants $a_0(x)$ and $b_0(x)$ {\it decrease} linearly with doping $x$.\cite{2} The reason is the small size of the $O$ atom, about half the size of the space between $Cu^{2+}$ ions. Figure 2 shows a double-hole hosting $O$ atom at an anion site and surrounding ions in the $CuO_2$ plane at ideal-lattice sites. Relaxation of near ions occurs due to the $O$ atom's small size and its lack of net charge. The latter shifts neighboring $Cu^{2+}$ ions away from $O$ but lets neighbor $O^{2-}$ ions come closer.  
When relaxation is taken into account, the {\it effective} radius of a relaxed oxygen atom, $\tilde{O}$---based on a simulation of quasi-tetragonal lattice

\includegraphics[width=3.5in]{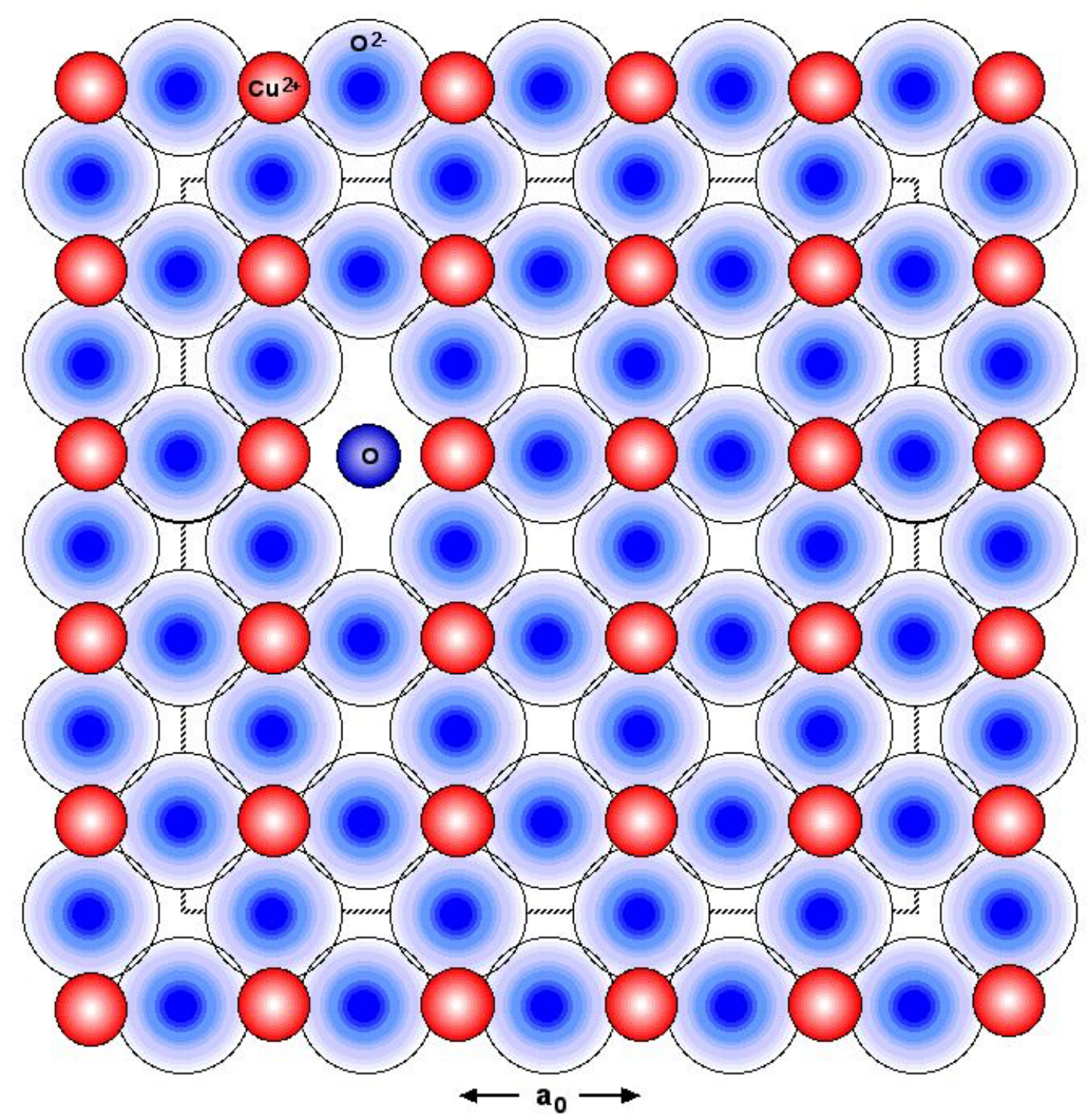}  

\noindent \footnotesize  FIG. 2. $CuO_2$ plane with double hole at $O \leftarrow O^{2-} + 2e^+$ at an anion ``b-site'' (being in line with only $O^{2-}$ ions in the $[010]$ direction. Conversely, an $O$ atom is at an anion ``$a$-site'' if in line with only $O^{2-}$ ions in the $[100]$ direction.) Ions are displayed at their ideal-crystal sites, not at relaxed positions due to displacements from size and charge effects. When relaxation is taken into account, the small size of $O$ compared to $O^{2-}$ dominates lattice spacing, resulting in lateral contraction. \normalsize  

\bigskip

\includegraphics[width=3.5in]{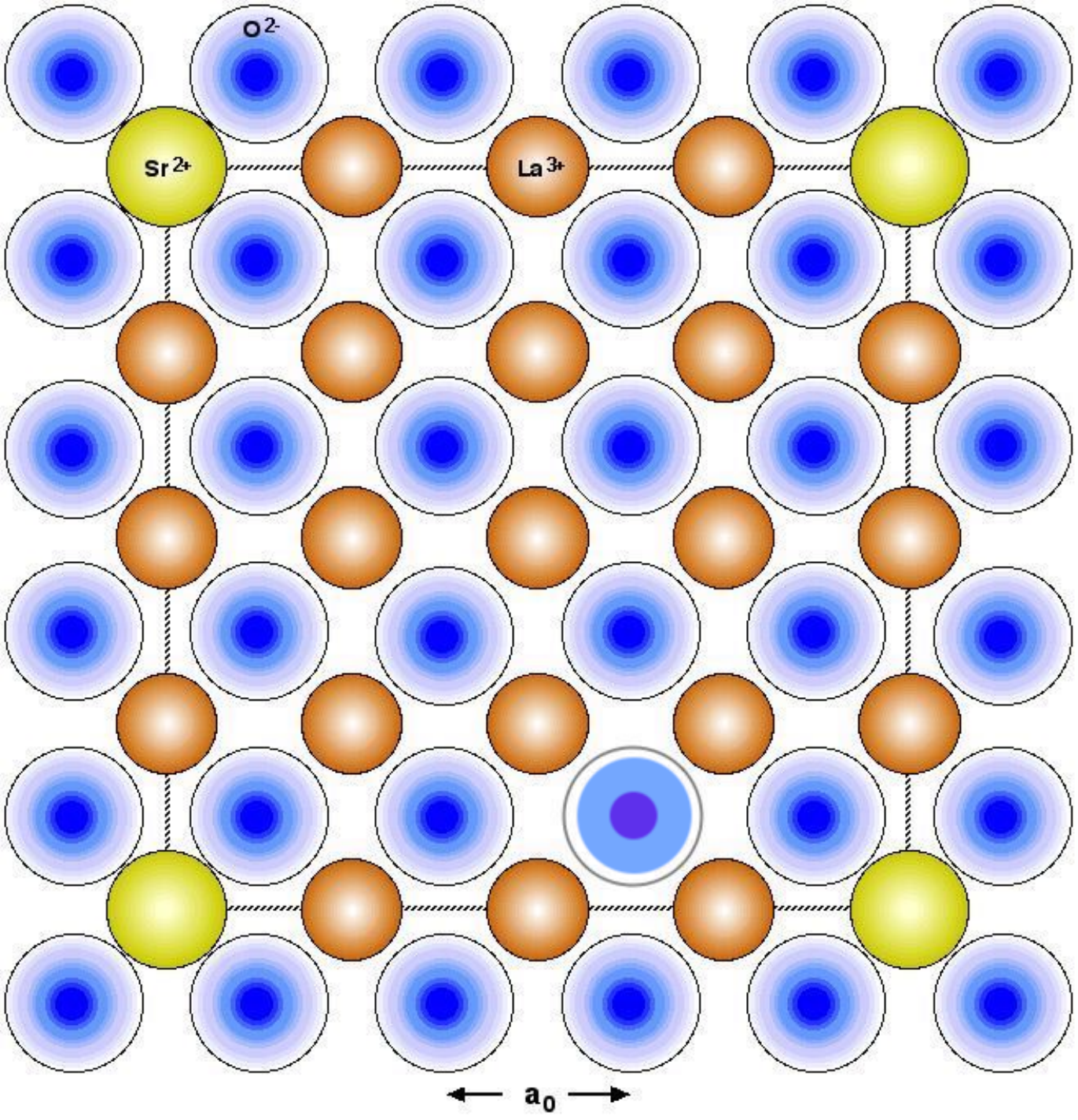}  

\noindent \footnotesize  FIG. 3. $La/SrO$ plane at doping $x = 1/8$. The size of $Sr^{2+} > La^{3+}$ affects the lattice spacing minimally. \normalsize  

\noindent  constants (see Appendix A)---is considerably larger than that of a bare oxygen atom, $O$, but still smaller than half the space between $Cu^{2+}$ ions, denoted $\overlinesegment{O^{2-}}$ (see Table I).
The net effect of the relaxation is a contraction of the planar lattice constants, $a_0(x)$ and $b_0(x)$, with increased doping $x$.\cite{2}


 \bigskip 

    \begin{table}[ht]

\begin{tabular}{|p{1.5cm}|p{1.5cm}|p{1.5cm}|p{1.5cm}|p{1.5cm}|p{1.5cm}|p{1.5cm}|}
 \hline  
$\;\;\;La^{3+}\;\;$ &$\;\;\;Sr^{2+}$&$\;\;\;Ba^{2+}$&$\;\;\;Cu^{2+}$&$\;\;\;\overlinesegment{O^{2-}}$&$\;\;\;\;\;O$&$\;\;\;\;\;\tilde{O}$\\
\hline
$\;\;\;\;1.06$&$\;\;\;\;1.12 $&$\;\;\;1.35$&$\;\;\;0.72$&$\;\;\;1.18^{*}$&$\;\;\;\;0.63^{\dagger}$&$\;\;\;\;1.09^{\ddagger}$\\ \hline 

\end{tabular}
\label{table:1}
\footnotesize \caption{Ionic and atomic radii in \r{A}.  $^*2r(\overlinesegment{O^{2-}})$
is the space between two $Cu^{2+}$ ions in $La_2CuO_4$, obtained by $ a_0 - 2r(Cu^{2+})$.
\textdagger Average of oxygen atomic radii from $0.60$ \r{A} to $0.66$ \r{A}, depending on the chemical environment in a molecule. \textdaggerdbl Effective radius of a relaxed oxygen atom, $\tilde{O}$ in $La_{2-x}Sr_xCuO_4$. } \normalsize
\end{table}

\includegraphics[width=2.7in]{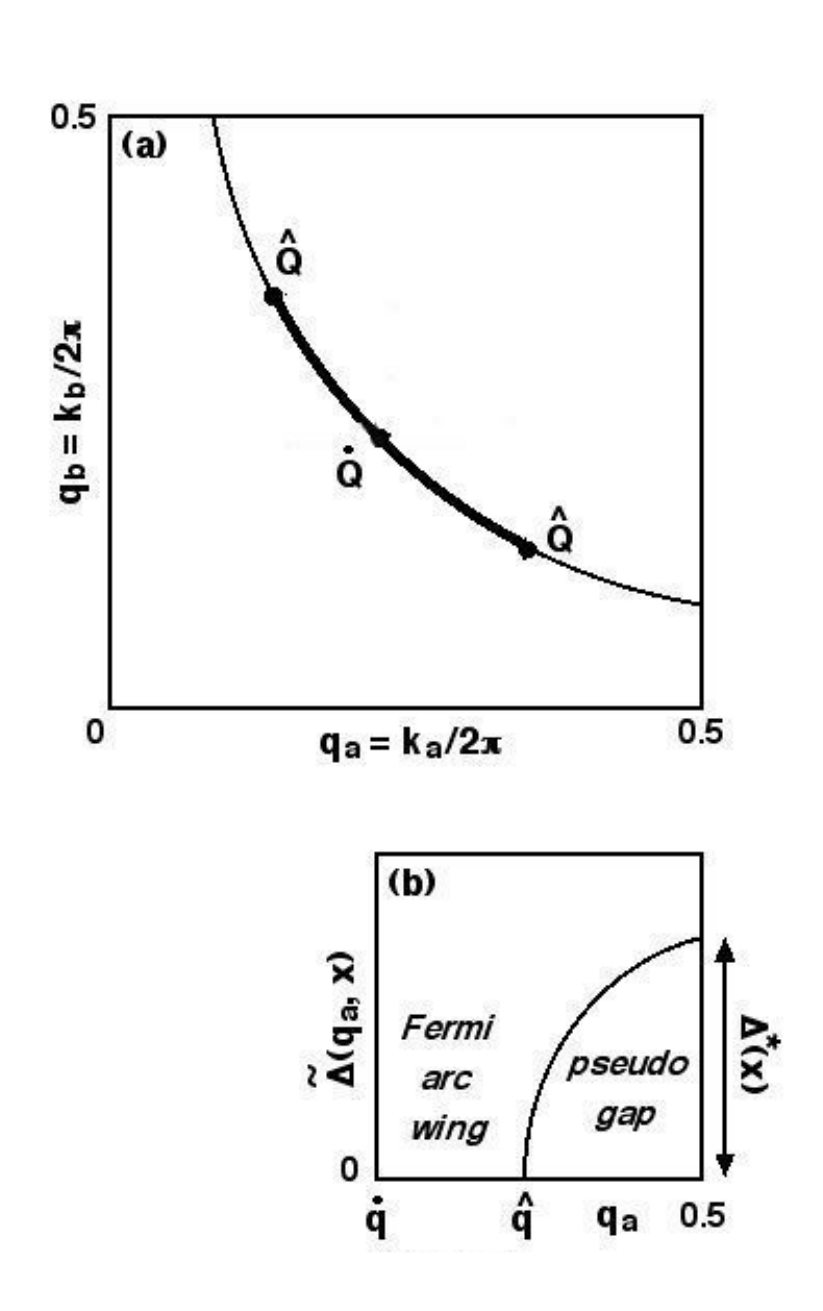}

\noindent \footnotesize FIG. 4. (a) First quadrant of the first Brillouin zone of $La_{2-x}Sr_{x}CuO_{4}$ with (approximate) Fermi arc (bold) and pseudogap states along the underlying (unoccupied) Fermi surface (thin curve).
(b) Dependence (schematic) of the pseudogap $\tilde{\Delta}(q_a, x=p)$ on a lateral $q$-space coordinate from the center of the Fermi arc, $\dot{q}$, along a Fermi-arc wing to the arc tip, $\hat{q}$, and further along the underlying Fermi surface to the boundary of the Brillouin zone at $q_a=0.5$ with maximum $\Delta^*(x)$---the anti-nodal pseudogap. \normalsize

\section{PSEUDOGAP AND FERMI ARCS}

Standard classification of solids is based on their electrical conductivity. Leaving semiconductors aside, this establishes a dichotomy of metals and insulators. In terms of quantum mechanics, metals are characterized by an occupied Fermi surface but insulators by an energy gap between a fully occupied valence band and an empty conduction band. A novel situation arises in hole-doped cuprates where {\it both} metallic and insulating properties are found in different parts of the Brillouin zone.
Pieces of a (2D) Fermi surface---called ``Fermi arcs''---extend about nodal points $\mathbf{\dot{Q}} =(\pm \dot{q}, \pm \dot{q})$ on the zone diagonals out to Fermi-arc  tips $\mathbf{\hat{Q}}$ (see Fig. 4).\cite{3} A gap function, $\tilde{\Delta}(q_j, p), \; j = a, b$, depending on reciprocal-space coordinates $q_j$ and hole doping $p$, continues from the arc tips to the boundary of the Brillouin zone at $q_j=\pm 0.5$ where the gap is maximal with the anti-nodal value $\Delta^*(p)$.

Increased hole doping extends each Fermi arc in length, with the arc tips $\mathbf{\hat{Q}}$ approaching the anti-nodal points $\mathbf{X} = (\pm 0.5, 0)$ and $\mathbf{Y} = (0, \pm 0.5)$, while the remaining energy gap shrinks.\cite{4} At zero temperature  and  a certain hole doping $p^*$---called ``quantum critical point''---the pseudogap closes as the separate Fermi arcs join to a complete (2D) Fermi surface. (Following practice of the relevant literature, the term {\it quantum critical point} is used here irrespective of it being a truely critical point, that is, the source of large quantum fluctuations, or merely an indicator of a phase transition. An assessment will be given after sufficient information.)

Due to its electron structure, $La^{3+}_2Cu^{2+}(3d^9)O^{2-}_4(2p^6)$, pristine lanthanum cuprate is in a ``half-filled'' state. The orbital of highest energy---hybridized of $Cu \; 3d_{x^2 - y^2}$ and $O \; 2p$---
is half occupied with an electron, which also means, half occupied with a hole.  By traditional band structure theory, the crystal  should be a conductor. However, this is not observed. What the band structure calculation doesn't take into account is  onsite Coulomb repulsion and superexchange coupling.
This condition causes the electrons in that quantum state to be localized, preventing them from moving freely. The result is called a ``Mott insulator.'' The spins of the half-filled electron states are ordered in an antiferromagnetic (AFM) arrangement where neighboring spins are aligned in opposite directions. This causes a lower energy level for electron localization than for delocalized electrons---hence the insulator. However, above the N\'{e}el temperature, $T > T_N \simeq 300$ K, thermal agitation delocalizes the half-filled $Cu  3d$-$O  2p$ holes, causing a transition to a metal.

Doping $La_2CuO_4$ with $Sr$ introduces holes in the $CuO_2$ planes of $La_{2-x}Sr_xCuO_4$. The holes, hosted pairwise at $O$ atoms, disturb and, at N\'{e}el doping $x_N = 0.02$, destroy 3D AFM. At very high doping, $x >0. 30$, the crystal is a metallic conductor with its $Cu  3d$-$O  2p$ holes completely delocalized, as observed with angular resolved photoemission spectroscopy (ARPES) by a closed Fermi surface 
in the first Brillouin zone. The transition from insulator to regular metal occurs at various temperatures in the doping range $x_N < x < x^* =0.18$.

\textbf{Assumption 2: An $\mathbf{O}$ atom, lacking a net charge, causes a ``\emph{demottification}'' of the ($\mathbf{Cu \;3d_{x^2 - y^2}}$-$\mathbf{O \; 2p}$) hole at both its $\mathbf{Cu^{2+}}$ neighbor ions (marked bold),}
\begin{equation}
\mathcolor{red}{Cu^{2+}} \;\; \mathcolor{blue}{O^{2-}} \;\;
\mathcolor{red}{\mathbf{Cu^{2+}}} \leftarrow \mathcolor{red}{Cu^{2+}} \Longleftarrow \mathbf{o} \Longrightarrow \mathcolor{red}{Cu^{2+}} \rightarrow  \mathcolor{red}{\mathbf{Cu^{2+}}}\ \;\; \mathcolor{blue}{O^{2-}} \;\; \mathcolor{red}{Cu^{2+}} \;
\end{equation} 
\textbf{due to reduced onsite Coulomb repulsion.} The flanking $Cu3d$-$O  2p$ holes delocalize at their hosts. However, those ``demotted'' holes are trapped between (motted) $Cu^{2+}$ ions, preventing them from moving freely.
Above a certain, doping-dependent temperature $T^*(x)$, the remaining $Cu3d$-$O  2p$ holes are {\it thermally demotted}, permitting free movement of all $Cu  3d$-$O  2p$ holes through the crystal along the $Cu$-$O$ bonds. This initiates the ``strange-metal'' phase.

\medskip

\includegraphics[width=4.8 in]{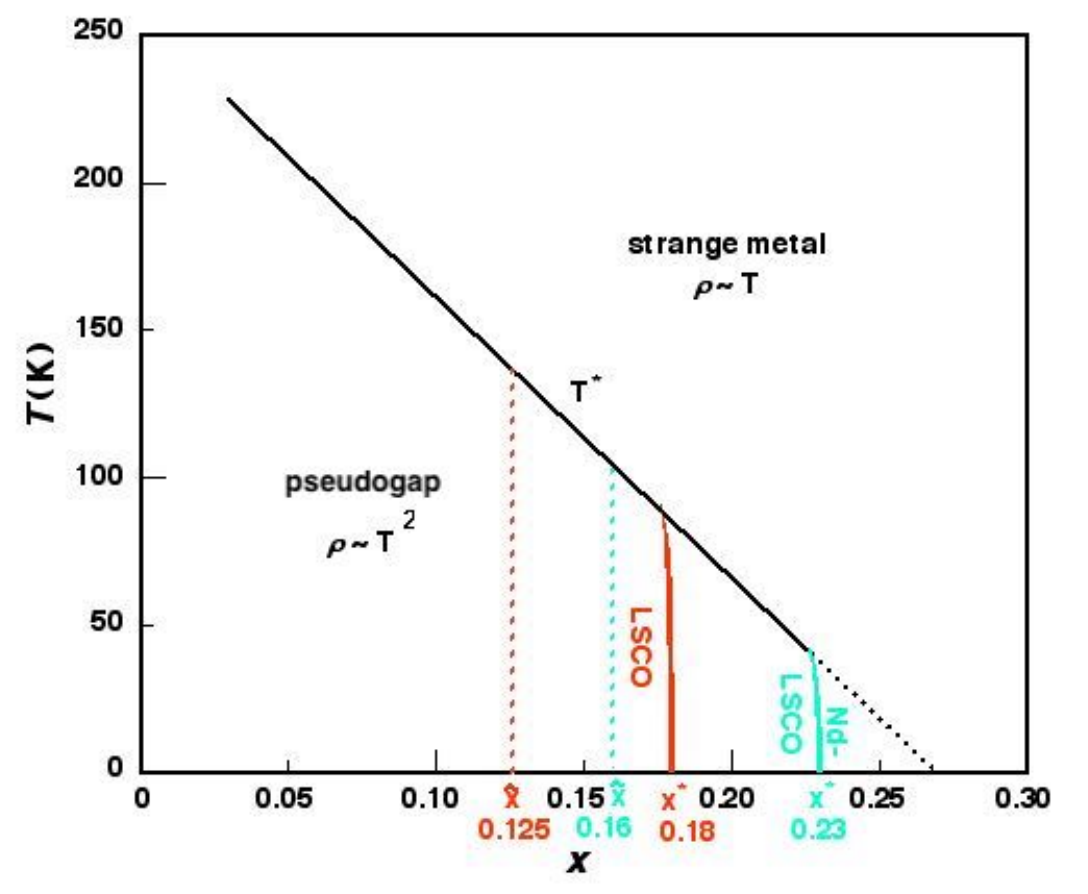} 

\noindent \footnotesize FIG. 5. Doping dependence of pseudogap temperature $T^*$ of $La_{2-x}Sr_xCuO_4$ and $La_{1.6-x}Nd_{0.4}Sr_xCuO_4$ (solid line) when superconductivity is suppressed by a magnetic field.
\normalsize

\section{PSEUDOGAP TEMPERATURE}
A prominent feature in the temperature $vs.$ doping
phase diagram of $La_{2-x}Sr_xCuO_4$ and co-doped $La_{2-z-x}Ln_zSr_xCuO_4 \; (Ln = Nd, Eu; z = 0.4, 0.2)$ is a straight line,
\begin{equation}
    T^*(x) = T^*_0 - sx \;, \;\;\; (s>0) \;,
\end{equation}
empirically observed and called ``pseudogap temperature,'' that slants essentially from the N\'{e}el temperature of the host crystal, $T^*_0 \equiv T^*(0) \approx T_N$, on the left, down until it drops abruptly at doping $x^*$---the putative quantum critical point  (see Fig. 5).\cite{5} 

 \textbf{The pseudogap temperature $\mathbf{T^*(x)}$ is the temperature necessary to demott the $\mathbf{Cu3d}$-$\mathbf{O  2p}$ holes that are not demotted from doping. At  $\mathbf{x^*}$ \emph{all} $\mathbf{Cu3d}$-$\mathbf{O  2p}$ holes are demotted by doping.} Then no \emph{thermal} demotting is necessary, $T^*(x^*) = 0$.

An extension of $T^*(x)$ beyond $x^*$ (dotted in Fig. 5) intercepts the abscissa, $T = 0$---accidentally coincident with the high-doping end of superconductivity but irrelevant for the latter. The $T^*$ line separates regions of different metallicity. Below $T^*(x)$ is the pseudogap regime with resistivity $\rho(T) \propto T^2$, familiar from common metals. Above $T^*(x)$ is the region of \emph{strange} metallicity with $\rho(T) \propto T$. In the very overdoped region, $x > 0.3$, resistivity $\rho(T) \propto T^2$ is observed again---here attributed to Fermi-liquid behavior. The pseudogap temperature $T_0^*(x)$ 
can be regarded as a temperature \emph{threshold}---lowered with increased doping $x$---above which strange metallicity becomes possible.


\section{CHARGE TRANSPORT IN THE $\mathbf{CuO_2}$ PLANE}

The square-dependence of the resistivity on temperature---$\rho(T) \propto T^2$ in the {\it pseudogap state} of $La_{2-x}Sr_xCuO_4$---does not imply that its charge carriers establish a Fermi liquid, as arises from electron-electron scattering in common metals. Emerging from an AFM Mott insulator at $x > 0.02$ as an ionic crystal, charge transport can here rather be viewed in terms of successive hopping of charges from $O$ atoms (Fig. 2) to neighbor $O^{2-}$ ions. 

\textbf{Assumption 3: In the pseudogap phase, a double hole hops in the $\mathbf{CuO_2}$ plane  from an $\mathbf{O}$ atom} \textbf{to a nearest neighbor} $\mathbf{O^{2-}}$ \textbf{ion}, \textbf{\emph{diagonal} to the $\mathbf{Cu}$-$\mathbf{O}$ bonds.} 

\noindent The elementary process is a $(\pm a_0/2, \pm b_0/2)$ hop that creates an $O$ atom at the new position while reverting the previous $O$ atom to an $O^{2-}$ ion, 
\begin{equation}
O^{2-} \leftarrow O-2e^+ \xLongrightarrow{2e^+ hop\;} 2e^+ + O^{2-} \rightarrow O \;.
\end{equation} 
At the quantum critical point $x^*$, ($T=0$), the carrier density $n$, measured by the Hall number $n_H \simeq n$, jumps from $n_H \simeq x$ at $x < x^*$ to $n_H \simeq 1+x$ at $x > x^*$.\cite{5}  The observation stimulates

\textbf{Assumption 4: Above the pseudogap temperature, $\mathbf{T > T^*(x)}$, all  $\mathbf{Cu3d}$-$\mathbf{O  2p}$ holes are {\it \textbf{demotted}}}  \textbf{(by doping and thermally) so that they can advance, by hops of \emph{one lattice-constant length}
along the $\mathbf{Cu}$-$\mathbf{O}$ bonds through the crystal.}

It is important to treat charge transport from one unit cell to the axial neighbor cell, that is, by an advancement of \emph{one lattice constant}, after which the process repeats periodically. In the pseudogap regime, progress of a double hole by one lattice constant, say along the positive $a$-direction, requires \emph{two} hops, namely by $( a_0/2, \pm b_0/2)$ succeeded by $( a_0/2, \mp b_0/2)$.
This is in contrast to the strange metal regime above the $T^*$ line where $Cu3d$-$O  2p$ holes advance by a lattice constant with {\it one}  hop.

\textbf{Assumption 5:}   \textbf{Increased doping} $\mathbf{x}$ \textbf{lowers the  energy gap,} $\Delta^*(x)= k_BT^*(x),$ \textbf{to activate lattice-constant hops of} $\mathbf{Cu3d}$-$\mathbf{O  2p}$   \textbf{holes 
along} $\mathbf{Cu}$-$\mathbf{O}$ \textbf{bonds,  as less such
holes need to be thermally \emph{demotted}.} (Here, $k_B$ is the Boltzmann constant.)
The assumption explains, on a qualitative level, the decrease of the pseudogap temperature $T^*(x)$ with increasing doping [Eq. (2) and Fig. 5].




\section{CHARGE TRANSPORT IN ELECTRON-DOPED CUPRATES}

When lanthanide cuprates are doped with cerium, $Ln_{2-x}Ce_xCuO_4, \; Ln = La, Nd, Pr$,  some of the lanthanide, $Ln \rightarrow Ln^{3+} + 3 e^-$, is replaced by $Ce \rightarrow Ce^{4+} + 4 e^-$, leaving each an extra electron behind.

\textbf{Assumption 6: In electron-doped cuprates, copper ions take up extra electrons,}
$\mathbf{Cu^{2+} + 2 e^- \rightarrow Cu}$, \textbf{reducing them to copper atoms} (see Fig. 8 below)\textbf{.}

Like the $Cu 3d$-$O2p$ holes in the strange-metal phase, an electron pair hosted at a $Cu$ atom can advance along the lines of  $Cu$-$O$ bonds with hops of lattice-constant length,
\begin{equation}
Cu^{2+} \leftarrow Cu - 2e^- \xLongrightarrow{2e^- hop\;} 2e^- + Cu^{2+} \rightarrow Cu \;.
\end{equation} 
Assumption 6 would qualitatively explain the {\it strange-metal} behavior of $Ln_{2-x}Ce_xCuO_4$, with resistivity $\rho \propto T$, for doping beyond the AFM regime, $x_N \approx 0.09$, as experimentally observed.\cite{6}

\section{RESISTIVITY}

In the pseudogap state the {\it planar} resistivity can be divided into three contributions,
\begin{equation}
\rho_{pg}(x, T) = \rho_0(x) + \rho_{ut}(x, T) + A_2T^2 \;. 
\end{equation}
Here, $\rho_0(x)$ is the residual resistivity, $\rho_{ut}(x, T)$ is a low-temperature {\it upturn}, and $A_2$ is the coefficient of the observed quadratic temperature dependence.\cite{7} It has been proposed that the upturn $\rho_{ut}(x,T)$---successfully modeled by a logarithmic temperature dependence---is caused by a loss of carrier density $n$.\cite{5}

By the Drude model, electric resistivity depends on the effective mass $m^*$, the carrier density $n$ and the scattering rate $1/\tau$,  
\begin{equation}
\rho = \frac{m^*}{n e^2} \; \frac{1}{\tau}  \; , 
\end{equation}
with the elementary charge $e$.
The Drude formula shows no {\it explicit} temperature dependence of the resistivity. The latter is attributed to the temperature dependence of the scattering rate, $\frac{1}{\tau}(T)$.
It has been noted that the scattering rate of the Drude model, Eq. (6), is limited by quantum theory,
 \begin{equation}
    \frac{1}{\tau} = \frac{k_B}{\hbar} T \;,
\end{equation}
called the ``Planckian limit'' of dissipation.\cite{8,9} It is the largest possible scattering rate---corresponding to the shortest possible relaxation time $\tau$---at a given temperature $T$. Here, $\hbar = h/(2\pi)$ denotes the reduced Planck constant. The concept of the Planckian limit has been corroborated by successful modeling of  strange-metal resistivity,\cite{9}
\begin{equation}
\rho_{sm}(x, T) = \rho_0(x) + A_1T \;, 
\end{equation}
of lanthanum cuprates and other high-$T_c$ superconductors using $1/\tau = \alpha_{PL} (k_B/\hbar) T$ with $\alpha_{PL} \approx 1$. Here, $A_1$ is the coefficient of the observed linear temperature dependence.

In either phase—--be it pseudogap or strange metal—--the charge carriers are scattered at \emph{each hop} by the extant scattering rate. For the strange-metal regime, the successful modeling of $\rho_{sm}(T,x)$ implies this to be the Planckian limit, Eq. (7). As a $Cu3d$-$O  2p$ hole jumps from demotted $Cu^{2+}$ ion to neighboring demotted $Cu^{2+}$ ion, it is scattered \emph{once} with each lattice-constant leap. In contrast, the \emph{zig-zag} hops of an $O$ atom's double hole, necessary for its advancement by one lattice constant, invoke \emph{two} subsequent scatterings. Assuming that their rate is also Planckian limited renders the pseudogap resistivity as
\begin{equation}
\rho_{pg} \propto (1/\tau)^2 \propto T^2 \;,
\end{equation} 
which would qualitatively explain the quadratic temperature dependence of the resistivity in the pseudogap state.



\section{OTHER HOLE-DOPED SUPERCONDUCTING CUPRATES}
Besides the lanthanum cuprates considered so far---called, for short, {\it ``214''} systems, as they are $La_\mathbf{2}Cu_\mathbf{1}O_\mathbf{4}$ crystals doped with $Ae = Sr, Ba$ and possibly co-doped with $Ln = Nd, Eu$---strange metallicity also occurs in oxygen-\emph{doped} $YBa_2Cu_3O_{6+y}$ and oxygen-{\it enriched} $LaCuO_{4+\delta}$, $HgBa_2CuO_{4+\delta}$, $Bi_2Sr_2CuO_{6+\delta}$, and $Bi_2Sr_2CaCu_2O_{8+\delta}$. Not all of the doped  or enriched  oxygen is implemented in the $CuO_2$ planes, though, leaving $p < y $ and $p \le \delta$, respectively. For instance, in $YBa_2Cu_3O_{6+y}$ extra oxygen is used to fill oxygen chains in a different plane (see Fig. 6). In $HgBa_2CuO_{4+\delta}$ some oxygen settles at interstitial sites in the $HgO$ layer (see Fig. 7).\cite{10,11,12} The enrichment process---say, by exposing the host crystal to a hot oxygen atmosphere---permits only a few direct assessments of $\delta$, such as thermogravimetric measurements or iodometry.\cite{13} One indirect determination assumes an analogy with the doping dependence of the superconducting transition temperature, $T_c(p)$, the so-called ``superconducting dome'' of $La_{2-x}Sr_xCuO_4$---which is a \emph{stoichiometric} compound. Based on an empirical formula for the superconducting dome,\cite{14} the hole density in the $CuO_2$ plane is given in terms of the observed $T_c$ and its maximum, $\wideparen{T}_c$ (see Appendix B). Related expressions for $p$ rely on empirical correlations with the lattice constant $c_0$ of $YBa_2Cu_3O_{6+y}$, perpendicular to the $CuO_2$ plane, or of $Bi\; 5d$ core-levels.\cite{15,16} Such $p$ values have been widely used to characterize  the doping dependence of charge order incommensurability $q_{co}(p)$ and of other properties in the pseudogap phase of oxygen-enriched cuprates. 

An immediate question concerns the position of the enriched oxygen in the host lattice. Based on the observed strange metallicity and superconductivity of those crystals, it is reasonable to assume that the extra oxygen enters the $CuO_2$ planes of the host crystal. But where exactly? One possibility that comes to mind would be for the extra oxygen atom to reside at an \emph{interstitial} site in the gap between four $O^{2-}$ ions. However, neither zig-zag hops 

\includegraphics[width=3.3in]{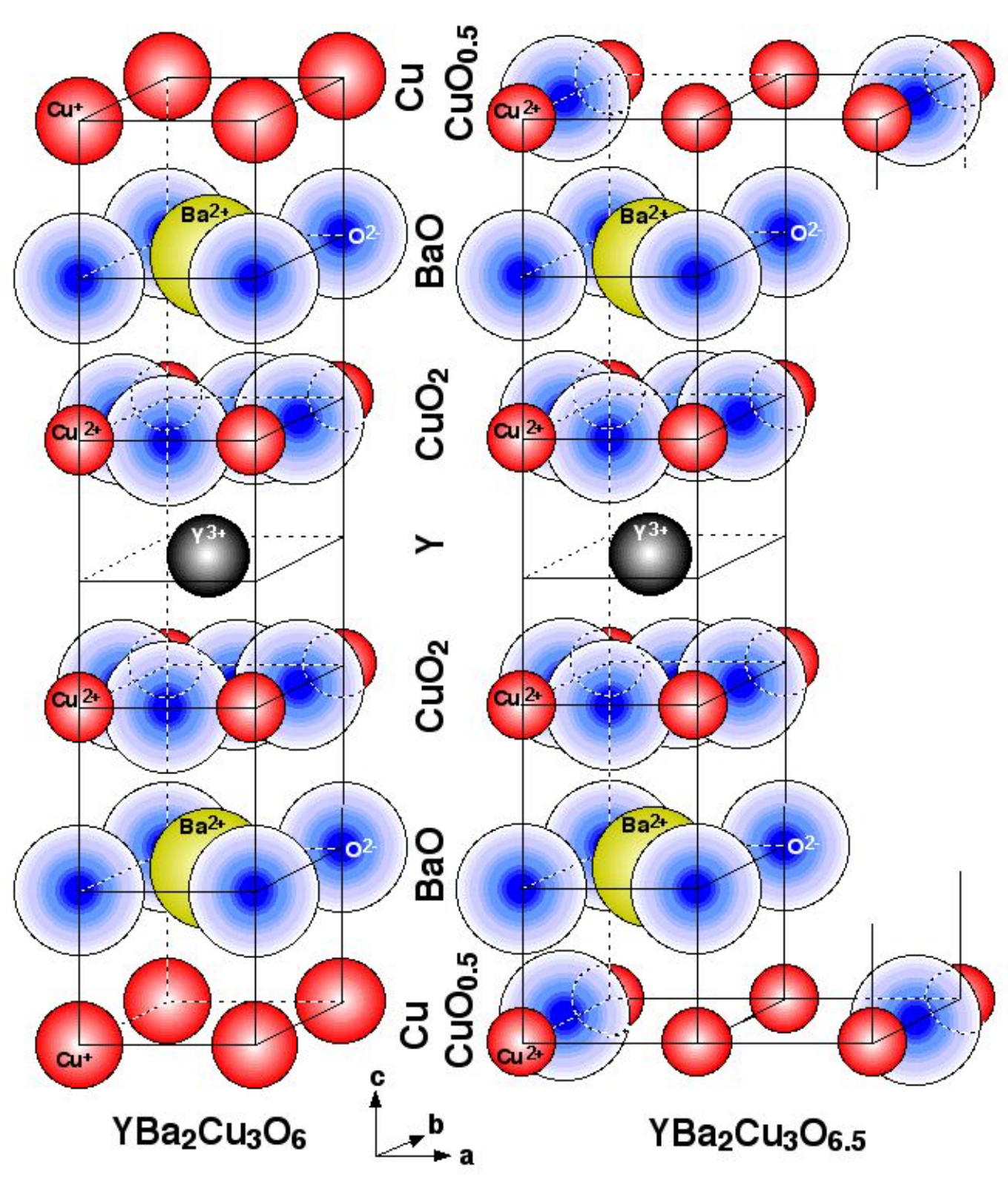}  

\noindent \footnotesize  FIG. 6.  Unit cell (exploded along $c$) of $YBa_2Cu_3O_6$ (left) and of $YBa_2Cu_3O_{6.5}$  with ortho-II ordered $CuO$ chains along the $b$ direction in the top and bottom plane (right). \normalsize

\bigskip

\includegraphics[width=2.2in]{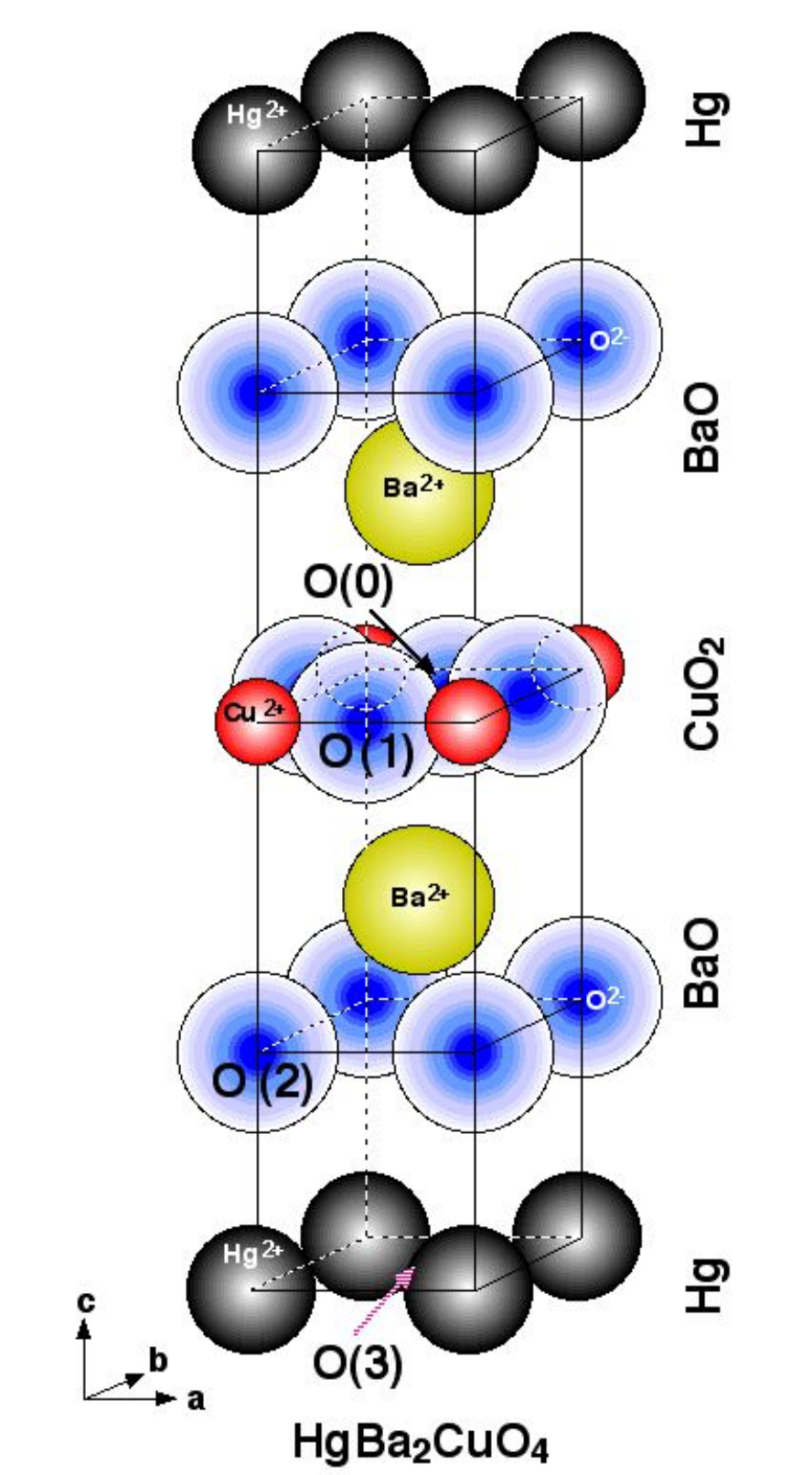}  

\noindent \footnotesize  FIG. 7.  Unit cell (exploded along $c$) of $HgBa_2Cu_3O_{4+\delta}$ with possible sites $O(0) - O(3)$ for enriched oxygen atoms. \normalsize

\pagebreak

\noindent nor lattice-constant hops of double holes in concert with advancing interstitial $O$ atoms are feasible. This {\it excludes} the possibility of charge transport by interstitial oxygen atoms.

\textbf{Assumption 7: Adoption of \emph{enriching} $\mathbf{O}$ atoms by cuprates gives rise to $\mathbf{O^{2-}}$ vacancies in the} $\mathbf{CuO_2}$ \textbf{plane, to be occupied by the enriching} $\mathbf{O}$ \textbf{atoms themselves.} 

The result are lattice-defect $O$ atoms on anion sites of the $CuO_2$ plane, very similar to those of the hole-doped {\it 214} systems (see Figs. 2, 8). There are, however, important differences where oxygen-enriched cuprates deviate from the stoichiometric {\it 214} systems:

(1) Different ranges of antiferromagnetism, with collapse of 3D AFM at N\'{e}el doping $p_N = 0.02$ in the {\it 214} systems but at $p_N = 0.08 \pm 0.03$ in various oxygen enriched cuprates. 

(2) Contrary dependence of incommensurability of charge-order stripes on hole doping, $q_{co}(p)$: In the {\it 214} systems, $q_{co}(x)$ rises in a square-root fashion from $q_{co}(0.02) = 0$ until leveling off to $q_{co}(0.125) = 0.235 \; r.l.u.$ for $La_{2-x}Sr_xCuO_4$ but to $q_{co}(0.16) = 0.266 \; r.l.u.$ for $La_{1.6-x}Nd_{0.4}Sr_xCuO_4$ (see Fig. 11 below). By contrast, oxygen enriched cuprates emerge from AFM with their highest incommensurabiliy, $q_{co}(p_N) \approx 0.30 \; r.l.u.$ (reciprocal lattice units) and drop moderately off to $q_{co}(0.18) \approx 0.23 \; r.l.u.$ (see Fig. 21 below).

(3) Accompaniment of charge-order stripes by (magnetic) spin-order stripes with incommensurabiltity $q_{so}(p) = \frac{1}{2} q_{co}(p)$ in the {\it 214} family but \emph{no} spin-order stripes in oxygen-enriched cuprates. The differences (1, 2, 3) have been known for a long time but were never satisfactorily explained. They are the consequence of the following two assumptions:

\textbf{Assumption 8:  The oxygen atom on the anion site of ``214'' cuprates, hosting a double hole, is in a \emph{triplet} state with spin quantum number} $\mathbf{S=1}$, here denoted by $ O^{\upuparrows}$ (marked with two spin arrows in Fig. 8). It is akin to the ground state of a {\it free} oxygen atom where the electrons in the $2p^4$ subshell are arranged, in agreement with Hund's rule of maximum multiplicity, with their spin as $[\uparrow\downarrow]$ $[\uparrow]$ $[\uparrow]$. This is in contrast to

\textbf{Assumption 9:  The oxygen atom on the anion site of oxygen-enriched cuprates, is in a \emph{singlet} state with spin quantum number} $\mathbf{S=0}$, here denoted as $O^{\uparrow \downarrow}$. It is similar to the {\it excited} state of a free oxygen atom with spins $[\uparrow\downarrow]$ $[\uparrow]$ $[\downarrow]$.

Assumptions 8 and 9 are not to be confused with the well-known Zhang-Rice singlet\cite{17}  where $Cu$-$O$ hybridization strongly binds a hole on each square of $O^{2-}$ ions to the central $Cu^{2+}$ ion to form a local singlet.

The ionic environment of an $O^{\uparrow \downarrow}$ ion on an anion site of oxygen-enriched cuprates is comparable to that of $O^{\upuparrows}$ in the lanthanum cuprates (both shown in Fig. 8). However, the net spin of the corresponding oxygen atoms differs. Accordingly, electrical properties of these two groups are similar but magnetic properties are very different. Thus, charge transport in the
 pseudogap phase of in oxygen-enriched cuprates is comparable to that in the stoichiometric {\it 214} cuprates, that is, by zig-zag hops of double holes. Properties affected by magnetic interaction are discussed next.

 \medskip

\includegraphics[width=5.6in]{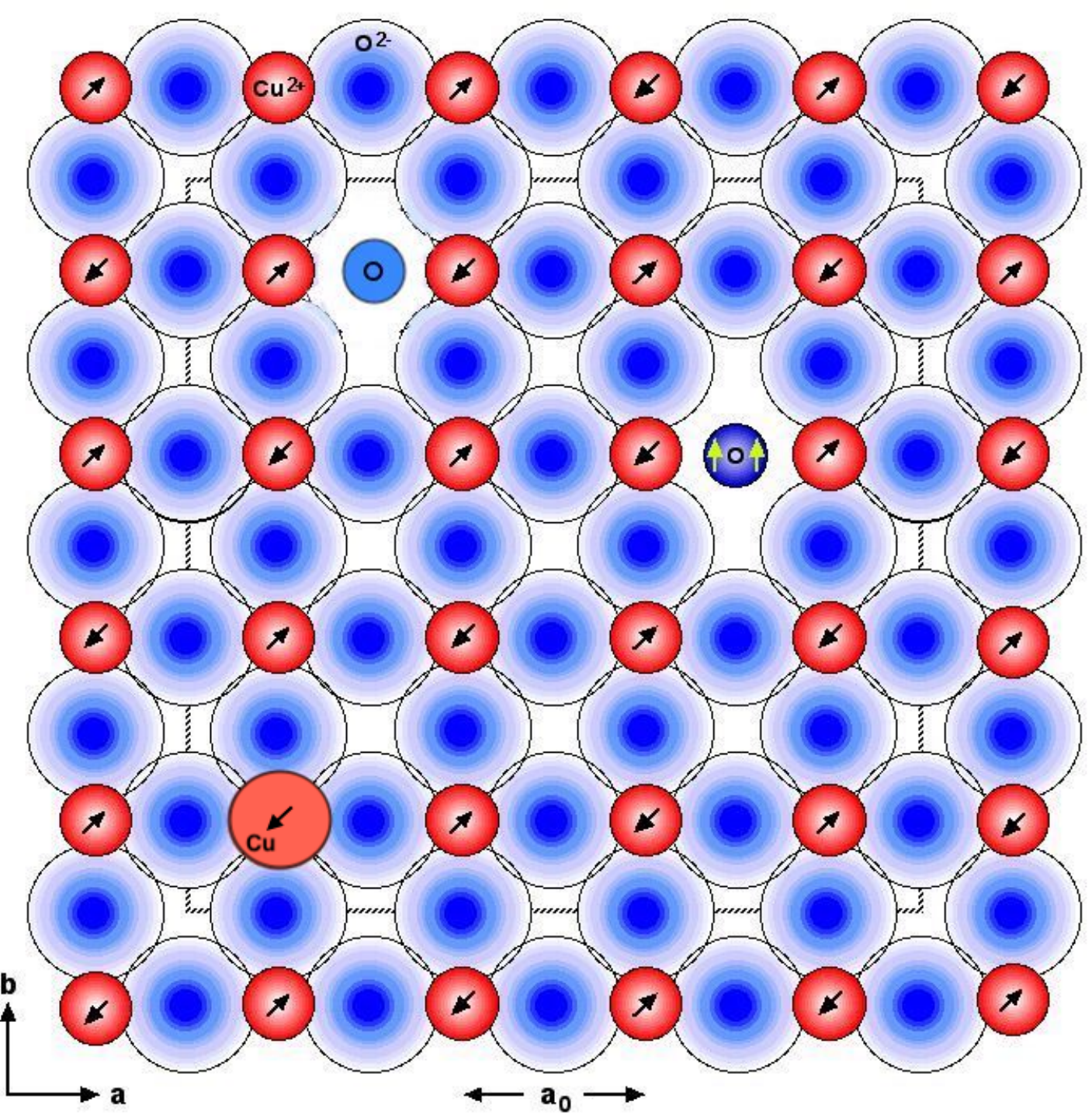}

\noindent \footnotesize FIG. 8. Non-relaxed $CuO_2$ plane with a hole pair residing (i) in an $^{\uparrow \downarrow}O$ atom (near top left) and (ii) in an $^{\uparrow \uparrow}O$ atom (on the right) at anion lattice ``b-sites.'' (iii) Electron pair residing in a $^{\uparrow}Cu$ atom on a cation lattice site (near bottom left). Arrows represent magnetic moments. The lattice-defect $Cu$ atom on the cation site is depicted with the standard atomic radius $r(Cu) = 1.28$ \AA, taken from the nearest-neighbor distance in face-centered cubic ({\it fcc}) metallic copper. It is not clear what its actual size in $Ln_{2-x}Ce_xCuO_4$ is. Nevertheless, the figure shows that the electron pair, residing on the $Cu$ atom, can by Eq. (4), travel {\it axially}, with hops of lattice-constant length, $a_0$ or $b_0$. \normalsize

\section{COLLAPSE OF ANTIFERROMAGNETISM}

Antiferromagnetism is the least stable in the {\it 214} family, collapsing at a small N\'{e}el doping, $x_N = 0.02$. In oxygen-enriched cuprates, AFM is more stable, with collapse in a doping range that extends from $p_N = 0.05$ for $HgBa_2CuO_{4+\delta}$ to $p_N = 0.11$ for $Bi_2Sr_{2-x}La_xCuO_{6+\delta}$.\cite{12,13,18} \emph{Electron}-doped cuprates also show more AFM stability than the {\it 214} systems, with $x_N \approx 0.09$ for $Ln_{2-x}Ce_xCuO_4, \; Ln = La, Nd, Pr$.\cite{6} 

A qualitative explanation of the different AFM collapses in the three cases can be inferred from Fig. 8: 
(i) The large magnetic dipole of the $O^{\upuparrows}$ atom on the anion site in {\it 214} systems is a major {\it magnetic} disturbance of the AFM of the host $^\uparrow Cu^{2+}$ ions, resulting in little AFM stability (small N\'{e}el doping $x_N = 0.02$). 
(ii) In oxygen-enriched cuprates,  an 
$O^{\uparrow \downarrow}$ atom ($S=0$) resides instead on the anion site. Lacking a net magnetic moment, it does not cause a direct magnetic disturbance. Differing in charge and size from $O^{2-}$, it merely displaces its surrounding $^\uparrow Cu^{2+}$ ions (with their magnetic moments) slightly from their lattice site. This amounts to an indirect, small magnetic disturbance, keeping considerable AFM stability in place ($p_N = 0.08 \pm 0.03$). 
(iii) The $^{\uparrow}Cu$ atom on a cation site, arising from electron doping, has the same spin quantum number, $S = 1/2$, as the host $^\uparrow Cu^{2+}$ ions. It continues the copper ions' AFM but, differing in charge and size from its surrounding $^\uparrow Cu^{2+}$ ions, pushes them slightly from their lattice sites. The result is again a small magnetic disturbance, retaining substantial AFM stability.
\textbf{The low AFM stability of the \emph{214} systems and the larger AFM stability of oxygen-doped/enriched cuprates are a direct consequence of the spin of their double-hole containing oxygen atoms,} $\mathbf{O^{\uparrow \uparrow}\;}$ \textbf{and} $\mathbf{\;O^{\uparrow \downarrow},}$ \textbf{respectively.}

\section{TILT of $\mathbf{CuO_6}$ OCTAHEDRA}
The low-temperature ground state of pristine $LaCuO_4$ has a monoclinic crystal structure.\cite{19} It can be regarded, in good approximation, as {\it orthorhombic}, having orthogonal lattice constants $a_0 < b_0 < c_0$, {\it but} with $CuO_6$ octahedra  {\it slightly tilted} diagonally from the $c$ axis orientation toward the planar axes. One and a half $CuO_6$ octahedra are shown in Fig. 9.  The base of each $CuO_6$ octahedron is formed by a quartet of $O^{2-}$ ions in the $CuO_2$ plane. The apical $O^{2-}$ ions are above and beneath the $Cu^{2+}$ ion. Why is the tilt diagonal? 
The $a_0 \times b_0$ rectangle ($a_0 \ne b_0$), with the octahedron's four $O^{2-}$ base ions at its sides, is {\it equally} divided by the diagonal line which, passing through the center $Cu^{2+}$ ion, serves as the octahedral tilt axis in the $[110]$ orientation.  Accordingly, two of the four $O^{2-}$ ions of each octahedron base tilt below the $CuO_2$ plane while the other two tilt above, denoted in subsequent configuration figures as $_\downarrow O^{2-}$ and $_\uparrow O^{2-}$, respectively. An example can be seen at the top left corner of Fig. 10 with $O^{2-}$ ions at positions ($\mathbf{\downarrow}\frac{-a_0}{2},0)$ and $(0, \mathbf{\downarrow}\frac{-b_0}{2})$ but  ($\mathbf{\uparrow}\frac{a_0}{2},0)$ and $(0, \mathbf{\uparrow}\frac{b_0}{2})$ relative to the center $Cu^{2+}$ ion and $CuO_2$ plane.  The resulting octahedron tilt is in the $[\overline{1},\overline{1}]$ direction, indicated by a slanted subarrow as $_{\swarrow}Cu^{2+}$. Analogously, an  octahedron tilt in the $[1,1]$ direction is indicated by a slanted subarrow as $_{\nearrow}Cu^{2+}$.

\medskip

\includegraphics[width= 2.18in]{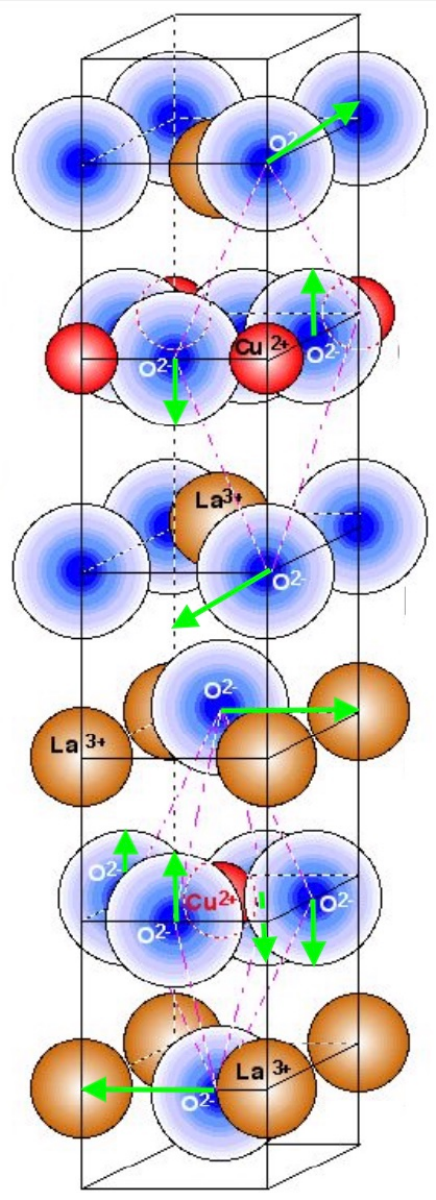} 

\noindent \footnotesize FIG. 9. Two unit cells of pristine $LaCuO_4$ (exploded along $c$), staggered by half the planar lattice constants, with a full and a half $CuO_6$ octahedron, indicated by dashed lines (bottom and top unit cell, respectively). Green arrows show {\it exaggerated} the displacement of $O^{2-}$ ions to tilt the $CuO_6$ octahedra diagonally. \normalsize 

\pagebreak

The tilt of the $CuO_6$ octahedra of $La_2CuO_4$ (Fig. 10) is repetitive in both $[100]$ and $[010]$ directions with a period $D_{100}^{010} = 2a_0$ (in quasi-tetragonal approximation) or  $D_{110} = 2\sqrt{2} a_0$ along $[110]$,  corresponding to a wave number $q = 1/D_{110} = \sqrt{2}/4 a_0^{-1} \simeq 0.354$ $r.l.u.$ (reciprocal lattice units). 
Neighboring $CuO_6$ octahedra (along the $Cu$-$O$ bonds) tilt past each other, making effective use of space availability.
We call the $CuO_6$ tilt pattern   ``host tilt.''

\medskip

\noindent $\mathcolor{blue}{\;\;\;\;\;\;\;\;\;\;\;\; _\uparrow O^{2-} \;\;\;\;\;\;\;\;\;\;\;\;\;\;_\downarrow O^{2-} \;\;\;\;\;\;\;\;\;\;\;\;\;\;_\uparrow O^{2-} \;\;\;\;\;\;\;\;\;\;\;\;\;\;\;_\downarrow O^{2-} \;\;\;\;\;\;\;\;\;\;\;\;\;\;\;_\uparrow O^{2-} \;\;\;\;\;\;\;\;\;\;\;\;\;\;\;_\downarrow O^{2-} \;\;\;\;\;\;\;\;\;\;\;\;\;\;\;_\uparrow O^{2-}}$

\noindent $ \mathcolor{blue}{_\downarrow O^{2-}} \; \mathcolor{red}{_{\swarrow}Cu^{2+}} \;\; \mathcolor{blue}{_\uparrow O^{2-}} \;\mathcolor{red}{_{\nearrow}  Cu^{2+}} \;\;\mathcolor{blue}{_\downarrow O^{2-}} \;\mathcolor{red}{_{\swarrow} {Cu^{2+}}} \;\;\mathcolor{blue}{_\uparrow O^{2-}}\mathcolor{blue}{ \;\mathcolor{red}{_{\nearrow}  {Cu^{2+}}} \;\;_\downarrow O^{2-}} \;\mathcolor{red}{_{\swarrow}  {Cu^{2+}}} \;\;\mathcolor{blue}{_\uparrow O^{2-}} \;\mathcolor{red}{_{\nearrow}  {Cu^{2+}}} \;\;\mathcolor{blue}{_\downarrow O^{2-}} \;\mathcolor{red}{_{\swarrow} {Cu^{2+}}} \;\;\mathcolor{blue}{_\uparrow O^{2-}} $ 

\noindent $\mathcolor{blue}{\;\;\;\;\;\;\;\;\;\;\;\; _\downarrow O^{2-} \;\;\;\;\;\;\;\;\;\;\;\;\;\;_\uparrow O^{2-} \;\;\;\;\;\;\;\;\;\;\;\;\;\;_\downarrow O^{2-} \;\;\;\;\;\;\;\;\;\;\;\;\;\;\;_\uparrow O^{2-} \;\;\;\;\;\;\;\;\;\;\;\;\;\;\;_\downarrow O^{2-} \;\;\;\;\;\;\;\;\;\;\;\;\;\;\;_\uparrow O^{2-} \;\;\;\;\;\;\;\;\;\;\;\;\;\;\;_\downarrow O^{2-}}$

\noindent $ \mathcolor{blue}{_\uparrow O^{2-}} \; \mathcolor{red}{_{\nearrow}Cu^{2+}} \;\; \mathcolor{blue}{_\downarrow O^{2-}} \;\mathcolor{red}{_{\swarrow}  Cu^{2+}} \;\;\mathcolor{blue}{_\uparrow O^{2-}} \;\mathcolor{red}{_{\nearrow} {Cu^{2+}}} \;\;\mathcolor{blue}{_\downarrow O^{2-}} \;\mathcolor{red}{_{\swarrow}  {Cu^{2+}}} \;\;\mathcolor{blue}{_\uparrow O^{2-}} \;\mathcolor{red}{_{\nearrow}  {Cu^{2+}}} \;\;\mathcolor{blue}{_\downarrow O^{2-}} \;\mathcolor{red}{_{\swarrow}  {Cu^{2+}}} \;\;\mathcolor{blue}{_\uparrow O^{2-}} \;\mathcolor{red}{_{\nearrow} {Cu^{2+}}} \;\;\mathcolor{blue}{_\downarrow O^{2-}} $ 

\noindent $\mathcolor{blue}{\;\;\;\;\;\;\;\;\;\;\;\; _\uparrow O^{2-} \;\;\;\;\;\;\;\;\;\;\;\;\;\;_\downarrow O^{2-} \;\;\;\;\;\;\;\;\;\;\;\;\;\;_\uparrow O^{2-} \;\;\;\;\;\;\;\;\;\;\;\;\;\;\;_\downarrow O^{2-} \;\;\;\;\;\;\;\;\;\;\;\;\;\;\;_\uparrow O^{2-} \;\;\;\;\;\;\;\;\;\;\;\;\;\;\;_\downarrow O^{2-} \;\;\;\;\;\;\;\;\;\;\;\;\;\;\;_\uparrow O^{2-}}$

\noindent $ \mathcolor{blue}{_\downarrow O^{2-}} \; \mathcolor{red}{_{\swarrow}Cu^{2+}} \;\; \mathcolor{blue}{_\uparrow O^{2-}} \;\mathcolor{red}{_{\nearrow}  Cu^{2+}} \;\;\mathcolor{blue}{_\downarrow O^{2-}} \;\mathcolor{red}{_{\swarrow} {Cu^{2+}}} \;\;\mathcolor{blue}{_\uparrow O^{2-}}\mathcolor{blue}{ \;\mathcolor{red}{_{\nearrow}  {Cu^{2+}}} \;\;_\downarrow O^{2-}} \;\mathcolor{red}{_{\swarrow}  {Cu^{2+}}} \;\;\mathcolor{blue}{_\uparrow O^{2-}} \;\mathcolor{red}{_{\nearrow}  {Cu^{2+}}} \;\;\mathcolor{blue}{_\downarrow O^{2-}} \;\mathcolor{red}{_{\swarrow} {Cu^{2+}}} \;\;\mathcolor{blue}{_\uparrow O^{2-}} $ 

\noindent $\mathcolor{blue}{\;\;\;\;\;\;\;\;\;\;\;\; _\downarrow O^{2-} \;\;\;\;\;\;\;\;\;\;\;\;\;\;_\uparrow O^{2-} \;\;\;\;\;\;\;\;\;\;\;\;\;\;_\downarrow O^{2-} \;\;\;\;\;\;\;\;\;\;\;\;\;\;\;_\uparrow O^{2-} \;\;\;\;\;\;\;\;\;\;\;\;\;\;\;_\downarrow O^{2-} \;\;\;\;\;\;\;\;\;\;\;\;\;\;\;_\uparrow O^{2-} \;\;\;\;\;\;\;\;\;\;\;\;\;\;\;_\downarrow O^{2-}}$

\noindent $ \mathcolor{blue}{_\uparrow O^{2-}} \; \mathcolor{red}{_{\nearrow}Cu^{2+}} \;\; \mathcolor{blue}{_\downarrow O^{2-}} \;\mathcolor{red}{_{\swarrow}  Cu^{2+}} \;\;\mathcolor{blue}{_\uparrow O^{2-}} \;\mathcolor{red}{_{\nearrow} {Cu^{2+}}} \;\;\mathcolor{blue}{_\downarrow O^{2-}} \;\mathcolor{red}{_{\swarrow}  {Cu^{2+}}} \;\;\mathcolor{blue}{_\uparrow O^{2-}} \;\mathcolor{red}{_{\nearrow}  {Cu^{2+}}} \;\;\mathcolor{blue}{_\downarrow O^{2-}} \;\mathcolor{red}{_{\swarrow}  {Cu^{2+}}} \;\;\mathcolor{blue}{_\uparrow O^{2-}} \;\mathcolor{red}{_{\nearrow} {Cu^{2+}}} \;\;\mathcolor{blue}{_\downarrow O^{2-}} $ 

\noindent $\mathcolor{blue}{\;\;\;\;\;\;\;\;\;\;\;\; _\uparrow O^{2-} \;\;\;\;\;\;\;\;\;\;\;\;\;\;_\downarrow O^{2-} \;\;\;\;\;\;\;\;\;\;\;\;\;\;_\uparrow O^{2-} \;\;\;\;\;\;\;\;\;\;\;\;\;\;\;_\downarrow O^{2-} \;\;\;\;\;\;\;\;\;\;\;\;\;\;\;_\uparrow O^{2-} \;\;\;\;\;\;\;\;\;\;\;\;\;\;\;_\downarrow O^{2-} \;\;\;\;\;\;\;\;\;\;\;\;\;\;\;_\uparrow O^{2-}}$

\noindent $ \mathcolor{blue}{_\downarrow O^{2-}} \; \mathcolor{red}{_{\swarrow}Cu^{2+}} \;\; \mathcolor{blue}{_\uparrow O^{2-}} \;\mathcolor{red}{_{\nearrow}  Cu^{2+}} \;\;\mathcolor{blue}{_\downarrow O^{2-}} \;\mathcolor{red}{_{\swarrow} {Cu^{2+}}} \;\;\mathcolor{blue}{_\uparrow O^{2-}}\mathcolor{blue}{ \;\mathcolor{red}{_{\nearrow}  {Cu^{2+}}} \;\;_\downarrow O^{2-}} \;\mathcolor{red}{_{\swarrow}  {Cu^{2+}}} \;\;\mathcolor{blue}{_\uparrow O^{2-}} \;\mathcolor{red}{_{\nearrow}  {Cu^{2+}}} \;\;\mathcolor{blue}{_\downarrow O^{2-}} \;\mathcolor{red}{_{\swarrow} {Cu^{2+}}} \;\;\mathcolor{blue}{_\uparrow O^{2-}} $  

\noindent $\mathcolor{blue}{\;\;\;\;\;\;\;\;\;\;\;\; _\downarrow O^{2-} \;\;\;\;\;\;\;\;\;\;\;\;\;\;_\uparrow O^{2-} \;\;\;\;\;\;\;\;\;\;\;\;\;\;_\downarrow O^{2-} \;\;\;\;\;\;\;\;\;\;\;\;\;\;\;_\uparrow O^{2-} \;\;\;\;\;\;\;\;\;\;\;\;\;\;\;_\downarrow O^{2-} \;\;\;\;\;\;\;\;\;\;\;\;\;\;\;_\uparrow O^{2-} \;\;\;\;\;\;\;\;\;\;\;\;\;\;\;_\downarrow O^{2-}}$

\noindent \footnotesize FIG. 10. $CuO_2$ plane of pristine  $La_2CuO_4$ with {\it diagonal} tilt of $CuO_6$ octahedra in the low-temperature orthorhombic (LTO) phase. Vertical subarrows $\mathcolor{blue}{_\downarrow}$ or $\mathcolor{blue}{_\uparrow}$ at $\mathcolor{blue}{O^{2-}}$ ions  indicate their location  below or above the $CuO_2$ plane. Slanted subarrows, $\mathcolor{red}{_\swarrow}$ or $\mathcolor{red}{_\nearrow}$  at $\mathcolor{red}{Cu^{2+}}$ indicate a {\it slight} diagonal slant of  $CuO_6$ octahedra. \normalsize

\section{CHARGE ORDER IN {\it \textbf{214}} LANTHANUM CUPRATES}
A common base member of neighboring $CuO_5O^{\upuparrows}$ octahedra, each  $O^{\upuparrows}$ atom, replacing a $O^{2-}$ ion, tilts the $CuO_5O^{\upuparrows}$ octahedra more than the slight tilt of the host crystal's $CuO_6$ octahedra. The heavier tilt is caused by the small size of the $O$ atom and the interaction of the magnetic moments of the triplet-state $O^{\upuparrows}$ atoms ($S=1$) and the $^{\uparrow}Cu^{2+}$ ions ($S=\frac{1}{2}$). The $O^{\upuparrows}$ atoms, spread to largest separation, form an incommensurable superlattice. Its wave number, called ``incommensurability,'' has been observed in diffraction experiments.\cite{20,21,22,23,24,25,26,27,28,29,30,31,32,33,34,35,36,37,38,39,40,41,42,43,44,45,46,47,48,49,50,51,52,53,54,55,56,57,58,59,60,61,62,63} 
Expressed in quasi-tetragonal approximation and {\it r.l.u.} units,
\begin{equation}
q_{co}(x)= \frac{\Omega^{\pm}}{{2}}\sqrt {x - \grave{t}}\;,\;\;\;\;\;x < \hat{x} \; ,
\end{equation}
it rises from zero in a square-root fashion. The orientation factor is $\Omega^{+}=\sqrt{2}$ 
\pagebreak
for  $x > x_6 \equiv 2/6^2  \simeq 0.056$  when the charge order is {\it axial}, that is, parallel to the $a$ or $b$  axis, but $\Omega^{-} = 1$ for $x < x_6$ when the charge order is {\it diagonal}. The offset value $\grave{t} = 0.02$/$\Omega^{\pm}$ under the radical is due to the tilt of the $CuO_5 O^{\upuparrows}$ octahedra. The change in charge-order orientation is caused by a change of tilt of  $CuO_5O^{\upuparrows}$ and neighboring $CuO_6$ octahedra which in turn is determined by two counteracting mechanisms. One mechanism causes the {\it diagonal} tilt of $CuO_6$ host octahedra (see Sec. X). The other mechanism tends an {\it axial} tilt of the neighboring $CuO_5 O^{\upuparrows}$ octahedra toward and away from each other due to the common $O^{\upuparrows}$ atom at their base.

At low doping, $0.02 = x_N < x < x_6 = 0.056$, where $O^{\upuparrows}$ and $O^{\downdownarrows}$ atoms are far separated, the diagonal tilt pervails. Shown on the left side of Fig. 12 is an oxygen atom, $O^{\upuparrows}$,  located slightly above the $CuO_2$ plane, at doping, $x < x_6$. The tendency of each $CuO_5O^{\upuparrows}$ octahedron to tilt axially, due to the shared $O^{\upuparrows}$ atom at its base, is overwhelmed by the majority of the   
diagonally tilted $CuO_6$ host octahedra (Fig. 10), forcing a heavy diagonally tilt (indicated

\includegraphics[width=5.9in]{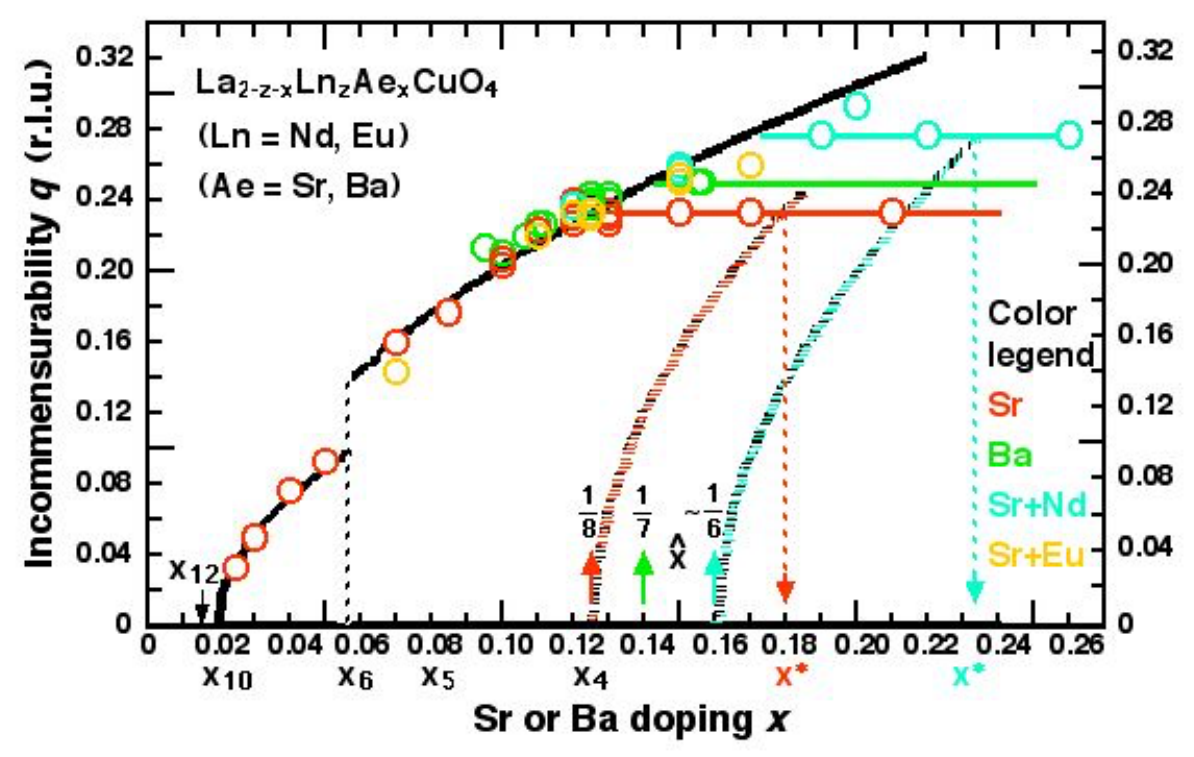}  

\noindent \footnotesize  FIG. 11. Superlattice incommensurability of $O^{\upuparrows}$ charge-order, $q = q_{co}$, and $O^{\uparrow \uparrow}$-$O^{\downarrow \downarrow}$ spin order, $q = 2q_{so}$, in $La_{2-z-x}Ln_zAe_{x}CuO_{4}$ ($z = 0, 0.4, 0.2$). Commensurate doping is denoted by $x_n \equiv 2/n^2$. Circles show data from  X-ray diffraction or neutron scattering. The broken black curve is a graph of $q_{co}(x)$, Eq. (10), calculated with tilt correction $\grave{t} = 0.02$. Its discontinuity at $x_6 \simeq 0.056$ is caused by a change of stripe orientation from diagonal for $x<x_6$ to axial for $x>x_6$. Colored horizontal lines are graphs of $q_{co}(\hat{x})$, Eq. (11), and hatched colored curves are graphs of $\sqrt{2} Q_{co}(x-\hat{x})$, Eq. (13), due to a secondary $O^{\upuparrows}$ superlattice. The condition for closing of the pseudogap, Eqs. (14, 15), is illustrated by dashed down-arrows. \normalsize 

\noindent by  $\mathcolor{red}{\Nearrow}$ and $\mathcolor{red}{\Swarrow}$). A moderate diagonal tilt of next-neighbor $CuO_6$ octahedra ($\mathcolor{red}{\nearrow}$ and $\mathcolor{red}{\swarrow}$) pro- vides a gradual transition to the slight tilt of the host $CuO_6$ octahedra (subarrow $\mathcolor{red}{_\nearrow}$ and $\mathcolor{red}{_\swarrow}$). 

The case of larger doping, $x > x_6$, is shown on the right side of Fig. 12. Because of stronger magnetic $^\uparrow Cu^{2+}$---$O^{\upuparrows}$ and $O^{\upuparrows}$---$O^{\downdownarrows}$ interaction at lesser superlattice spacing $A_0(x)$ and $B_0(x)$, heavy (moderate) {\it axial} tilts, indicated by $\mathcolor{red}{\overrightarrow{long}}$ ($\mathcolor{red}{\overrightarrow{sh}}ort$) overarrows at $Cu^{2+}$, overwhelm the diagonal tilt tendency, rendering them {\it axially} tilted. 

Equation (10) is valid for doping up to a ``watershed'' doping $\hat{x}$, which depends on the doping species: $\hat{x} = 0.125$ for $La_{2-x}Sr_{x}CuO_{4}$,  $\hat{x} = 0.14$ for  $La_{2-x}Ba_{x}CuO_{4}$, and  $\hat{x} = 0.16$ for $La_{2-z-x}Ln_zSr_xCuO_{4}$, ($Ln = Nd, Eu \; z = 0.4, 0.2$).

Macroscopically, doping of $La_{2-x}Ae_xCuO_4$ is a continuous quantity. The observed incommensurability, $q_{co}(x)$, is also continuous. On the microscopic level, the crystal lattice sites are discrete. In the few cases of {\it commensurate} doping, $x_n = 2/n^2$, where $n$ is an integer, the $n \times n$ superlattice spacing relates to the lattice constants as $A_0 = n a_0$ and $B_0 = n b_0$. The $O^{\upuparrows}$

\medskip

\noindent $\mathcolor{blue}{\;\;\;\;\;\;\;\;\;\;\; 
_\downarrow O^{2-}} \;\;\;\;\;\;\;\;\;\;\;\;\;\;
\mathcolor{blue}{_\uparrow O^{2-}} \;\;\;\;\;\;\;\;\;\;\;\;\;\;
\mathcolor{blue}{_\downarrow O^{2-}} \;\;\;\;\;\;\;\;\;\;\;\;\;\;\;
\;\;\;\;| \;\;\;\;\;\;\;\;\;\;\;\;\;\;\;\;\;\;
\mathcolor{blue}{_\downarrow O^{2-}} \;\;\;\;\;\;\;\;\;\;\;\;\;\;\;
\mathcolor{blue}{_\uparrow O^{2-} \;\;\;\;\;\;\;\;\;\;\;\;\;\; _\downarrow O^{2-}}$

\noindent $\mathcolor{blue}{_\uparrow O^{2-}}\;\mathcolor{red}{_{\nearrow}Cu^{2+}}\;\; 
\mathcolor{blue}{_\downarrow O^{2-}} \;\mathcolor{red}{{\swarrow}  Cu^{2+}} \;\;
\mathcolor{blue}{_\uparrow O^{2-}} \;\mathcolor{red}{_{\nearrow} {Cu^{2+}}} \;\;\mathcolor{blue}{_\downarrow O^{2-}} \; \;\;\;\;\; | \;\;\;\;\;\;
\;\mathcolor{blue}{_\uparrow O^{2-}} \mathcolor{red}{_{\nearrow} {Cu^{2+}}} \;\;
\mathcolor{blue}{_\downarrow O^{2-}} \;  \mathcolor{red}{_{\swarrow}{Cu}^{2+}} \;\;\mathcolor{blue}{_\uparrow O^{2-}} \;\mathcolor{red}{_{\nearrow} {Cu}^{2+}} \;\;\mathcolor{blue}{_\downarrow O^{2-}}$ 

\noindent $\mathcolor{blue}{\;\;\;\;\;\;\;\;\;\;\;\; _\uparrow O^{2-}} 
\;\;\;\;\;\;\;\;\;\;\;\;\;\;\mathcolor{green}{_\downarrow O^{2-}} 
\;\;\;\;\;\;\;\;\;\;\;\;\;\;\mathcolor{blue}{_\uparrow O^{2-}}
\;\;\;\;\;\;\;\;\;\;\;\;\;\;\;\;\;\;\;|\;\;\;
\;\;\;\;\;\;\;\;\;\;\;\;\;\;\mathcolor{blue}{_\uparrow O^{2-}} 
\;\;\;\;\;\;\;\;\;\;\;\;\;\;\;\;\mathcolor{blue}{O^{2-}}
\;\;\;\;\;\;\;\;\;\;\;\;\;\;\;\;\mathcolor{blue}{O^{2-}}$

\noindent $ \mathcolor{blue}{_\downarrow O^{2-}} \; \mathcolor{red}{{\swarrow}Cu^{2+}} \; \mathcolor{green}{_\uparrow O^{2-}} \mathcolor{red}{\mathbf{\Nearrow}  \mathbf{Cu^{2+}}} \; \mathcolor{green}{_\downarrow O^{2-}} \;\mathcolor{red}{{\swarrow} {Cu^{2+}}} \;\;\mathcolor{blue}{_\uparrow O^{2-}} \;
\;\;\;\;\; | \;\;\;\;\;\;
\;\mathcolor{blue}{_\downarrow O^{2-}} \mathcolor{red}{_{\swarrow} {Cu^{2+}}} \;\;\mathcolor{blue}{_\uparrow O^{2-}} \;\;\;\mathcolor{red}{\overrightarrow{Cu}^{2+}} \;\;\mathcolor{green}{_\downarrow O^{2-}} \;\;\;
\;\mathcolor{red}{\overleftarrow{Cu}^{2+}}\;\;\; \mathcolor{blue}{_\uparrow O^{2-}} $ 

\noindent $\mathcolor{green}{\;\;\;\;\;\;\;\;\;\;\;\; _\downarrow O^{2-}} \;\;\;\;\;\;\;\;\;\;\;\;\;\;_\uparrow\mathbf{o}^{\upuparrows} \;\;\;\;\;\;\;\;\;\;\;\;\;\;\;\;\mathcolor{green}{_\downarrow O^{2-}} 
\;\;\;\;\;\;\;\;\;\;\;\;\;\;\;\;\;\;\;|\;\;\;\;
\;\;\;\;\;\;\;\;\;\;\;\;\;\;\;\;\mathcolor{blue}{O^{2-}} \;\;\;\;\;\;\;\;\;\;\;\;\;\;\; \mathcolor{green}{O^{2-} \;\;\;\;\;\;\;\;\;\;\;\;\;\;\;\; O^{2-}}$

\noindent $\mathcolor{blue}{ _\uparrow O^{2-}} \; \mathcolor{red}{{\nearrow}Cu^{2+}} \; \mathcolor{green}{_\downarrow O^{2-}} \mathcolor{red}{\mathbf{\Swarrow}  \mathbf{Cu^{2+}}} \;\mathcolor{green}{_\uparrow O^{2-}} \;\mathcolor{red}{{\nearrow} {Cu^{2+}}} \;\;\mathcolor{blue}{_\downarrow O^{2-}} \;\;\;\;\;\; | \;\;\;\;\;\;
\;\mathcolor{blue}{_\uparrow O^{2-}}  \;\;\;\mathcolor{red}{\overrightarrow{Cu}^{2+}}\; \;\;\mathcolor{green}{\mathbf{\downarrow} O^{2-}} \;\;  \mathcolor{red}{\overleftarrow{\mathbf{Cu^{2+}}}} \;\;\; _\uparrow\mathbf{o}^{\downdownarrows} \;\;\;\;  \mathcolor{red}{\overrightarrow{\mathbf{Cu^{2+}}}} \;\;\mathcolor{green}{\mathbf{\downarrow} O^{2-}}$ 

\noindent $\mathcolor{blue}{\;\;\;\;\;\;\;\;\;\;\;\; 
_\uparrow O^{2-}} \;\;\;\;\;\;\;\;\;\;\;\;\;\;
\mathcolor{green}{_\downarrow O^{2-}} \;\;\;\;\;\;\;\;\;\;\;\;\;\;\;
\mathcolor{blue}{_\uparrow O^{2-}} \;\;\;\;\;\;\;\;\;\;\;\;\;\;
\;\;\;\;|\;\;\;\;\;\;\;\;\;\;\;\;\;\;\;\;\;\;\;\;
\mathcolor{blue}{ O^{2-}} \;\;\;\;\;\;\;\;\;\;\;\;\;\;\;\;\;
\mathcolor{green}{O^{2-} \;\;\;\;\;\;\;\;\;\;\;\;\;\;\;\; O^{2-}}$

\noindent $\mathcolor{blue}{_\downarrow O^{2-}}\;\mathcolor{red}{_{\swarrow}Cu^{2+}}\;\; 
\mathcolor{blue}{_\uparrow O^{2-}} \;\mathcolor{red}{{\nearrow}  Cu^{2+}} \;\;
\mathcolor{blue}{_\downarrow O^{2-}} \;\mathcolor{red}{_{\swarrow} {Cu^{2+}}} \;\;\mathcolor{blue}{_\uparrow O^{2-}} \;\;\;\;\;\;\; | \;\;\;\;
\;\;\;\mathcolor{blue}{_\downarrow O^{2-}} \;\;\mathcolor{red}{_{\swarrow} {Cu^{2+}}} \;\;
\mathcolor{blue}{_\uparrow O^{2-}} \;\;\;\;  \mathcolor{red}{\overrightarrow{Cu}^{2+}} \;\;\mathcolor{green}{_\downarrow O^{2-}} \;\;\;\mathcolor{red}{\overleftarrow{Cu}^{2+}} \;\;\mathcolor{blue}{_\uparrow O^{2-}}$

\noindent $\mathcolor{blue}{\;\;\;\;\;\;\;\;\;\;\;\;\;\; _\downarrow O^{2-} \;\;\;\;\;\;\;\;\;\;\;\;\;\;_\uparrow O^{2-}} \;\;\;\;\;\;\;\;\;\;\;\;\;\;\mathcolor{blue}{_\downarrow O^{2-}} \;\;\;\;\;\;\;\;\;\;\;\;\;\;\;
\;\;\;|\;\;\;\;\;\;\;\;\;\;\;\;\;\;\;\;\;\;
\mathcolor{blue}{_\downarrow O^{2-}} \;\;\;\;\;\;\;\;\;\;\;\;\;\;\;\;\mathcolor{blue}{O^{2-} \;\;\;\;\;\;\;\;\;\;\;\;\;\;\; O^{2-}}$

\noindent $ \mathcolor{blue}{_\uparrow O^{2-}} \;\; \mathcolor{red}{{_\nearrow}Cu^{2+}} \;\;
\mathcolor{blue}{_\downarrow O^{2-}} \; \mathcolor{red}{_{\swarrow}Cu^{2+}} \;\;\mathcolor{blue}{_\uparrow O^{2-}} \;\mathcolor{red}{_{\nearrow}  Cu^{2+}} \;\;\;\mathcolor{blue}{_\downarrow O^{2-}} \;\;\;\;\;\;\; | \;\;\;\;
\;\;\;\mathcolor{blue}{_\uparrow O^{2-}} \;\;\mathcolor{red}{_{\nearrow} {Cu^{2+}}} \;
\mathcolor{blue}{_\downarrow O^{2-}} \;  \mathcolor{red}{_{\swarrow}{Cu}^{2+}} \;\;\mathcolor{blue}{_\uparrow O^{2-}} \;\mathcolor{red}{_{\nearrow} {Cu}^{2+}} \;\;\mathcolor{blue}{_\downarrow O^{2-}}$  

\noindent $
\mathcolor{blue}{\;\;\;\;\;\;\;\;\;\;\;\;\;_\uparrow O^{2-}}
\mathcolor{blue}{\;\;\;\;\;\;\;\;\;\;\;\;\;\;_\downarrow O^{2-}}
\mathcolor{blue}{\;\;\;\;\;\;\;\;\;\;\;\;\;\;\;_\uparrow O^{2-}}
\;\;\;\;\;\;\;\;\;\;\;\;\;\;\;\;\;\;\;|\;\;\;\;
\mathcolor{blue}{\;\;\;\;\;\;\;\;\;\;\;\;\;\;_\uparrow O^{2-}}
\mathcolor{blue}{\;\;\;\;\;\;\;\;\;\;\;\;\;\;\;_\downarrow O^{2-}} 
\mathcolor{blue}{\;\;\;\;\;\;\;\;\;\;\;\;\;\;\;_\uparrow O^{2-}}  $

\noindent \footnotesize FIG. 12. $CuO_2$ plane at low doping, $x < x_6 = 0.056$ (left side), and at larger doping, $x > x_6$ (right side). 
Vertical subarrows  $\mathcolor{blue}{_\downarrow}$ or $\mathcolor{blue}{_\uparrow}$ indicate $O^{2-}$ or $O^{\upuparrows}$ below and above the $CuO_2$ plane.
The $O^{\upuparrows}$ atom is at an ``$a$''-site (left) and a ``$b$''-site (right).
At low doping, the diagonal tilt of host $CuO_6$ octahedra ($\mathcolor{red}{_\swarrow}|\mathcolor{red}{_\nearrow}$) overwhelms the axial tilt tendency next to the small $O^{\upuparrows}$ atoms, causing a heavy {\it diagonal} tilt of $CuO_5O^{\upuparrows}$ octahedra ($\mathcolor{red}{\mathbf{\Swarrow}}|\mathcolor{red}{\mathbf{\Nearrow}}$) and a moderate {\it diagonal} tilt of neighboring $CuO_6$ octahedra ($\mathcolor{red}{\mathbf{\swarrow}}|\mathcolor{red}{\mathbf{\nearrow}}$).
At larger doping, stronger magnetic  $^\uparrow Cu^{2+}$---$O^{\upuparrows}$ and $O^{\upuparrows}$-$O^{\downdownarrows}$ interaction enforces the axial tilt tendency next to  $O^{\upuparrows}$ atoms, overwhelming the diagonal host-lattice tilt. This causes heavy (moderate) {\it axial} tilts of $CuO_5O^{\upuparrows}$ and neighboring $CuO_6$ octahedra, indicated by $\mathcolor{red}{\overrightarrow{long}}$ ($\mathcolor{red}{\overrightarrow{sh}}ort$) overarrows. 
Marked are {\it stressed} $\mathcolor{green}{O^{2-}}$ ions and {\it demotted} $\mathcolor{red}{\mathbf{Cu^{2+}}}$ ions.
\normalsize

\noindent   atoms then reside exactly on the ideal lattice sites (like in Fig. 13). In all other cases, the  $O^{\upuparrows}$ atom and the ions in its surrounding are correspondingly displaced---complicated situations that are hard to display and difficult to comprehend visually. One can formally generalize $n = \sqrt{2/x_n}$ by a non-integer $\tilde{n} = \sqrt{2/x}$ for the spacing of an incommensurate $\tilde{n} \times \tilde{n}$ superlattice, $A(x) = \tilde{n} a_0$ and $B(x) = \tilde{n} b_0$, with the understanding that this represents an {\it average} value, taken over a large crystal region. However, it does not lend itself to a local display on a small scale. Visualization rests mostly on the commensurate cases.



The regularity of axial $Cu^{2+}O^{2-}_5O^{\upuparrows}$ octahedra tilts at commensurate doping $x_n$ provides for maximal constructive or destructive interference in diffraction experiments (Sec. XVIII).  

A saturation of the growth of the $O^{\upuparrows}$ superlattice is reached when the  heavily and moderately tilted octahedra around the $O^{\upuparrows}$ atoms come in contact. The doping value where this happens is the {\it watershed} doping, $\hat{x}$. 
For $La_{2-x}Sr_xCuO_4$ we have $\hat{x} = x_4$, shown in Fig. 13. 
No further growth of the $O^{\upuparrows}$ superlattice is possible---i. e., no spacing $A(x) < 4a_0, B(x) < 4b_0$ for doping $x > \hat{x}$---as the tilted octahedra would get in each other's way. Accordingly, the incommensurability of the primary superlattice, $q_{co}(x)$, levels off from the square-root curve of Eq. (10) to a constant plateau (depending on dopant and co-dopant species, see Fig. 11),
\begin{equation}
q_{co}(x) = \frac{\sqrt{2}}{2} \sqrt{\hat{x} - \grave{t}} \;,\;\;\;\; x > \hat{x} \;.
\end{equation}


\noindent $\mathcolor{blue}{\;\;\;\;\;\;\;\;\;\;\;\;O^{2-}} \;\;\;\;\;\;\;\;\;\;\;\;\;\;\;\; 
\mathcolor{green}{O^{2-} \;\;\;\;\;\;\;\;\;\;\;\;\;\;\;\; O^{2-}} \;\;\;\;\;\;\;\;\;\;\;\;\;\;\;\;\mathcolor{blue}{ O^{2-} \;\;\;\;\;\;\;\;\;\;\;\;\;\;\;\;\; O^{2-}} \;\;\;\;\;\;\;\;\;\;\;\;\;\;\;\; \mathcolor{green}{O^{2-} \;\;\;\;\;\;\;\;\;\;\;\;\;\;\;\; O^{2-}}$

\noindent $ \mathcolor{green}{_\downarrow O^{2-}} \;\;\; \mathcolor{red}{\overleftarrow{Cu}^{2+}} \;\; \mathcolor{green}{\mathbf{\uparrow} O^{2-}} \;\;\;\mathcolor{red}{\overrightarrow{\mathbf{Cu^{2+}}}} \;\;\;_\downarrow\mathbf{o}^\mathbf{\upuparrows} \;\;\;\;\mathcolor{red}{\overleftarrow{\mathbf{Cu^{2+}}}} \;\mathcolor{green}{\mathbf{\uparrow} O^{2-}} \;\;\;\mathcolor{red}{\overrightarrow{Cu}^{2+}} \;\;\mathcolor{green}{_\downarrow O^{2-}} \;\;\;\;\mathcolor{red}{\overleftarrow{Cu}^{2+}} \;\;\;\mathcolor{green}{\mathbf{\uparrow} O^{2-}} \;\;\;\mathcolor{red}{\overrightarrow{\mathbf{Cu^{2+}}}} \;\;_\downarrow\mathbf{o}^\mathbf{\downdownarrows} \;\;\;\;\mathcolor{red}{\overleftarrow{\mathbf{Cu^{2+}}}} \;\mathcolor{green}{\mathbf{\uparrow} O^{2-}}$  

\noindent $\mathcolor{blue}{\;\;\;\;\;\;\;\;\;\;\;\; O^{2-}} \;\;\;\;\;\;\;\;\;\;\;\;\;\;\;\; \mathcolor{green}{O^{2-} \;\;\;\;\;\;\;\;\;\;\;\;\;\;\;\; O^{2-}} \;\;\;\;\;\;\;\;\;\;\;\;\;\;\;\;\mathcolor{blue}{ O^{2-} \;\;\;\;\;\;\;\;\;\;\;\;\;\;\;\; O^{2-}}\mathcolor{green}{ \;\;\;\;\;\;\;\;\;\;\;\;\;\;\;\; O^{2-} \;\;\;\;\;\;\;\;\;\;\;\;\;\;\;\; O^{2-}}$

\noindent $ \mathcolor{blue}{_\uparrow O^{2-}} \; \mathcolor{red}{_{\nearrow}{Cu^{2+}}} \;\; \mathcolor{blue}{_\downarrow O^{2-}} \;\;\;\;\mathcolor{red}{\overleftarrow{Cu}^{2+}} \;\;\mathcolor{green}{_\uparrow O^{2-}}\mathcolor{red}{ \;\;\; \mathcolor{red}{\overrightarrow{Cu}^{2+}}} \;\;\mathcolor{blue}{_\downarrow O^{2-}} \;\mathcolor{red}{_{\swarrow}  {Cu^{2+}}} \;\;\mathcolor{blue}{_\uparrow O^{2-}}\mathcolor{red}{\mathcolor{red}{ \;_{\nearrow}  {Cu^{2+}}}} \;\;\mathcolor{blue}{_\downarrow O^{2-}} \;\;\;\;\mathcolor{red}{\overleftarrow{Cu}^{2+}} \;\;\mathcolor{green}{_\uparrow O^{2-}} \;\;\;\;\mathcolor{red}{\overrightarrow{Cu}^{2+}} \;\;\mathcolor{blue}{_\downarrow O^{2-}}$ 

\noindent $\mathcolor{blue}{\;\;\;\;\;\;\;\;\;\;\;\; _\uparrow O^{2-}} \;\;\;\;\;\;\;\;\;\;\;\;\;\;\;\mathcolor{blue}{ O^{2-} \;\;\;\;\;\;\;\;\;\;\;\;\;\;\; O^{2-}} \;\;\;\;\;\;\;\;\;\;\;\;\;\;\;\mathcolor{blue}{_\downarrow O^{2-} \;\;\;\;\;\;\;\;\;\;\;\;\;\;\;_\uparrow O^{2-}} \;\;\;\;\;\;\;\;\;\;\;\;\;\;\;\;\mathcolor{blue}{O^{2-}\;\;\;\;\;\;\;\;\;\;\;\;\;\;\;\; O^{2-}}$

\noindent $ \mathcolor{blue}{_\downarrow O^{2-}} \; \mathcolor{red}{_{\swarrow}Cu^{2+}} \;\; \mathcolor{blue}{_\uparrow O^{2-}} \;\mathcolor{red}{_{\nearrow}  Cu^{2+}} \;\;\mathcolor{blue}{_\downarrow O^{2-}} \;\mathcolor{red}{_{\swarrow} {Cu^{2+}}} \;\;\mathcolor{blue}{_\uparrow O^{2-}} \;\mathcolor{red}{_{\nearrow}  {Cu^{2+}}} \;\;\mathcolor{blue}{_\downarrow O^{2-}} \;\mathcolor{red}{_{\swarrow}  {Cu^{2+}}} \;\;\mathcolor{blue}{_\uparrow O^{2-}} \;\mathcolor{red}{_{\nearrow}  {Cu^{2+}}} \;\;\mathcolor{blue}{_\downarrow O^{2-}} \;\mathcolor{red}{_{\swarrow} {Cu^{2+}}} \;\;\mathcolor{blue}{_\uparrow O^{2-}} $ 

    \noindent $\mathcolor{blue}{\;\;\;\;\;\;\;\;\;\;\;\; _\downarrow O^{2-}} \;\;\;\;\;\;\;\;\;\;\;\;\;\;\;\mathcolor{blue}{ O^{2-} \;\;\;\;\;\;\;\;\;\;\;\;\;\;\; O^{2-}} \;\;\;\;\;\;\;\;\;\;\;\;\;\;\;\mathcolor{blue}{_\uparrow O^{2-} \;\;\;\;\;\;\;\;\;\;\;\;\;\;\;_\downarrow O^{2-}} \;\;\;\;\;\;\;\;\;\;\;\;\;\;\;\;\mathcolor{blue}{ O^{2-} \;\;\;\;\;\;\;\;\;\;\;\;\;\;\;\; O^{2-}}$

\noindent $ \mathcolor{blue}{_\uparrow O^{2-}} \; \mathcolor{red}{_{\nearrow}{Cu^{2+}}} \;\; \mathcolor{blue}{_\downarrow O^{2-}} \;\;\;\;\mathcolor{red}{\overleftarrow{Cu}^{2+}} \;\;\mathcolor{green}{_\uparrow O^{2-}}\mathcolor{red}{ \;\;\; \mathcolor{red}{\overrightarrow{Cu}^{2+}}} \;\;\mathcolor{blue}{_\downarrow O^{2-}} \;\mathcolor{red}{_{\swarrow}  {Cu^{2+}}} \;\;\mathcolor{blue}{_\uparrow O^{2-}}\mathcolor{red}{\mathcolor{red}{ \;_{\nearrow}  {Cu^{2+}}}} \;\;\mathcolor{blue}{_\downarrow O^{2-}} \;\;\;\;\mathcolor{red}{\overleftarrow{Cu}^{2+}} \;\;\mathcolor{green}{_\uparrow O^{2-}} \;\;\;\;\mathcolor{red}{\overrightarrow{Cu}^{2+}} \;\;\mathcolor{blue}{_\downarrow O^{2-}}$ 

\noindent $\mathcolor{blue}{\;\;\;\;\;\;\;\;\;\;\;\;O^{2-}} \;\;\;\;\;\;\;\;\;\;\;\;\;\;\;\; 
\mathcolor{green}{O^{2-} \;\;\;\;\;\;\;\;\;\;\;\;\;\;\;\; O^{2-}} \;\;\;\;\;\;\;\;\;\;\;\;\;\;\;\;\mathcolor{blue}{ O^{2-} \;\;\;\;\;\;\;\;\;\;\;\;\;\;\;\;\; O^{2-}} \;\;\;\;\;\;\;\;\;\;\;\;\;\;\;\; \mathcolor{green}{O^{2-} \;\;\;\;\;\;\;\;\;\;\;\;\;\;\;\; O^{2-}}$

\noindent $ \mathcolor{green}{_\downarrow O^{2-}} \;\;\; \mathcolor{red}{\overleftarrow{Cu}^{2+}} \;\; \mathcolor{green}{\mathbf{\uparrow} O^{2-}} \;\;\;\mathcolor{red}{\overrightarrow{\mathbf{Cu^{2+}}}} \;\;\;_\downarrow\mathbf{o}^\mathbf{\downdownarrows} \;\;\;\;\mathcolor{red}{\overleftarrow{\mathbf{Cu^{2+}}}} \;\mathcolor{green}{\mathbf{\uparrow} O^{2-}} \;\;\;\mathcolor{red}{\overrightarrow{Cu}^{2+}} \;\;\mathcolor{green}{_\downarrow O^{2-}} \;\;\;\;\mathcolor{red}{\overleftarrow{Cu}^{2+}} \;\;\;\mathcolor{green}{\mathbf{\uparrow} O^{2-}} \;\;\;\mathcolor{red}{\overrightarrow{\mathbf{Cu^{2+}}}} \;\;_\downarrow\mathbf{o}^\mathbf{\upuparrows} \;\;\;\;\mathcolor{red}{\overleftarrow{\mathbf{Cu^{2+}}}} \;\mathcolor{green}{\mathbf{\uparrow} O^{2-}}$  

\noindent $\mathcolor{blue}{\;\;\;\;\;\;\;\;\;\;\;\; O^{2-}} \;\;\;\;\;\;\;\;\;\;\;\;\;\;\;\; \mathcolor{green}{O^{2-} \;\;\;\;\;\;\;\;\;\;\;\;\;\;\;\; O^{2-}} \;\;\;\;\;\;\;\;\;\;\;\;\;\;\;\;\mathcolor{blue}{ O^{2-} \;\;\;\;\;\;\;\;\;\;\;\;\;\;\;\; O^{2-}}\mathcolor{green}{ \;\;\;\;\;\;\;\;\;\;\;\;\;\;\;\; O^{2-} \;\;\;\;\;\;\;\;\;\;\;\;\;\;\;\; O^{2-}}$

\noindent \footnotesize FIG. 13. $CuO_2$ plane at commensurate doping $x_4 = 1/8$ in the LTO phase. Vertical arrows indicate $O^{2-}$ ions slightly ($\mathcolor{blue}{_\downarrow}|\mathcolor{blue}{_\uparrow}$) or considerably ($\mathcolor{green}{\uparrow}$) off the $CuO_2$ plane.
$\mathcolor{red}{\overrightarrow{Long}}$ ($\mathcolor{red}{\overrightarrow{sh}}ort$) overarrows indicate heavy (medium) axial tilts. {\it Stressed} $O^{2-}$ ions are marked green ($\mathcolor{green}{O^{2-}}$); {\it demotted} $Cu^{2+}$ ions are marked bold ($\mathcolor{red}{\mathbf{Cu^{2+}}}$).  
\normalsize

The charge order is unidirectional within domains of the crystal. For doping $x > x_6$ charge order is along the crystal's $a$-axis in the $a$-domains but along the $b$-axis in  the $b$-domains. The magnetic moments of the triplet-state oxygen atoms align antiferromagnetically, $-\;O^{\upuparrows}-O^{\downdownarrows}-O^{\upuparrows}-O^{\downdownarrows}\;-$, along the direction of the charge order. This doubles the wave length of the spin order over the charge order---collectively called ``stripes''---but halves its incommensurability,
\begin{equation}
    q_{so}(x) = \frac{1}{2} \; q_{co}(x) \;.
\end{equation}


What happens with the $O^{\upuparrows}$ atoms in the doping range beyond the watershed, $x > \hat{x}$? As Fig. 13 shows, there is still space for {\it one} additional $O^{\upuparrows}$ atom---and its entourage of tilted octahedra---in the {\it center} region of each $4 \times 4$ square spanned by its four $O^{\upuparrows/ \downdownarrows}$ atoms. Now a {\it secondary} $O^{\upuparrows}$ superlattice commences with an additional $O^{\upuparrows}$ ion at the center of 
 $O^{\upuparrows}$ quartets of the primary superlattice. The incommensurability of the secondary superlattice, 
 \begin{equation}
Q_{co}(x) = \frac{\sqrt{2}}{2} \sqrt{x-\hat{x}}  \;,\;\;\;\; x > \hat{x} \;,
\end{equation}

\bigskip  

\noindent $ \mathcolor{blue}{_\uparrow O^{2-}} \ \mathcolor{red}{_{\nearrow}{Cu^{2+}}} \; \mathcolor{blue}{_\downarrow O^{2-}} \;\;\;\;\mathcolor{red}{\overleftarrow{Cu}^{2+}} \;\;\mathcolor{green}{_\uparrow O^{2-}}\mathcolor{red}{ \;\;\; \mathcolor{red}{\overrightarrow{Cu}^{2+}}} \;\;\mathcolor{blue}{_\downarrow O^{2-}}\; \;\mathcolor{red}{_{\swarrow}  {Cu^{2+}}} \;\;\mathcolor{blue}{_\uparrow O^{2-}}\mathcolor{red}{\mathcolor{red}{ \;_{\nearrow}  {Cu^{2+}}}} \;\mathcolor{blue}{_\downarrow O^{2-}} \;\;\;\;\mathcolor{red}{\overleftarrow{Cu}^{2+}} \;\;\mathcolor{green}{_\uparrow O^{2-}} \;\;\mathcolor{red}{\overrightarrow{Cu}^{2+}} \;\;\mathcolor{blue}{_\downarrow O^{2-}}$ 

\noindent $\mathcolor{blue}{\;\;\;\;\;\;\;\;\;\;\;\; _\uparrow O^{2-}} \;\;\;\;\;\;\;\;\;\;\;\;\;\;\; \mathcolor{green}{O^{2-} \;\;\;\;\;\;\;\;\;\;\;\;\;\;\; O^{2-}} \;\;\;\;\;\;\;\;\;\;\;\;\;\;\;\mathcolor{blue}{_\downarrow O^{2-} \;\;\;\;\;\;\;\;\;\;\;\;\;\;\;_\uparrow O^{2-}} \;\;\;\;\;\;\;\;\;\;\;\;\;\;\;\; \mathcolor{green}{O^{2-} \;\;\;\;\;\;\;\;\;\;\;\;\;\;\;\; O^{2-}}$

\noindent $ \mathcolor{blue}{_\downarrow O^{2-}} \;\;\; \mathcolor{red}{\overleftarrow{Cu}^{2+}} \;\; \mathcolor{green}{\mathbf{\uparrow} O^{2-}} \;\;\;\mathcolor{red}{\overrightarrow{\mathbf{Cu^{2+}}}} \;\;\;_\downarrow\mathbf{o}^\mathbf{\upuparrows} \;\;\;\;\mathcolor{red}{\overleftarrow{\mathbf{Cu^{2+}}}} \;\mathcolor{green}{\mathbf{\uparrow} O^{2-}} \;\;\;\mathcolor{red}{\overrightarrow{\mathbf{Cu}}^{2+}} \;\;\mathcolor{green}{_\downarrow O^{2-}} \;\;\;\;\mathcolor{red}{\overleftarrow{\mathbf{Cu}}^{2+}} \;\mathcolor{green}{\mathbf{\uparrow} O^{2-}} \;\;\;\mathcolor{red}{\overrightarrow{\mathbf{Cu^{2+}}}} \;\;_\downarrow\mathbf{o}^\mathbf{\downdownarrows} \;\;\;\;\mathcolor{red}{\overleftarrow{\mathbf{Cu^{2+}}}} \;\mathcolor{green}{\mathbf{\uparrow} O^{2-}}$  

\noindent $\mathcolor{blue}{\;\;\;\;\;\;\;\;\;\;\;\; _\downarrow O^{2-}} \;\;\;\;\;\;\;\;\;\;\;\;\;\;\; \mathcolor{green}{O^{2-} \;\;\;\;\;\;\;\;\;\;\;\;\;\;\; O^{2-}} \;\;\;\;\;\;\;\;\;\;\;\;\;\;\;\;\mathcolor{green}{ O^{2-} \;\;\;\;\;\;\;\;\;\;\;\;\;\;\;\; O^{2-}}\mathcolor{green}{ \;\;\;\;\;\;\;\;\;\;\;\;\;\;\;\; O^{2-} \;\;\;\;\;\;\;\;\;\;\;\;\;\;\;\; O^{2-}}$

\noindent $ \mathcolor{blue}{_\uparrow O^{2-}} \; \mathcolor{red}{_{\nearrow}{Cu^{2+}}} \;\; \mathcolor{blue}{_\downarrow O^{2-}} \;\;\;\;\mathcolor{red}{\overleftarrow{Cu}^{2+}} \;\;\mathcolor{green}{_\uparrow O^{2-}}\mathcolor{red}{ \;\;\; \mathcolor{red}{\overrightarrow{\mathbf{Cu}}^{2+}}} \;\mathcolor{green}{_\downarrow O^{2-}} \;\;\;\;\mathcolor{red}{\overleftarrow {\mathbf{Cu}}^{2+}} \;\;\mathcolor{green}{_\uparrow O^{2-}}\;\;\;\mathcolor{red}{ \overrightarrow{\mathbf{Cu}}^{2+}} \;\mathcolor{green}{_\downarrow O^{2-}} \;\;\;\;\mathcolor{red}{\overleftarrow{\mathbf{Cu}}^{2+}} \;\mathcolor{green}{_\uparrow O^{2-}} \;\;\;\;\mathcolor{red}{\overrightarrow{Cu}^{2+}} \;\;\mathcolor{blue}{_\downarrow O^{2-}}$ 

\noindent $\;\;\;\;\;\;\;\;\;\;\;\;\mathcolor{blue}{ _\uparrow O^{2-}} 
\;\;\;\;\;\;\;\;\;\;\;\;\;\;\mathcolor{blue}{_\downarrow O^{2-}} 
\mathcolor{green}{\;\;\;\;\;\;\;\;\;\;\;\;\;\;\; O^{2-} 
\;\;\;\;\;\;\;\;\;\;\;\;\;\;\;\;O^{2-} 
\;\;\;\;\;\;\;\;\;\;\;\;\;\;\;\; O^{2-} 
    \;\;\;\;\;\;\;\;\;\;\;\;\;\;\;\; O^{2-}} 
\;\;\;\;\;\;\;\;\;\;\;\;\;\;\;\mathcolor{blue}{_\uparrow O^{2-}}$

\noindent $ \mathcolor{blue}{_\downarrow O^{2-}} \;\; \mathcolor{red}{_{\swarrow}Cu^{2+}} \;\; \mathcolor{blue}{_\uparrow O^{2-}} \;\mathcolor{red}{_{\nearrow}  Cu^{2+}} \;\;\mathcolor{blue}{_\downarrow O^{2-}} \;\; \mathcolor{red}{\overleftarrow{\mathbf{Cu}}^{2+}} \;\; \mathcolor{green}{\mathbf{\uparrow} O^{2-}} \;\;\;\mathcolor{red}{\overrightarrow{\mathbf{Cu^{2+}}}} \;\;\;_\downarrow\mathbf{o}^\mathbf{\upuparrows} \;\;\;\;\mathcolor{red}{\overleftarrow{\mathbf{Cu^{2+}}}} \;\mathcolor{green}{\mathbf{\uparrow} O^{2-}} \;\;\;\mathcolor{red}{\overrightarrow{\mathbf{Cu}}^{2+}} \;\;\mathcolor{blue}{_\downarrow O^{2-}} \;\mathcolor{red}{_{\swarrow}  Cu^{2+}} \;\;\mathcolor{blue}{_\uparrow O^{2-}} $

    \noindent $\mathcolor{blue}{\;\;\;\;\;\;\;\;\;\;\;\;\; _\downarrow O^{2-}}
    \;\;\;\;\;\;\;\;\;\;\;\;\;\;\; \mathcolor{blue}{O^{2-}} 
    \mathcolor{green}{\;\;\;\;\;\;\;\;\;\;\;\;\;\;\;O^{2-}
    \;\;\;\;\;\;\;\;\;\;\;\;\;\;\;\; O^{2-} 
    \;\;\;\;\;\;\;\;\;\;\;\;\;\;\;\;\; O^{2-}} 
    \;\;\;\;\;\;\;\;\;\;\;\;\;\;\;\;\mathcolor{green}{ O^{2-}} 
    \;\;\;\;\;\;\;\;\;\;\;\;\;\;\;\mathcolor{blue}{_\downarrow O^{2-}}$

\noindent $ \mathcolor{blue}{_\uparrow O^{2-}} \; \mathcolor{red}{_{\nearrow}{Cu^{2+}}} \;\; \mathcolor{blue}{_\downarrow O^{2-}} \;\;\;\;\mathcolor{red}{\overleftarrow{Cu}^{2+}} \;\;\mathcolor{green}{_\uparrow O^{2-}}\mathcolor{red}{ \;\;\; \mathcolor{red}{\overrightarrow{\mathbf{Cu}}^{2+}}} \;\mathcolor{green}{_\downarrow O^{2-}} \;\;\;\;\mathcolor{red}{\overleftarrow {\mathbf{Cu}}^{2+}} \;\;\mathcolor{green}{_\uparrow O^{2-}}\;\;\;\mathcolor{red}{ \overrightarrow{\mathbf{Cu}}^{2+}} \;\mathcolor{green}{_\downarrow O^{2-}} \;\;\;\;\mathcolor{red}{\overleftarrow{\mathbf{Cu}}^{2+}} \;\mathcolor{green}{_\uparrow O^{2-}} \;\;\;\;\mathcolor{red}{\overrightarrow{Cu}^{2+}} \;\;\mathcolor{blue}{_\downarrow O^{2-}}$ 

\noindent $\mathcolor{blue}{\;\;\;\;\;\;\;\;\;\;\;\; _\uparrow O^{2-}} \;\;\;\;\;\;\;\;\;\;\;\;\;\;\;\; \mathcolor{green}{O^{2-} \;\;\;\;\;\;\;\;\;\;\;\;\;\;\; O^{2-}} \;\;\;\;\;\;\;\;\;\;\;\;\;\;\;\;\mathcolor{green}{ O^{2-} \;\;\;\;\;\;\;\;\;\;\;\;\;\;\;\; O^{2-}} \;\;\;\;\;\;\;\;\;\;\;\;\;\;\;\; \mathcolor{green}{O^{2-} \;\;\;\;\;\;\;\;\;\;\;\;\;\;\;\; O^{2-}}$

\noindent $ \mathcolor{blue}{_\downarrow O^{2-}} \;\;\; \mathcolor{red}{\overleftarrow{Cu}^{2+}} \;\; \mathcolor{green}{\mathbf{\uparrow} O^{2-}} \;\;\;\mathcolor{red}{\overrightarrow{\mathbf{Cu^{2+}}}} \;\;\;_\downarrow\mathbf{o}^\mathbf{\downdownarrows} \;\;\;\;\mathcolor{red}{\overleftarrow{\mathbf{Cu^{2+}}}} \;\mathcolor{green}{\mathbf{\uparrow} O^{2-}} \;\;\;\mathcolor{red}{\overrightarrow{\mathbf{Cu}}^{2+}} \;\;\mathcolor{green}{_\downarrow O^{2-}} \;\;\;\;\mathcolor{red}{\overleftarrow{\mathbf{Cu}}^{2+}} \;\mathcolor{green}{\mathbf{\uparrow} O^{2-}} \;\;\;\mathcolor{red}{\overrightarrow{\mathbf{Cu^{2+}}}} \;\;_\downarrow\mathbf{o}^\mathbf{\upuparrows} \;\;\;\;\mathcolor{red}{\overleftarrow{\mathbf{Cu^{2+}}}} \;\mathcolor{green}{\mathbf{\uparrow} O^{2-}}$  

\noindent $\mathcolor{blue}{\;\;\;\;\;\;\;\;\;\;\;\; _\downarrow O^{2-}} \;\;\;\;\;\;\;\;\;\;\;\;\;\;\;\; \mathcolor{green}{O^{2-} \;\;\;\;\;\;\;\;\;\;\;\;\;\;\; O^{2-}} \;\;\;\;\;\;\;\;\;\;\;\;\;\;\;\mathcolor{blue}{_\uparrow O^{2-} \;\;\;\;\;\;\;\;\;\;\;\;\;\;\;_\downarrow O^{2-}} \;\;\;\;\;\;\;\;\;\;\;\;\;\;\;\; \mathcolor{green}{O^{2-} \;\;\;\;\;\;\;\;\;\;\;\;\;\;\;\; O^{2-}}$

\noindent $ \mathcolor{blue}{_\uparrow O^{2-}} \ \mathcolor{red}{_{\nearrow}{Cu^{2+}}} \; \mathcolor{blue}{_\downarrow O^{2-}} \;\;\;\;\mathcolor{red}{\overleftarrow{Cu}^{2+}} \;\;\mathcolor{green}{_\uparrow O^{2-}}\mathcolor{red}{ \;\;\; \mathcolor{red}{\overrightarrow{Cu}^{2+}}} \;\;\mathcolor{blue}{_\downarrow O^{2-}}\; \;\mathcolor{red}{_{\swarrow}  {Cu^{2+}}} \;\;\mathcolor{blue}{_\uparrow O^{2-}}\mathcolor{red}{\mathcolor{red}{ \;_{\nearrow}  {Cu^{2+}}}} \;\mathcolor{blue}{_\downarrow O^{2-}} \;\;\;\;\mathcolor{red}{\overleftarrow{Cu}^{2+}} \;\;\mathcolor{green}{_\uparrow O^{2-}} \;\;\mathcolor{red}{\overrightarrow{Cu}^{2+}} \;\;\mathcolor{blue}{_\downarrow O^{2-}}$ 

\noindent \footnotesize FIG. 14. $CuO_2$ plane of $La_{2-x}Sr_xCuO_4$ beyond watershed doping, $x> \hat{x}$, with a primary $4 \times 4$ $O^{\upuparrows}$ superlattice and an additional $O^{\upuparrows}$ atom inside \textbf{\emph{one}} $4 \times 4$ $O^{\upuparrows}$ quartet, here called the ``$\mathbf{4 \times 4 + 1}$'' island.
{\it Demotted} copper ions, $\mathcolor{red}{\mathbf{Cu^{2+}}}$---each flanked by at least three {\it stressed} oxygen ions, $\mathcolor{green}{O^{2-}}$---fill the island but don't extend into the surrounding $4 \times 4$ quartets of the primary superlattice. 
\normalsize 

\noindent is graphed as $\sqrt2 Q_{co(x)}$ by hatched curves in Fig. 11. Its intersection with the constant  line of $q_{co}(x)$ for $x > \hat{x}$, Eq. (11), determines the quantum critical point $x^*$. 



\textbf{Assumption 10: When a copper ion is adjacent to least three stressed $\mathcolor{green}{O^{2-}}$  ions, the hosted $\mathbf{Cu3d}$-$\mathbf{O2p}$ hole is \emph{demotted}.} With this meaning in mind, we want to say, for simplicity's sake,  that  \emph{``the copper ion is demotted''} and denote it by $\mathcolor{red}{\mathbf{Cu^{2+}}}$.


As Fig. 14 illustrates, a $4 \times 4 + 1$ island is completely {\it demotted}, but surrounded by $4 \times 4$ quartets of the primary superlattice filled with {\it motted} $Cu^{2+}$ ions (as in Fig. 13). Thus the demotted $Cu  3d$-$O2p$ holes are trapped inside the $4 \times 4 + 1$ island. Only at doping $x \ge x^*$, when $4 \times 4 + 1$ islands become contiguous at their corners, do lines of demotted $\mathcolor{red}{\mathbf{Cu^{2+}}}$ ions extend through the crystal, as shown in Fig. 16.

\section{QUANTUM CRITICAL POINT OF HOLE-DOPED {\it \textbf{214}} CUPRATES}

Inserting the observed data of watershed doping $\hat{x}$, tilt correction $\grave{t}$, and quantum critical point $x^*$ of $La_{2-x}Sr_xCuO_4$ and co-doped $La_{1.6-z-x}Nd_{0.4}Ae_xCuO_4$ into Eqs. (11, 13) gives numerically equal values of the incommensurabilities,   
\begin{equation}
 q_{co}(\hat{x})= q_{co}(x^*) = \sqrt{2} Q_{co}(x^*) \;, 
\end{equation}
all listed in Table II. This surprising result warrants closer inspection. Solving Eq. (14) for $x^*$ relates the $Sr$-doping that causes closing of the pseudogap (at $T=0$) with  watershed doping $\hat{x}$ and  tilt correction $\grave{t}$,
\begin{equation}
   x^* = \frac{3\hat{x} - \grave{t}}{2} \;.
\end{equation}

\begin{table}[ht]

\begin{tabular}{|p{4cm}|p{1.8cm}|p{2cm}|p{1.7cm}|p{1.8cm}|p{2cm}|p{2.2cm}|}
 \hline  \hline
Compound &$\;\;\;\;\;\;\hat{x}$&$\;\;\;\;\;\;\grave{t}$&$\;\;\;\;x^*$&$\;\;\;q_{co}(\hat{x}) $&$\;\;\; q_{co}(x^*) $&$\;\;\sqrt{2} Q_{co}(x^*) $\\
 \hline  \hline
$La_{2-x}Sr_xCuO_4$&$\;\;\;\;0.125 $&$\;\;\;0.015 $&$\;\;\;0.18 $&$\;\;\;\mathbf{0.235} $&$\;\;\;\mathbf{0.235} $&$\;\;\;\mathbf{0.235} $\\ \hline
$La_{1.6-x}Nd_{0.4}Sr_xCuO_4$&$\;\;\;\;0.16 $&$\;\;\;0.019 $&$\;\;\;0.23$&$\;\;\;\mathbf{0.266}$&$\;\;\;\mathbf{0.266}$&$\;\;\;\mathbf{0.266}$\\ \hline
 \hline
\end{tabular}
\label{table:2}
\footnotesize \caption{Watershed concentration $\hat{x}$, tilt correction $\grave{t}$, quantum critical point $x^*$, incommensurabilty of the saturated primary and secondary superlattice, $q_{co}$ and $Q_{co}$, respectively.} \normalsize
\end{table}

\section{SELECTION OF DATA}
The condition of Eq. (14) for the three parameters $\hat{x}$, $\grave{t}$ and $x^*$ is satisfied, within error bars, for the data of $La_{2-x}Sr_xCuO_4$, listed in Table II. The tilt correction $\grave{t} = 0.015 \pm 0.002$ is confirmed, using Eq. (10), by many measurements of charge-order and spin-order stripes in $La_{2-x}Ae_xCuO_4$ near $x = 1/8 = 0.125$.\cite{20,21,22,23,24,25,26,27,28,29,30,31,32,33,34,35,36,37,38,39,40,41,42,43,44,45,46,47,48,49,50,51,52,53,54,55,56,57,58,59,60,61,62,63}  Visually, it can be inferred from the accumulation of data points \textit{slightly above} the square-root curve in Fig. 11 (drawn with constant $\grave{t} = 0.02$) for $x > 0.09$ but coincident with a slightly upshifted curve in that range (not shown) when $\grave{t} = 0.02$/$\Omega^+ = 0.014$ is used. The value of the watershed doping in Table II, $\hat{x} = 0.125$, is obtained by Eq. (12) from a \textit{constant} charge-order incommensurability, $q_{co} = 0.235, x \ge \hat{x}$,  being within error bars of $q_{co} = 0.235 \pm 0.005$ r.l.u., observed with non-resonant hard X-rays, and $q_{co} = 0.231 \pm 0.005$ r.l.u., observed with resonant inelastic soft X-ray spectroscopy.\cite {31,32} The doping level where the pseudogap closes (at $T=0$), $x^* = 0.18 \pm 0.01$, has been determined with resistivity and Nernst-effect measurements.\cite{64,65}

Turning to the data for $La_{1.6-x}Nd_{0.4}Sr_xCuO_4$, a tilt correction $\grave{t} = 0.019$ is obtained with Eq. (10) from $q_{co} = 0.266 \pm  0.005$ r.l.u. of $La_{1.475}Nd_{0.4}Sr_{0.125}CuO_4$.\cite{66} 
Measurements of resistivity, Hall effect, Nernst effect, and thermoelectric power give $x^* = 0.23 \pm 0.01$.\cite{67,68,69} A \textit{constant} incommensurability of spin-order stripes, $q_{so} = 0.139 \pm 0.01$ $r.l.u.$ has been observed, at temperature $T = 1.5$ K and doping $x = 0.19, 0.24, 0.26$, with neutron scattering.\cite{70} With Eq. (12), 
$q_{co}=0.278 \pm 0.02$. This gives, by Eq. (11), a value of $\hat{x}=0.15 \pm 0.01$. 

\section{COLLAPSE OF THE PSEUDOGAP TEMPERATURE LINE $\mathbf{T^*(x)}$}

With doping beyond the watershed value, $x > \hat{x}$, {\it two} $O^{\upuparrows}$ superlattices interlace in the $CuO_2$ plane---the primary with incommensurability $q_{co}(x)$, Eq. (11), and the secondary  with $Q_{co}(x)$, Eq. (13).
At the quantum critical point $x^*$, Eq. (15) holds. Neglecting the  tilt correction $\grave{t} \ll \hat{x}$,
this gives the ratio of $O$ atoms  of the secondary and primary charge order,
\begin{equation}
\frac{x^* - \hat{x}}{\hat{x}} = \frac{(3 \hat{x} - \grave{t}) - 2\hat{x}}{2\hat{x}} =
\frac{\hat{x} - \grave{t}}{2 \hat{x}} \simeq \frac{1}{2} \;.
\end{equation}

Disregarding {\it all host ions}, Fig. 15 shows the $O^{\upuparrows/\downdownarrows}$ atoms in the $CuO_2$ plane where those of the primary superlattice are marked in blue and those of the secondary one in yellow and green. The doping level is at the quantum critical point, $x^*$. Every {\it other} primary superunit cell (blue) hosts one $O^{\upuparrows}$ atom (yellow or green) of the secondary superlattice, in agreement with Eq. (16). 

The marked rectangle in Fig. 15 contains a $4 \times 4 + 1$ island, shown in detail by Fig. 16. The center $O^{\upuparrows}$ atom in the island corresponds to the yellow dot in the marked rectangle while the four $O^{\upuparrows}$ atoms at the island's corners correspond to the blue corner dots.  
The condition for closing of the pseudogap, Eq. (14),
\begin{equation} \frac{q_{co}(x^*)}{Q_{co}(x^*)} =
    \frac{\mathbb{A}_0^{dia}(x^*)}{A_0(x^*)} = \sqrt{2} \;,
\end{equation}
is geometrically illustrated in Fig. 15 by the spacing (lattice constants) of the primary and (diagonal) secondary superlattice, $A_0(x^*)$ and $\mathbb{A}_0^{dia}(x^*)$
in terms of the blue-blue  and yellow-green distances, respectively.

At doping $x^*$, the $4 \times 4 + 1$ islands become contiguous at their corners (Figs. 15, 16). Now lines of {\it demotted} copper ions $\mathcolor{red}{\mathbf{Cu^{2+}}}$, bracketing the islands, extend throughout crystal domains. Charge transport at $T = 0$ is fully carried by lattice-constant hops of $Cu3d$-$O2p$ holes. The scenario coincides with the sudden drop of the $T$*$(x)$ line to  $T$*$(x^*) = 0$ (Fig. 5).

\medskip

\includegraphics[width=4.2in]{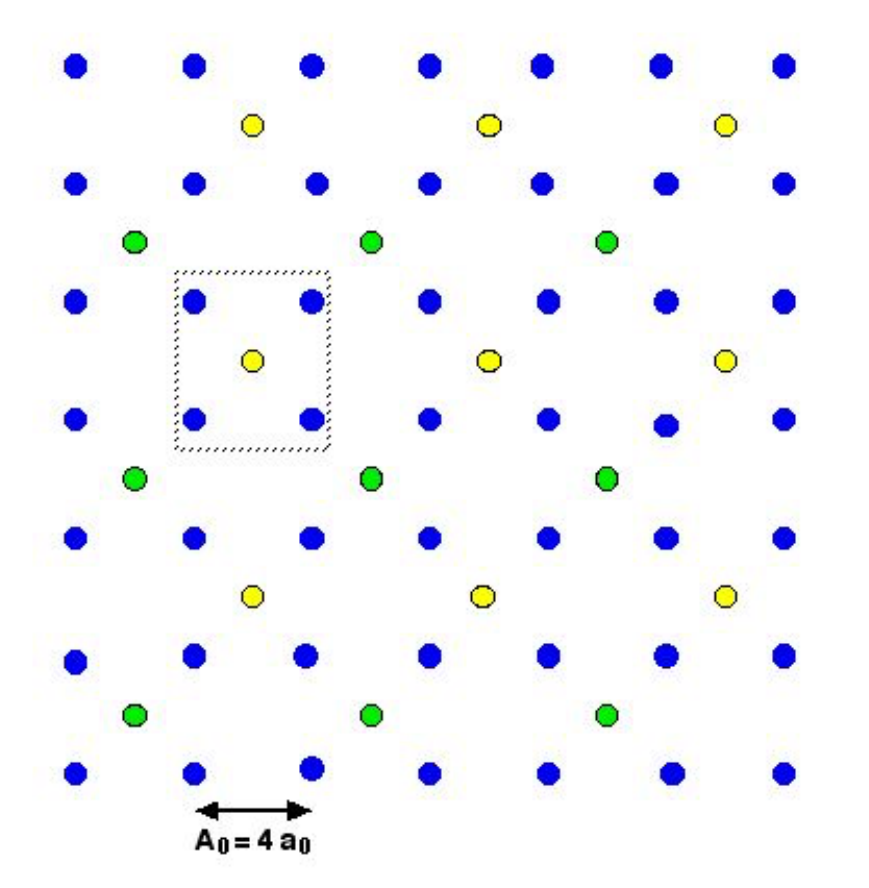} 

\noindent \footnotesize FIG. 15. Primary superlattice of $O^{\upuparrows}$ atoms (size exaggerated) in the $CuO_2$ plane (blue) and interlaced secondary $O^{\upuparrows}$ superlattice of diagonal orientation (yellow, green) at the quantum critical point $x^*$.  \normalsize

\noindent $\mathcolor{green}{\;\;\;\;\;\;\;\;\;\;\;\; _\uparrow O^{2-}} \;\;\;\;\;\;\;\;\;\;\;\;\;\;\; \mathcolor{green}{O^{2-} \;\;\;\;\;\;\;\;\;\;\;\;\;\;\; O^{2-}} \;\;\;\;\;\;\;\;\;\;\;\;\;\;\;\mathcolor{blue}{_\downarrow O^{2-} \;\;\;\;\;\;\;\;\;\;\;\;\;\;\;_\uparrow O^{2-}} \;\;\;\;\;\;\;\;\;\;\;\;\;\;\;\; \mathcolor{green}{O^{2-} \;\;\;\;\;\;\;\;\;\;\;\;\;\;\;\; O^{2-}}$

\noindent $ \mathcolor{green}{_\downarrow O^{2-}} \;\;\; \mathcolor{red}{\overleftarrow{\mathbf{Cu}}^{2+}} \; \mathcolor{green}{\mathbf{\uparrow} O^{2-}} \;\;\;\mathcolor{red}{\overrightarrow{\mathbf{Cu^{2+}}}} \;\;\;_\downarrow\mathbf{o}^\mathbf{\upuparrows} \;\;\;\;\mathcolor{red}{\overleftarrow{\mathbf{Cu^{2+}}}} \;\mathcolor{green}{\mathbf{\uparrow} O^{2-}} \;\;\;\mathcolor{red}{\overrightarrow{\mathbf{Cu}}^{2+}} \;\;\mathcolor{green}{_\downarrow O^{2-}} \;\;\;\;\mathcolor{red}{\overleftarrow{\mathbf{Cu}}^{2+}} \;\mathcolor{green}{\mathbf{\uparrow} O^{2-}} \;\;\;\mathcolor{red}{\overrightarrow{\mathbf{Cu^{2+}}}} \;\;_\downarrow\mathbf{o}^\mathbf{\downdownarrows} \;\;\;\;\mathcolor{red}{\overleftarrow{\mathbf{Cu^{2+}}}} \;\;\;\mathcolor{green}{\mathbf{\uparrow} O^{2-}}$  

\noindent $\mathcolor{blue}{\;\;\;\;\;\;\;\;\;\;\;\;\; O^{2-}} \;\;\;\;\;\;\;\;\;\;\;\;\;\;\; \mathcolor{green}{O^{2-} \;\;\;\;\;\;\;\;\;\;\;\;\;\;\; O^{2-}} \;\;\;\;\;\;\;\;\;\;\;\;\;\;\;\;\mathcolor{green}{ O^{2-} \;\;\;\;\;\;\;\;\;\;\;\;\;\;\;\; O^{2-}}\mathcolor{green}{ \;\;\;\;\;\;\;\;\;\;\;\;\;\;\;\; O^{2-} \;\;\;\;\;\;\;\;\;\;\;\;\;\;\;\; O^{2-}}$

\noindent $ \mathcolor{blue}{_\uparrow O^{2-}} \; \mathcolor{red}{_{\nearrow}{Cu^{2+}}} \;\; \mathcolor{blue}{_\downarrow O^{2-}} \;\;\;\;\mathcolor{red}{\overleftarrow{Cu}^{2+}} \;\;\mathcolor{green}{_\uparrow O^{2-}}\mathcolor{red}{ \;\;\; \mathcolor{red}{\overrightarrow{\mathbf{Cu}}^{2+}}} \;\mathcolor{green}{_\downarrow O^{2-}} \;\;\;\;\mathcolor{red}{\overleftarrow {\mathbf{Cu}}^{2+}} \;\;\mathcolor{green}{_\uparrow O^{2-}}\;\;\;\mathcolor{red}{ \overrightarrow{\mathbf{Cu}}^{2+}} \;\mathcolor{green}{_\downarrow O^{2-}} \;\;\;\;\mathcolor{red}{\overleftarrow{\mathbf{Cu}}^{2+}} \;\mathcolor{green}{_\uparrow O^{2-}} \;\;\;\;\mathcolor{red}{\overrightarrow{Cu}^{2+}} \;\;\mathcolor{blue}{_\downarrow O^{2-}}$ 

\noindent $\;\;\;\;\;\;\;\;\;\;\;\;\mathcolor{blue}{ _\uparrow O^{2-}} 
\;\;\;\;\;\;\;\;\;\;\;\;\;\;\mathcolor{blue}{_\downarrow O^{2-}} 
\mathcolor{green}{\;\;\;\;\;\;\;\;\;\;\;\;\;\;\; O^{2-} 
\;\;\;\;\;\;\;\;\;\;\;\;\;\;\;\;O^{2-} 
\;\;\;\;\;\;\;\;\;\;\;\;\;\;\;\; O^{2-} 
    \;\;\;\;\;\;\;\;\;\;\;\;\;\;\;\; O^{2-}} 
\;\;\;\;\;\;\;\;\;\;\;\;\;\;\;\mathcolor{blue}{_\uparrow O^{2-}}$

\noindent $ \mathcolor{blue}{_\downarrow O^{2-}} \;\; \mathcolor{red}{_{\swarrow}Cu^{2+}} \;\; \mathcolor{blue}{_\uparrow O^{2-}} \;\mathcolor{red}{_{\nearrow}  Cu^{2+}} \;\;\mathcolor{blue}{_\downarrow O^{2-}} \;\; \mathcolor{red}{\overleftarrow{\mathbf{Cu}}^{2+}} \;\; \mathcolor{green}{\mathbf{\uparrow} O^{2-}} \;\;\;\mathcolor{red}{\overrightarrow{\mathbf{Cu^{2+}}}} \;\;\;_\downarrow\mathbf{o}^\mathbf{\upuparrows} \;\;\;\;\mathcolor{red}{\overleftarrow{\mathbf{Cu^{2+}}}} \;\mathcolor{green}{\mathbf{\uparrow} O^{2-}} \;\;\;\mathcolor{red}{\overrightarrow{\mathbf{Cu}}^{2+}} \;\;\mathcolor{blue}{_\downarrow O^{2-}} \;\mathcolor{red}{_{\swarrow}  Cu^{2+}} \;\;\mathcolor{blue}{_\uparrow O^{2-}} $

    \noindent $\mathcolor{blue}{\;\;\;\;\;\;\;\;\;\;\;\;\; _\downarrow O^{2-}}
    \;\;\;\;\;\;\;\;\;\;\;\;\;\;\; \mathcolor{blue}{O^{2-}} 
    \mathcolor{green}{\;\;\;\;\;\;\;\;\;\;\;\;\;\;\;O^{2-}
    \;\;\;\;\;\;\;\;\;\;\;\;\;\;\;\; O^{2-} 
    \;\;\;\;\;\;\;\;\;\;\;\;\;\;\;\;\; O^{2-}} 
    \;\;\;\;\;\;\;\;\;\;\;\;\;\;\;\;\mathcolor{green}{ O^{2-}} 
    \;\;\;\;\;\;\;\;\;\;\;\;\;\;\;\mathcolor{blue}{_\downarrow O^{2-}}$

\noindent $ \mathcolor{blue}{_\uparrow O^{2-}} \; \mathcolor{red}{_{\nearrow}{Cu^{2+}}} \;\; \mathcolor{blue}{_\downarrow O^{2-}} \;\;\;\;\mathcolor{red}{\overleftarrow{Cu}^{2+}} \;\;\mathcolor{green}{_\uparrow O^{2-}}\mathcolor{red}{ \;\;\; \mathcolor{red}{\overrightarrow{\mathbf{Cu}}^{2+}}} \;\mathcolor{green}{_\downarrow O^{2-}} \;\;\;\;\mathcolor{red}{\overleftarrow {\mathbf{Cu}}^{2+}} \;\;\mathcolor{green}{_\uparrow O^{2-}}\;\;\;\mathcolor{red}{ \overrightarrow{\mathbf{Cu}}^{2+}} \;\mathcolor{green}{_\downarrow O^{2-}} \;\;\;\;\mathcolor{red}{\overleftarrow{\mathbf{Cu}}^{2+}} \;\mathcolor{green}{_\uparrow O^{2-}} \;\;\;\;\mathcolor{red}{\overrightarrow{Cu}^{2+}} \;\;\mathcolor{blue}{_\downarrow O^{2-}}$ 

\noindent $\mathcolor{blue}{\;\;\;\;\;\;\;\;\;\;\;\;\; O^{2-}} \;\;\;\;\;\;\;\;\;\;\;\;\;\;\;\; \mathcolor{green}{O^{2-} \;\;\;\;\;\;\;\;\;\;\;\;\;\;\; O^{2-}} \;\;\;\;\;\;\;\;\;\;\;\;\;\;\;\;\mathcolor{green}{ O^{2-} \;\;\;\;\;\;\;\;\;\;\;\;\;\;\;\; O^{2-}} \;\;\;\;\;\;\;\;\;\;\;\;\;\;\;\; \mathcolor{green}{O^{2-} \;\;\;\;\;\;\;\;\;\;\;\;\;\;\;\; O^{2-}}$

\noindent $ \mathcolor{green}{_\downarrow O^{2-}} \;\;\; \mathcolor{red}{\overleftarrow{\mathbf{Cu}}^{2+}} \; \mathcolor{green}{\mathbf{\uparrow} O^{2-}} \;\;\;\mathcolor{red}{\overrightarrow{\mathbf{Cu^{2+}}}} \;\;\;_\downarrow\mathbf{o}^\mathbf{\upuparrows} \;\;\;\;\mathcolor{red}{\overleftarrow{\mathbf{Cu^{2+}}}} \;\mathcolor{green}{\mathbf{\uparrow} O^{2-}} \;\;\;\mathcolor{red}{\overrightarrow{\mathbf{Cu}}^{2+}} \;\;\mathcolor{green}{_\downarrow O^{2-}} \;\;\;\;\mathcolor{red}{\overleftarrow{\mathbf{Cu}}^{2+}} \;\mathcolor{green}{\mathbf{\uparrow} O^{2-}} \;\;\;\mathcolor{red}{\overrightarrow{\mathbf{Cu^{2+}}}} \;\;_\downarrow\mathbf{o}^\mathbf{\downdownarrows} \;\;\;\;\mathcolor{red}{\overleftarrow{\mathbf{Cu^{2+}}}} \;\;\;\mathcolor{green}{\mathbf{\uparrow} O^{2-}}$  

\noindent $\mathcolor{green}{\;\;\;\;\;\;\;\;\;\;\;\;\; O^{2-}} \;\;\;\;\;\;\;\;\;\;\;\;\;\;\;\; \mathcolor{green}{O^{2-} \;\;\;\;\;\;\;\;\;\;\;\;\;\;\; O^{2-}} \;\;\;\;\;\;\;\;\;\;\;\;\;\;\;\mathcolor{blue}{_\uparrow O^{2-} \;\;\;\;\;\;\;\;\;\;\;\;\;\;\;_\downarrow O^{2-}} \;\;\;\;\;\;\;\;\;\;\;\;\;\;\;\; \mathcolor{green}{O^{2-} \;\;\;\;\;\;\;\;\;\;\;\;\;\;\;\; O^{2-}}$

\noindent \footnotesize FIG. 16. $CuO_2$ plane of $La_{2-x}Sr_xCuO_4$ 
at the quantum critical point $x^*$ where $4 \times 4 + 1$ islands become continguous at their corners. The environment of the $O^{\upuparrows}$ ($O^{\downdownarrows}$) atom at the top left (right) corner is equivalent to that of the $O^{\downdownarrows}$ ($O^{\upuparrows}$) atom at the bottom right (left) corner. Thus, demotted copper ions, $\mathcolor{red}{\mathbf{Cu^{2+}}}$, along the top and bottom $Cu$-$O$ rows, extend throughout crystal domains. \normalsize

\section{LOW-TEMPERATURE CRYSTAL PHASES OF $\mathbf{p}$-DOPED {\it \textbf{214}} CUPRATES}

Table II gives the watershed doping $\hat{x} = 0.125$ for $La_{2-x}Sr_xCuO_4$ but $\hat{x} = 0.16$ for $La_{1.6-x}Nd_{0.4}Sr_xCuO_4$. By Eq. (15), the quantum critical points are shifted,  accordingly, i. e., to $x^* = 0.18$ for the former and to $x^* = 0.23$ for the latter. What causes the difference? The difference is caused by different low-temperature crystal phases due to the co-dopant $Nd$. In the doping range, $x < 0.21$, $La_{2-x}Sr_xCuO_4$ is in the low-temperature orthogonal (LTO) phase. $La_{1.6-x}Nd_{0.4}Sr_xCuO_4$ is also in the LTO phase for $0.01 < x < 0.07$ but in the low-temperature tetragonal (LTT) phase for $0.12 < x < 0.24$, followed by the high-temperature tetragonal (HTT) phase for $0.27 < x < 0.40$, with transitional ion configurations between. 

In the LTT phase, where $a_0 = b_0$, it is energetically favorable that the host octahedra $CuO_6$ tilt {\it axially}, keeping two base $O^{2-}$ ions on the $CuO_2$ plane while the other two tilt above and beneath. The tilt axis passes through the {\it stationary} base ions and the center $Cu^{2+}$ ion. The orientation of the tilt axis alternates in subsequent $CuO_2$ planes (along the  $c$ axis), that is, in the [010] and [100] direction (Figs. 17, 18). As an $O^{\upuparrows}$ atom is always shared in the base of neighboring $CuO_5O^{\upuparrows}$ octahedra, the latter tilt toward or away from each other. The result is an axial tilt like that of the host $CuO_6$ octahedra in the LTT phase, but heavier. The common tilt direction of both kind of octahedra in the LTT phase accommodates the $CuO_5O^{\upuparrows}$ octahedra. This is in contrast to the LTO phase where the diagonal tilt of the host $CuO_6$ octahedra impedes the axial tilt of $CuO_5O^{\upuparrows}$ octahedra. Thus a higher $O^{\upuparrows}$ concentration (higher watershed $\hat{x}$) is possible in the LTT phase before octahedra tilts get in each other's way---for $La_{1.6-x}Nd_{0.4}Sr_xCuO_4$ at $\hat{x} = 0.16$.

\medskip

\includegraphics[width= 2.4in]{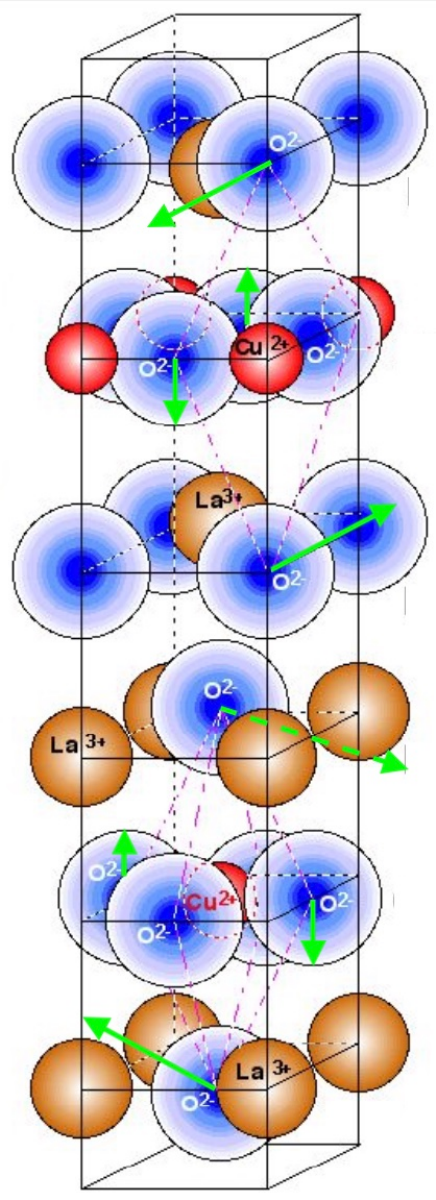} 

\noindent \footnotesize FIG. 17. Full (bottom) and half (top) $CuO_6$ octahedron, indicated by dashed lines, in two $LaCuO_4$ unit cells (exploded along $c$). Green arrows show {\it exaggerated} the displacement of $O^{2-}$ ions to tilt the $CuO_6$ octahedra {\it axially} in the (doped) LTT phase. \normalsize 

\pagebreak

For a quantitative assessment we resort to the generalized formula for the {\it average} spacing of an incommensurate $\tilde {n} \times \tilde {n}$ superlattice, $\tilde{n} = \sqrt{2/x} = A_0(x)/a_0$. For $La_{1.6-x}Nd_{0.4}Sr_xCuO_4$ with $\hat{x} = 0.16$, this gives $\tilde{n} = \sqrt{2/0.16} \simeq 3.5$, that is, an average spacing of the saturated  primary $O^{\upuparrows}$ superlattice $A_0(x) \simeq 3.5 a_0$ ($x>\hat{x})$. Doubling this spacing leads to a {\it composite commensurate}  $(4 + 3) \times (4 + 3)$  superlattice of $O^{\upuparrows}$ atoms (see Appendix C),  shown in Fig. 19.  It is an arrangement of $4 \times 4$  and $3 \times 3$ squares, joined by $4 \times 3$ rectangles. At doping $\hat{x} < x < x^*$, a secondary superlattice grows by filling the  centers of the squares with $O^{\upuparrows}$ atoms. When the quantum critical point $x^* = 0.23$ is reached (Fig. 20), lines of demotted $\mathcolor{red}{\mathbf{Cu^{2+}}}$ ions extend along each $(4 + 3) \times (4 + 3)$ superlattice and throughout the crystal domain.

\bigskip

\noindent $\mathcolor{blue}{\;\;\;\;\;\;\;\;\;\;\; _\uparrow O^{2-} \;\;\;\;\;\;\;\;\;\;\;\;\;_\downarrow O^{2-} \;\;\;\;\;\;\;\;\;\;\;\;\;\;_\uparrow O^{2-} \;\;\;\;\;\;\;\;\;\;\;\;\;\;_\downarrow O^{2-} \;\;\;\;\;\;\;\;\;\;\;\;\;\;\;_\uparrow O^{2-} \;\;\;\;\;\;\;\;\;\;\;\;\;\;_\downarrow O^{2-} \;\;\;\;\;\;\;\;\;\;\;\;\;\;_\uparrow O^{2-}}$

\noindent $\mathcolor{blue}{ \; O^{2-}} \;\;\; \mathcolor{red}{\mathbf{_\downarrow} {Cu}^{2+}} \;\; \mathcolor{blue}{O^{2-}}\;\;\; \mathcolor{red}{\mathbf{_\uparrow}{Cu}^{2+}} \;\; \mathcolor{blue}{O^{2-}} \;\;\;\mathcolor{red}{\mathbf{_\downarrow} {Cu}^{2+}} \;\; \mathcolor{blue}{O^{2-}} \;\;\;\mathcolor{red}{\mathbf{_\uparrow}{Cu}^{2+}} \;\; \mathcolor{blue}{O^{2-}} \;\;\;\;\mathcolor{red}{\mathbf{_\downarrow} {Cu}^{2+}} \;\; \mathcolor{blue}{O^{2-}} \;\;\mathcolor{red}{\mathbf{\uparrow}{Cu}^{2+}} \;\;\; \mathcolor{blue}{O^{2-}} \;\;\;\mathcolor{red}{\mathbf{_\downarrow} {Cu}^{2+}} \;\; \mathcolor{blue}{O^{2-}} $ 

    \noindent $\mathcolor{blue}{\;\;\;\;\;\;\;\;\;\;\; _\downarrow O^{2-} \;\;\;\;\;\;\;\;\;\;\;\;\;_\uparrow O^{2-} \;\;\;\;\;\;\;\;\;\;\;\;\;\;_\downarrow O^{2-} \;\;\;\;\;\;\;\;\;\;\;\;\;\;_\uparrow O^{2-} \;\;\;\;\;\;\;\;\;\;\;\;\;\;\;_\downarrow O^{2-}} \;\;\;\;\;\;\;\;\;\;\;\;\;\;\mathcolor{green}{_\uparrow O^{2-}} \;\;\;\;\;\;\;\;\;\;\;\;\;\;\mathcolor{blue}{_\downarrow O^{2-}}$

\noindent $\mathcolor{blue}{ \; O^{2-}} \;\;\; \mathcolor{red}{\mathbf{_\uparrow} {Cu}^{2+}} \;\; \mathcolor{blue}{O^{2-}}\;\;\; \mathcolor{red}{\mathbf{_\downarrow}{Cu}^{2+}} \;\; \mathcolor{blue}{O^{2-}} \;\;\;\mathcolor{red}{\mathbf{_\uparrow} {Cu}^{2+}} \;\; \mathcolor{blue}{O^{2-}} \;\;\;\mathcolor{red}{\mathbf{_\downarrow}{Cu}^{2+}} \;\; \mathcolor{blue}{O^{2-}} \;\;\;\;\mathcolor{red}{\mathbf{\uparrow} {Cu}^{2+}} \;\; \mathcolor{green}{O^{2-}} \;\;\mathcolor{red}{\mathbf{\Downarrow} \mathbf{Cu^{2+}}} \;\; \mathcolor{green}{O^{2-}} \;\;\mathcolor{red}{\mathbf{\uparrow} {Cu}^{2+}} \;\; \mathcolor{blue}{O^{2-}} $ 

    \noindent $\mathcolor{blue}{\;\;\;\;\;\;\;\;\;\;\; _\uparrow O^{2-} \;\;\;\;\;\;\;\;\;\;\;\;\;_\downarrow O^{2-} \;\;\;\;\;\;\;\;\;\;\;\;\;\;_\uparrow O^{2-} \;\;\;\;\;\;\;\;\;\;\;\;\;\;_\downarrow O^{2-}} \;\;\;\;\;\;\;\;\;\;\;\;\;\;\;\mathcolor{green}{_\uparrow O^{2-}} \;\;\;\;\;\;\;\;\;\;\;\;\;\;\;_\downarrow \mathbf{o}^{\upuparrows} \;\;\;\;\;\;\;\;\;\;\;\;\;\;\;\mathcolor{green}{_\uparrow O^{2-}}$

\noindent $\mathcolor{blue}{ \; O^{2-}} \;\;\; \mathcolor{red}{\mathbf{_\downarrow} {Cu}^{2+}} \;\; \mathcolor{blue}{O^{2-}}\;\;\; \mathcolor{red}{\mathbf{_\uparrow}{Cu}^{2+}} \;\; \mathcolor{blue}{O^{2-}} \;\;\;\mathcolor{red}{\mathbf{_\downarrow} {Cu}^{2+}} \;\; \mathcolor{blue}{O^{2-}} \;\;\;\mathcolor{red}{\mathbf{_\uparrow}{Cu}^{2+}} \;\; \mathcolor{blue}{O^{2-}} \;\;\;\;\mathcolor{red}{\mathbf{\downarrow} {Cu}^{2+}} \;\; \mathcolor{green}{O^{2-}} \;\;\mathcolor{red}{\mathbf{\Uparrow}\mathbf{Cu^{2+}}} \;\; \mathcolor{green}{O^{2-}} \;\;\mathcolor{red}{\mathbf{\downarrow} {Cu}^{2+}} \;\; \mathcolor{blue}{O^{2-}}$ 

    \noindent $\mathcolor{blue}{\;\;\;\;\;\;\;\;\;\;\; _\downarrow O^{2-} \;\;\;\;\;\;\;\;\;\;\;\;\;_\uparrow O^{2-} \;\;\;\;\;\;\;\;\;\;\;\;\;\;_\downarrow O^{2-} \;\;\;\;\;\;\;\;\;\;\;\;\;\;_\uparrow O^{2-} \;\;\;\;\;\;\;\;\;\;\;\;\;\;\;_\downarrow O^{2-}} \;\;\;\;\;\;\;\;\;\;\;\;\;\;\mathcolor{green}{_\uparrow O^{2-}} \;\;\;\;\;\;\;\;\;\;\;\;\;\;\mathcolor{blue}{_\downarrow O^{2-}}$

\noindent $\mathcolor{blue}{ \; O^{2-}} \;\;\; \mathcolor{red}{\mathbf{_\uparrow} {Cu}^{2+}} \;\; \mathcolor{blue}{O^{2-}}\;\;\; \mathcolor{red}{\mathbf{_\downarrow}{Cu}^{2+}} \;\; \mathcolor{blue}{O^{2-}} \;\;\;\mathcolor{red}{\mathbf{_\uparrow} {Cu}^{2+}} \;\; \mathcolor{blue}{O^{2-}} \;\;\;\mathcolor{red}{\mathbf{_\downarrow}{Cu}^{2+}} \;\; \mathcolor{blue}{O^{2-}} \;\;\;\;\mathcolor{red}{\mathbf{_\uparrow} {Cu}^{2+}} \;\; \mathcolor{blue}{O^{2-}} \;\;\mathcolor{red}{\mathbf{\downarrow}{Cu}^{2+}} \;\;\; \mathcolor{blue}{O^{2-}} \;\;\;\mathcolor{red}{\mathbf{_\uparrow} {Cu}^{2+}} \;\; \mathcolor{blue}{O^{2-}} $ 

    \noindent $\mathcolor{blue}{\;\;\;\;\;\;\;\;\;\;\; _\uparrow O^{2-} \;\;\;\;\;\;\;\;\;\;\;\;\;_\downarrow O^{2-} \;\;\;\;\;\;\;\;\;\;\;\;\;\;_\uparrow O^{2-} \;\;\;\;\;\;\;\;\;\;\;\;\;\;_\downarrow O^{2-} \;\;\;\;\;\;\;\;\;\;\;\;\;\;\;_\uparrow O^{2-} \;\;\;\;\;\;\;\;\;\;\;\;\;\;_\downarrow O^{2-} \;\;\;\;\;\;\;\;\;\;\;\;\;\;_\uparrow O^{2-}}$

\noindent |||||||||||||||||||||||||||||||||||||||||

    \noindent $\mathcolor{blue}{
    O^{2-} \;\;\;\;\;\;\;\;\;\;\;\;\;\;\;\;\; 
    O^{2-} \;\;\;\;\;\;\;\;\;\;\;\;\;\;\;\; 
    O^{2-} \;\;\;\;\;\;\;\;\;\;\;\;\;\;\;\; 
    O^{2-} \;\;\;\;\;\;\;\;\;\;\;\;\;\;\;\;\; 
    O^{2-} \;\;\;\;\;\;\;\;\;\;\;\;\;\;\;\;
    O^{2-} \;\;\;\;\;\;\;\;\;\;\;\;\;\;\;\; 
    O^{2-}  \;\;\;\;\;\;\;\;\;\;\;\;\;\;\;\; O^{2-}}$

\noindent $\mathcolor{red}{\overrightarrow{C}u^{2+}} \;\; \mathcolor{blue}{_\downarrow O^{2-}} \;\;\;\mathcolor{red}{\overleftarrow{Cu}^{2+}} \;\;\mathcolor{green}{_\uparrow O^{2-}} \;\;\;\mathcolor{red}{\overrightarrow {Cu}^{2+}} \;\;\mathcolor{blue}{_\downarrow O^{2-}} \;\;\;\mathcolor{red}{\overleftarrow {C}u^{2+}} \;\;\mathcolor{blue}{_\uparrow O^{2-}} \;\;\;\;\mathcolor{red}{\overrightarrow {C}u^{2+}} \;\;\mathcolor{blue}{_\downarrow O^{2-}} \;\;\;\mathcolor{red}{\overleftarrow   {C}u^{2+}} \;\;\;\mathcolor{blue}{_\uparrow O^{2-}} \;\;\;\mathcolor{red}{\overrightarrow {C}u^{2+}} \;\;\mathcolor{blue}{_\downarrow O^{2-}} \;\;\;\mathcolor{red}{\overleftarrow {C}u^{2+}}$ 

  \noindent $\mathcolor{blue}{O^{2-}} \;\;\;\;\;\;\;\;\;\;\;\;\;\;\;\;\; 
    \mathcolor{green}{O^{2-} \;\;\;\;\;\;\;\;\;\;\;\;\;\;\;\; 
    O^{2-}} \;\;\;\;\;\;\;\;\;\;\;\;\;\;\;\; 
    \mathcolor{blue}{O^{2-} \;\;\;\;\;\;\;\;\;\;\;\;\;\;\;\;\; 
    O^{2-} \;\;\;\;\;\;\;\;\;\;\;\;\;\;\;\;
    O^{2-} \;\;\;\;\;\;\;\;\;\;\;\;\;\;\;\; 
    O^{2-}  \;\;\;\;\;\;\;\;\;\;\;\;\;\;\;\; O^{2-}}$

\noindent $ \mathcolor{red}{\overleftarrow{Cu}^{2+}} \;\;\mathcolor{green}{ _\uparrow O^{2-}} \;\;\;\;\mathcolor{red}{\overrightarrow{\mathbf{Cu^{2+}}}} \;\;\;\mathcolor{black}{_\downarrow} \mathbf{o}^{\upuparrows} \;\;\;\mathcolor{red}{\overleftarrow {\mathbf{Cu^{2+}}}} \;\mathcolor{green}{_\uparrow O^{2-}} \;\;\;\mathcolor{red}{\overrightarrow{Cu}^{2+}} \;\;\mathcolor{blue}{_\downarrow O^{2-}} \;\;\;\;\mathcolor{red}{\overleftarrow {C}u^{2+}} \;\;\mathcolor{blue}{_\uparrow O^{2-}} \;\;\;\mathcolor{red}{\overrightarrow{C}u^{2+}} \;\;\;\mathcolor{blue}{_\downarrow O^{2-}} \;\;\;\mathcolor{red}{\overleftarrow {C}u^{2+}} \;\;\mathcolor{blue}{_\uparrow O^{2-}}  \;\;\;\mathcolor{red}{\overrightarrow {C}u^{2+}} $ 

  \noindent $\mathcolor{blue}{O^{2-}} \;\;\;\;\;\;\;\;\;\;\;\;\;\;\;\;\; 
    \mathcolor{green}{O^{2-} \;\;\;\;\;\;\;\;\;\;\;\;\;\;\;\; 
    O^{2-}} \;\;\;\;\;\;\;\;\;\;\;\;\;\;\;\; 
    \mathcolor{blue}{O^{2-} \;\;\;\;\;\;\;\;\;\;\;\;\;\;\;\;\; 
    O^{2-} \;\;\;\;\;\;\;\;\;\;\;\;\;\;\;\;
    O^{2-} \;\;\;\;\;\;\;\;\;\;\;\;\;\;\;\; 
    O^{2-}  \;\;\;\;\;\;\;\;\;\;\;\;\;\;\;\; O^{2-}}$

\noindent $\mathcolor{red}{\overrightarrow{C}u^{2+}} \;\; \mathcolor{blue}{_\downarrow O^{2-}} \;\;\;\mathcolor{red}{\overleftarrow{Cu}^{2+}} \;\;\mathcolor{green}{_\uparrow O^{2-}} \;\;\;\mathcolor{red}{\overrightarrow {Cu}^{2+}} \;\;\mathcolor{blue}{_\downarrow O^{2-}} \;\;\;\mathcolor{red}{\overleftarrow {C}u^{2+}} \;\;\mathcolor{blue}{_\uparrow O^{2-}} \;\;\;\;\mathcolor{red}{\overrightarrow {C}u^{2+}} \;\;\mathcolor{blue}{_\downarrow O^{2-}} \;\;\;\mathcolor{red}{\overleftarrow   {C}u^{2+}} \;\;\;\mathcolor{blue}{_\uparrow O^{2-}} \;\;\;\mathcolor{red}{\overrightarrow {C}u^{2+}} \;\;\mathcolor{blue}{_\downarrow O^{2-}} \;\;\;\mathcolor{red}{\overleftarrow {C}u^{2+}}$ 

   \noindent $\mathcolor{blue}{
    O^{2-} \;\;\;\;\;\;\;\;\;\;\;\;\;\;\;\;\; 
    O^{2-} \;\;\;\;\;\;\;\;\;\;\;\;\;\;\;\; 
    O^{2-} \;\;\;\;\;\;\;\;\;\;\;\;\;\;\;\; 
    O^{2-} \;\;\;\;\;\;\;\;\;\;\;\;\;\;\;\;\; 
    O^{2-} \;\;\;\;\;\;\;\;\;\;\;\;\;\;\;\;
    O^{2-} \;\;\;\;\;\;\;\;\;\;\;\;\;\;\;\; 
    O^{2-}  \;\;\;\;\;\;\;\;\;\;\;\;\;\;\;\; O^{2-}}$

\noindent \footnotesize FIG. 18. Two subsequent $CuO_2$ planes (along the $c$ axis) in the LTT phase with {\it slight axial} tilt of $CuO_6$ octahedra along [010] (top part) and along [100] (bottom part), indicated by vertical subarrow $\mathcolor{red}{_\uparrow}|\mathcolor{red}{_\downarrow}$ or $\mathcolor{red}{\overrightarrow{s}}hort$ overarrow at $\mathcolor{red}{Cu^{2+}}$; {\it heavy axial} tilt (vertical double arrow $\mathcolor{red}{\mathbf{\Uparrow}}$ or $\mathcolor{red}{\overrightarrow{long}}$ overarrow) of $CuO_5O^{\upuparrows}$ octahedra; and moderate tilt of neighboring $CuO_6$ octahedra (single vertical arrow $\mathcolor{red}{\mathbf{\uparrow}}$ or $\mathcolor{red}{\overrightarrow{me}}dium$ overarrow). 
Vertical subarrows $\mathcolor{blue}{_\uparrow}|\mathcolor{blue}{_\downarrow}$ at $\mathcolor{blue}{O^{2-}}$ or $\mathbf{o}$ indicate their position above or below the $CuO_2$ plane.
\normalsize

\pagebreak

Barium-doped $La_{2-x}Ba_xCuO_4$   is mostly in the LTO phase except for the LTT phase in a  region around $ x =0.125$. This shifts the watershed doping slightly up to $\hat{x} = 0.14$ instead of $\hat{x} = 0.125$ for LTO-based $La_{2-x}Sr_xCuO_4$  (see Fig. 11).\cite{59,60} Quadrupling $\tilde{n} = \sqrt{2/0.14} = 3.78$  for the average spacing of an incommensurate $\tilde{n} \times \tilde{n}$ $O^{\upuparrows}$ of the  primary superlattice, $4\tilde{n} \simeq 15$, leads to a {\it commensurate} $(4+4+4+3)\times(4+4+4+3)$ primary $O^{\upuparrows}$ superlattice (not shown).

\bigskip
 
\noindent $\mathcolor{green}{\;\;\;\;\;\;\;\;\;\;\;\;O^{2-}}\;\;\;\;\;\;\;\;\;\;\;\;\;\;\; \mathcolor{blue}{O^{2-} \;\;\;\;\;\;\;\;\;\;\;\;\;\;\;\; O^{2-}} \;\;\;\;\;\;\;\;\;\;\;\;\;\;\; 
    \mathcolor{green}{O^{2-} \;\;\;\;\;\;\;\;\;\;\;\;\;\;\;\; O^{2-}} 
    \mathcolor{blue}{ \;\;\;\;\;\;\;\;\;\;\;\;\;\;\;
    O^{2-}} \;\;\;\;\;\;\;\;\;\;\;\;\;\;\;\; \mathcolor{green}{O^{2-}}$

\noindent $
\;\;\mathcolor{black}{_\downarrow} \mathbf{o}^{\upuparrows}\;\;\;
\mathcolor{red}{\overleftarrow{\mathbf{Cu^{2+}}}} \;\mathcolor{green}{_\uparrow O^{2-}} \;\;\;
\mathcolor{red}{\overrightarrow {Cu}^{2+}}\;\;\mathcolor{blue}{_\downarrow O^{2-}}\;\;\;
\mathcolor{red}{\overleftarrow {Cu}^{2+}} \;\;\mathcolor{green}{_\uparrow O^{2-}}\;\;\;\;
\mathcolor{red}{\overrightarrow{\mathbf{Cu^{2+}}}}\;\mathcolor{black}{_\downarrow} \mathbf{o}^{\downdownarrows}\;\;\;\;
\mathcolor{red}{\overleftarrow{\mathbf{Cu^{2+}}}}\;\;\mathcolor{green}{_\uparrow O^{2-}} \;\;\;
\mathcolor{red}{\overrightarrow {Cu}^{2+}}\;\;\mathcolor{green}{_\downarrow O^{2-}}\;\;\;
\mathcolor{red}{\overleftarrow {\mathbf{Cu^{2+}}}}\;\;\;\mathcolor{black}{_\uparrow} \mathbf{o}^{\upuparrows}$ 

\noindent $\mathcolor{green}{\;\;\;\;\;\;\;\;\;\;\;\;O^{2-}}\;\;\;\;\;\;\;\;\;\;\;\;\;\;\; \mathcolor{blue}{O^{2-} \;\;\;\;\;\;\;\;\;\;\;\;\;\;\;\; O^{2-}} \;\;\;\;\;\;\;\;\;\;\;\;\;\;\; 
    \mathcolor{green}{O^{2-} \;\;\;\;\;\;\;\;\;\;\;\;\;\;\;\; O^{2-}} 
    \mathcolor{blue}{ \;\;\;\;\;\;\;\;\;\;\;\;\;\;\;
    O^{2-}} \;\;\;\;\;\;\;\;\;\;\;\;\;\;\;\; \mathcolor{green}{O^{2-}}$

\noindent $\mathcolor{green}{_\uparrow O^{2-}}\;\;\;
\mathcolor{red}{\overrightarrow{Cu}^{2+}} \;\;\mathcolor{blue}{_\downarrow O^{2-}} \;\;\;
\mathcolor{red}{\overleftarrow {C}u^{2+}}\;\;\mathcolor{blue}{_\uparrow O^{2-}}\;\;\;
\mathcolor{red}{\overrightarrow{C}u^{2+}}\;\;\mathcolor{blue}{_\downarrow O^{2-}}\;\;\;
\mathcolor{red}{\overleftarrow{Cu}^{2+}}\;\;\mathcolor{green}{_\uparrow O^{2-}}\;\;\;
\mathcolor{red}{\overrightarrow{Cu}^{2+}}\;\;\;\mathcolor{blue}{_\downarrow O^{2-}}\;\;
\mathcolor{red}{\overleftarrow {C}u^{2+}}\;\;\mathcolor{blue}{_\uparrow O^{2-}}\;\;\;
\mathcolor{red}{\overrightarrow {Cu}^{2+}}\;\;\mathcolor{green}{_\downarrow O^{2-}}$ 

    \noindent $\mathcolor{blue}{
    \;\;\;\;\;\;\;\;\;\;\;\;
    O^{2-} \;\;\;\;\;\;\;\;\;\;\;\;\;\;\; 
    O^{2-} \;\;\;\;\;\;\;\;\;\;\;\;\;\;\;\; 
    O^{2-} \;\;\;\;\;\;\;\;\;\;\;\;\;\;\; 
    O^{2-} \;\;\;\;\;\;\;\;\;\;\;\;\;\;\;\; 
    O^{2-} \;\;\;\;\;\;\;\;\;\;\;\;\;\;\;
    O^{2-} \;\;\;\;\;\;\;\;\;\;\;\;\;\;\;\; O^{2-}}$

\noindent $\mathcolor{blue}{_\downarrow O^{2-}}\;\;\;
\mathcolor{red}{\overleftarrow{C}u^{2+}} \;\;\mathcolor{blue}{_\uparrow O^{2-}} \;\;
\mathcolor{red}{\overrightarrow {C}u^{2+}}\;\;\mathcolor{blue}{_\downarrow O^{2-}}\;\;\;
\mathcolor{red}{\overleftarrow {C}u^{2+}} \;\;\mathcolor{blue}{_\uparrow O^{2-}}\;\;
\mathcolor{red}{\overrightarrow{C}u^{2+}}\;\;\mathcolor{blue}{_\downarrow O^{2-}}\;\;\;
\mathcolor{red}{\overleftarrow{C}u^{2+}} \;\;\mathcolor{blue}{_\uparrow O^{2-}} \;\;
\mathcolor{red}{\overrightarrow {C}u^{2+}}\;\;\mathcolor{blue}{_\downarrow O^{2-}}\;\;\;
\mathcolor{red}{\overleftarrow {C}u^{2+}}\;\;\mathcolor{blue}{_\downarrow O^{2-}}$ 

    \noindent $\mathcolor{blue}{
    \;\;\;\;\;\;\;\;\;\;\;\;
    O^{2-} \;\;\;\;\;\;\;\;\;\;\;\;\;\;\; 
    O^{2-} \;\;\;\;\;\;\;\;\;\;\;\;\;\;\;\; 
    O^{2-} \;\;\;\;\;\;\;\;\;\;\;\;\;\;\; 
    O^{2-} \;\;\;\;\;\;\;\;\;\;\;\;\;\;\;\; 
    O^{2-} \;\;\;\;\;\;\;\;\;\;\;\;\;\;\;
    O^{2-} \;\;\;\;\;\;\;\;\;\;\;\;\;\;\;\; O^{2-}}$

\noindent $\mathcolor{green}{_\uparrow O^{2-}}\;\;\;
\mathcolor{red}{\overrightarrow{Cu}^{2+}} \;\;\mathcolor{blue}{_\downarrow O^{2-}} \;\;\;
\mathcolor{red}{\overleftarrow {C}u^{2+}}\;\;\mathcolor{blue}{_\uparrow O^{2-}}\;\;\;
\mathcolor{red}{\overrightarrow{C}u^{2+}}\;\;\mathcolor{blue}{_\downarrow O^{2-}}\;\;\;
\mathcolor{red}{\overleftarrow{Cu}^{2+}}\;\;\mathcolor{green}{_\uparrow O^{2-}}\;\;\;
\mathcolor{red}{\overrightarrow{Cu}^{2+}}\;\;\;\mathcolor{blue}{_\downarrow O^{2-}}\;\;
\mathcolor{red}{\overleftarrow {C}u^{2+}}\;\;\mathcolor{blue}{_\uparrow O^{2-}}\;\;\;
\mathcolor{red}{\overrightarrow {Cu}^{2+}}\;\;\mathcolor{green}{_\downarrow O^{2-}}$ 

    \noindent $\mathcolor{green}{\;\;\;\;\;\;\;\;\;\;\;\;O^{2-}} 
    \mathcolor{blue}{\;\;\;\;\;\;\;\;\;\;\;\;\;\;\; 
    O^{2-} \;\;\;\;\;\;\;\;\;\;\;\;\;\;\;\; 
    O^{2-}} \;\;\;\;\;\;\;\;\;\;\;\;\;\;\; 
    \mathcolor{green}{O^{2-} \;\;\;\;\;\;\;\;\;\;\;\;\;\;\;\; 
    O^{2-}} \;\;\;\;\;\;\;\;\;\;\;\;\;\;\;
    \mathcolor{blue}{O^{2-}} \;\;\;\;\;\;\;\;\;\;\;\;\;\;\;\; \mathcolor{green}{O^{2-}}$

\noindent $
\;\;\mathcolor{black}{_\downarrow} \mathbf{o}^{\downdownarrows}\;\;\;
\mathcolor{red}{\overleftarrow{\mathbf{Cu^{2+}}}} \;\mathcolor{green}{_\uparrow O^{2-}} \;\;\;
\mathcolor{red}{\overrightarrow {Cu}^{2+}}\;\;\mathcolor{blue}{_\downarrow O^{2-}}\;\;\;
\mathcolor{red}{\overleftarrow {Cu}^{2+}} \;\;\mathcolor{green}{_\uparrow O^{2-}}\;\;\;\;
\mathcolor{red}{\overrightarrow{\mathbf{Cu^{2+}}}}\;\mathcolor{black}{_\downarrow} \mathbf{o}^{\upuparrows}\;\;\;\;
\mathcolor{red}{\overleftarrow{\mathbf{Cu^{2+}}}}\;\;\mathcolor{green}{_\uparrow O^{2-}} \;\;\;
\mathcolor{red}{\overrightarrow {Cu}^{2+}}\;\;\mathcolor{green}{_\downarrow O^{2-}}\;\;\;
\mathcolor{red}{\overleftarrow {\mathbf{Cu^{2+}}}}\;\;\;\;\mathcolor{black}{_\uparrow} \mathbf{o}^{\downdownarrows}$ 

    \noindent $\mathcolor{green}{\;\;\;\;\;\;\;\;\;\;\;\;O^{2-}} 
    \mathcolor{blue}{\;\;\;\;\;\;\;\;\;\;\;\;\;\;\; 
    O^{2-} \;\;\;\;\;\;\;\;\;\;\;\;\;\;\;\; 
    O^{2-}} \;\;\;\;\;\;\;\;\;\;\;\;\;\;\; 
    \mathcolor{green}{O^{2-} \;\;\;\;\;\;\;\;\;\;\;\;\;\;\;\; 
    O^{2-}} \;\;\;\;\;\;\;\;\;\;\;\;\;\;\;
    \mathcolor{blue}{O^{2-}} \;\;\;\;\;\;\;\;\;\;\;\;\;\;\;\; \mathcolor{green}{O^{2-}}$

\noindent $\mathcolor{green}{_\uparrow O^{2-}}\;\;\;
\mathcolor{red}{\overrightarrow{Cu}^{2+}} \;\;\mathcolor{blue}{_\downarrow O^{2-}} \;\;\;
\mathcolor{red}{\overleftarrow {C}u^{2+}}\;\;\mathcolor{blue}{_\uparrow O^{2-}}\;\;\;
\mathcolor{red}{\overrightarrow{C}u^{2+}}\;\;\mathcolor{blue}{_\downarrow O^{2-}}\;\;\;
\mathcolor{red}{\overleftarrow{Cu}^{2+}}\;\;\mathcolor{green}{_\uparrow O^{2-}}\;\;\;
\mathcolor{red}{\overrightarrow{Cu}^{2+}}\;\;\;\mathcolor{blue}{_\downarrow O^{2-}}\;\;\;
\mathcolor{red}{\overleftarrow {C}u^{2+}}\;\;\mathcolor{blue}{_\uparrow O^{2-}}\;\;\;
\mathcolor{red}{\overrightarrow {Cu}^{2+}}\;\;\mathcolor{blue}{_\downarrow O^{2-}}$ 

    \noindent $\mathcolor{blue}{
    \;\;\;\;\;\;\;\;\;\;\;\;
    O^{2-} \;\;\;\;\;\;\;\;\;\;\;\;\;\;\; 
    O^{2-} \;\;\;\;\;\;\;\;\;\;\;\;\;\;\;\; 
    O^{2-} \;\;\;\;\;\;\;\;\;\;\;\;\;\;\; 
    O^{2-} \;\;\;\;\;\;\;\;\;\;\;\;\;\;\;\; 
    O^{2-} \;\;\;\;\;\;\;\;\;\;\;\;\;\;\;
    O^{2-} \;\;\;\;\;\;\;\;\;\;\;\;\;\;\;\; O^{2-}}$

\noindent $
\mathcolor{green}{_\downarrow O^{2-}}\;\;\;
\mathcolor{red}{\overleftarrow{Cu}^{2+}} \;\;\mathcolor{blue}{_\uparrow O^{2-}} \;\;\;
\mathcolor{red}{\overrightarrow {C}u^{2+}}\;\;\mathcolor{blue}{_\downarrow O^{2-}}\;\;\;
\mathcolor{red}{\overleftarrow {C}u^{2+}} \;\;\mathcolor{blue}{_\uparrow O^{2-}}\;\;\;
\mathcolor{red}{\overrightarrow{Cu}^{2+}}\;\;\mathcolor{green}{_\downarrow O^{2-}}\;\;\;
\mathcolor{red}{\overleftarrow{Cu}^{2+}} \;\;\;\mathcolor{blue}{_\uparrow O^{2-}} \;\;\;
\mathcolor{red}{\overrightarrow {C}u^{2+}}\;\;\mathcolor{blue}{_\downarrow O^{2-}}\;\;\;
\mathcolor{red}{\overleftarrow {Cu}^{2+}}\;\;\mathcolor{blue}{_\uparrow O^{2-}}$ 

\noindent $\mathcolor{green}{\;\;\;\;\;\;\;\;\;\;\;\;O^{2-}}\;\;\;\;\;\;\;\;\;\;\;\;\;\;\; \mathcolor{blue}{O^{2-} \;\;\;\;\;\;\;\;\;\;\;\;\;\;\;\; O^{2-}} \;\;\;\;\;\;\;\;\;\;\;\;\;\;\; 
    \mathcolor{green}{O^{2-} \;\;\;\;\;\;\;\;\;\;\;\;\;\;\;\; O^{2-}} 
    \mathcolor{blue}{ \;\;\;\;\;\;\;\;\;\;\;\;\;\;\;
    O^{2-}} \;\;\;\;\;\;\;\;\;\;\;\;\;\;\;\; \mathcolor{green}{O^{2-}}$

\noindent $\;\;\mathcolor{black}{_\uparrow} \mathbf{o}^{\upuparrows}\;\;\;
\mathcolor{red}{\overrightarrow{\mathbf{Cu^{2+}}}} \;\mathcolor{green}{_\downarrow O^{2-}}\;\;\;
\mathcolor{red}{\overleftarrow {Cu}^{2+}}\;\;\mathcolor{blue}{_\uparrow O^{2-}}\;\;\;
\mathcolor{red}{\overrightarrow{Cu}^{2+}}\;\;\mathcolor{green}{_\downarrow O^{2-}}\;\;\;\;
\mathcolor{red}{\overleftarrow{\mathbf{Cu^{2+}}}}\;\;\mathcolor{black}{_\uparrow} \mathbf{o}^{\downdownarrows}\;\;\;\;\;
\mathcolor{red}{\overrightarrow{\mathbf{Cu^{2+}}}}\;\;\mathcolor{green}{_\downarrow O^{2-}}\;\;\;
\mathcolor{red}{\overleftarrow {Cu}^{2+}}\;\;\mathcolor{green}{_\uparrow O^{2-}}\;\;\;
\mathcolor{red}{\overrightarrow {\mathbf{Cu^{2+}}}}\;\;\;\mathcolor{black}{_\downarrow} \mathbf{o}^{\upuparrows}$   

\noindent $\mathcolor{green}{\;\;\;\;\;\;\;\;\;\;\;\;O^{2-}}\;\;\;\;\;\;\;\;\;\;\;\;\;\;\; \mathcolor{blue}{O^{2-} \;\;\;\;\;\;\;\;\;\;\;\;\;\;\;\; O^{2-}} \;\;\;\;\;\;\;\;\;\;\;\;\;\;\; 
    \mathcolor{green}{O^{2-} \;\;\;\;\;\;\;\;\;\;\;\;\;\;\;\;\; O^{2-}} 
    \mathcolor{blue}{ \;\;\;\;\;\;\;\;\;\;\;\;\;\;\;
    O^{2-}} \;\;\;\;\;\;\;\;\;\;\;\;\;\;\;\; \mathcolor{green}{O^{2-}}$

\noindent \footnotesize FIG. 19. Primary $\mathbf{(4 + 3) \times (4+3)}$ composite superlattice of $O^{\upuparrows}$ atoms in $La_{1.6-x}Nd_{0.4}Sr_xCuO_4$ at watershed doping $\hat{x} = 0.16$.  It is an arrangement of $4 \times 4$ and  $3 \times 3$ squares (top left and bottom right), joined by 4 × 3 rectangles between.
Long, medium, and short overarrows at $Cu^{2+}$ indicate $\mathcolor{red}{\overrightarrow{heavy}}$,  $\mathcolor{red}{\overrightarrow{mo}}derate$, and $\mathcolor{red}{\overrightarrow{s}}light$ axial octahedra tilts.  
 The depicted section of the $CuO_2$ plane contains four complete $O^{\upuparrows/ \downdownarrows}$ atoms (counted $1/4$ at the corners and $1/2$ on the sides). \normalsize

\pagebreak

\noindent $\mathcolor{green}{\;\;\;\;\;\;\;\;\;\;\;\;O^{2-}}\;\;\;\;\;\;\;\;\;\;\;\;\;\;\; \mathcolor{blue}{O^{2-} \;\;\;\;\;\;\;\;\;\;\;\;\;\;\;\; O^{2-}} \;\;\;\;\;\;\;\;\;\;\;\;\;\;\; 
    \mathcolor{green}{O^{2-} \;\;\;\;\;\;\;\;\;\;\;\;\;\;\;\; O^{2-}} 
    \mathcolor{blue}{ \;\;\;\;\;\;\;\;\;\;\;\;\;\;\;
    O^{2-}} \;\;\;\;\;\;\;\;\;\;\;\;\;\;\;\; \mathcolor{green}{O^{2-}}$

\noindent $
\;\;\mathcolor{black}{_\downarrow} \mathbf{o}^{\upuparrows}\;\;\;
\mathcolor{red}{\overleftarrow{\mathbf{Cu^{2+}}}} \;\mathcolor{green}{_\uparrow O^{2-}}\;\;\;
\mathcolor{red}{\overrightarrow {\mathbf{Cu}}^{2+}}\;\mathcolor{green}{_\downarrow O^{2-}}\;\;\;
\mathcolor{red}{\overleftarrow {\mathbf{Cu}}^{2+}} \;\mathcolor{green}{_\uparrow O^{2-}}\;\;\;\;
\mathcolor{red}{\overrightarrow{\mathbf{Cu^{2+}}}}\;\mathcolor{black}{_\downarrow} \mathbf{o}^{\downdownarrows}\;\;\;\;
\mathcolor{red}{\overleftarrow{\mathbf{Cu^{2+}}}}\;\;\mathcolor{green}{_\uparrow O^{2-}} \;\;\;
\mathcolor{red}{\overrightarrow {{Cu}}^{2+}}\;\mathcolor{green}{_\downarrow O^{2-}}\;\;\;
\mathcolor{red}{\overleftarrow {\mathbf{Cu^{2+}}}}\;\;\mathcolor{black}{_\uparrow} \mathbf{o}^{\upuparrows}$ 

\noindent $\mathcolor{green}{\;\;\;\;\;\;\;\;\;\;\;\;O^{2-}}\;\;\;\;\;\;\;\;\;\;\;\;\;\;\; \mathcolor{green}{O^{2-} \;\;\;\;\;\;\;\;\;\;\;\;\;\;\;\; O^{2-}} \;\;\;\;\;\;\;\;\;\;\;\;\;\;\; 
    \mathcolor{green}{O^{2-} \;\;\;\;\;\;\;\;\;\;\;\;\;\;\;\; O^{2-}} 
    \mathcolor{blue}{ \;\;\;\;\;\;\;\;\;\;\;\;\;\;\;
    O^{2-}} \;\;\;\;\;\;\;\;\;\;\;\;\;\;\;\; \mathcolor{green}{O^{2-}}$

\noindent $\mathcolor{green}{_\uparrow O^{2-}}\;\;\;
\mathcolor{red}{\overrightarrow{\mathbf{Cu}}^{2+}} \;\mathcolor{green}{_\downarrow O^{2-}} \;\;\;
\mathcolor{red}{\overleftarrow{\mathbf{Cu}}^{2+}}\;\mathcolor{green}{_\uparrow O^{2-}}\;\;
\mathcolor{red}{\overrightarrow{\mathbf{Cu}}^{2+}}\;\mathcolor{green}{_\downarrow O^{2-}}\;\;\;\;
\mathcolor{red}{\overleftarrow{\mathbf{Cu}}^{2+}}\;\;\mathcolor{green}{_\uparrow O^{2-}}\;\;\;
\mathcolor{red}{\overrightarrow{Cu}^{2+}}\;\;\;\mathcolor{blue}{_\downarrow O^{2-}}\;\;
\mathcolor{red}{\overleftarrow {C}u^{2+}}\;\;\mathcolor{blue}{_\uparrow O^{2-}}\;\;\;
\mathcolor{red}{\overrightarrow {Cu}^{2+}}\;\;\mathcolor{green}{_\downarrow O^{2-}}$ 

    \noindent $\mathcolor{green}{ \;\;\;\;\;\;\;\;\;\;\;\;
    O^{2-}} \;\;\;\;\;\;\;\;\;\;\;\;\;\;\; 
    \mathcolor{green}{O^{2-} \;\;\;\;\;\;\;\;\;\;\;\;\;\;\;\; 
    O^{2-}} \;\;\;\;\;\;\;\;\;\;\;\;\;\;\; 
    \mathcolor{green}{O^{2-} \;\;\;\;\;\;\;\;\;\;\;\;\;\;\;\; 
    O^{2-}} \;\;\;\;\;\;\;\;\;\;\;\;\;\;\;
    \mathcolor{blue}{O^{2-}} \;\;\;\;\;\;\;\;\;\;\;\;\;\;\;\; \mathcolor{blue}{O^{2-}}$

\noindent $\mathcolor{green}{_\downarrow O^{2-}}\;\;\;
\mathcolor{red}{\overleftarrow{\mathbf{Cu}}^{2+}} \;\mathcolor{green}{_\uparrow O^{2-}} \;\;\;
\mathcolor{red}{\overrightarrow {\mathbf{Cu^{2+}}}}\;\mathcolor{black}{_\downarrow} \mathbf{o}^{\downdownarrows}\;\;\;\;
\mathcolor{red}{\overleftarrow {\mathbf{Cu^{2+}}}} \;\mathcolor{green}{_\uparrow O^{2-}}\;\;\;\;
\mathcolor{red}{\overrightarrow{\mathbf{Cu}}^{2+}}\;\mathcolor{green}{_\downarrow O^{2-}}\;\;\;
\mathcolor{red}{\overleftarrow{C}u^{2+}} \;\;\;\mathcolor{blue}{_\uparrow O^{2-}} \;\;\;
\mathcolor{red}{\overrightarrow {C}u^{2+}}\;\;\mathcolor{blue}{_\downarrow O^{2-}}\;\;\;
\mathcolor{red}{\overleftarrow {C}u^{2+}}\;\;\mathcolor{blue}{_\uparrow O^{2-}}$ 

    \noindent $\mathcolor{green}{ \;\;\;\;\;\;\;\;\;\;\;\;
    O^{2-}} \;\;\;\;\;\;\;\;\;\;\;\;\;\;\; 
    \mathcolor{green}{O^{2-} \;\;\;\;\;\;\;\;\;\;\;\;\;\;\;\; 
    O^{2-} \;\;\;\;\;\;\;\;\;\;\;\;\;\;\; 
    O^{2-}} \;\;\;\;\;\;\;\;\;\;\;\;\;\;\;\; 
    \mathcolor{blue}{O^{2-} \;\;\;\;\;\;\;\;\;\;\;\;\;\;\;
    O^{2-} \;\;\;\;\;\;\;\;\;\;\;\;\;\;\;\; O^{2-}}$
    
\noindent $\mathcolor{green}{_\uparrow O^{2-}}\;\;\;\;
\mathcolor{red}{\overrightarrow{\mathbf{Cu}}^{2+}} \;\mathcolor{green}{_\downarrow O^{2-}} \;\;\;
\mathcolor{red}{\overleftarrow{\mathbf{Cu}}^{2+}}\;\mathcolor{green}{_\uparrow O^{2-}}\;\;
\mathcolor{red}{\overrightarrow{\mathbf{Cu}}^{2+}}\;\mathcolor{green}{_\downarrow O^{2-}}\;\;\;\;
\mathcolor{red}{\overleftarrow{\mathbf{Cu}}^{2+}}\;\;\mathcolor{green}{_\uparrow O^{2-}}\;\;\;
\mathcolor{red}{\overrightarrow{Cu}^{2+}}\;\;\;\mathcolor{blue}{_\downarrow O^{2-}}\;\;
\mathcolor{red}{\overleftarrow {C}u^{2+}}\;\;\mathcolor{blue}{_\uparrow O^{2-}}\;\;\;
\mathcolor{red}{\overrightarrow {Cu}^{2+}}\;\;\mathcolor{green}{_\downarrow O^{2-}}$ 

    \noindent $\mathcolor{green}{\;\;\;\;\;\;\;\;\;\;\;\;O^{2-}} 
    \mathcolor{green}{\;\;\;\;\;\;\;\;\;\;\;\;\;\;\; O^{2-} 
    \;\;\;\;\;\;\;\;\;\;\;\;\;\;\;\; O^{2-}} 
    \;\;\;\;\;\;\;\;\;\;\;\;\;\;\; \mathcolor{green}{O^{2-} 
    \;\;\;\;\;\;\;\;\;\;\;\;\;\;\;\; O^{2-}} 
    \mathcolor{blue}{\;\;\;\; \;\;\;\;\;\;\;\;\;\;\;\;O^{2-}} 
    \;\;\;\;\;\;\;\;\;\;\;\;\;\;\;\; \mathcolor{green}{O^{2-}}$

\noindent $\;\mathcolor{black}{_\downarrow} \mathbf{o}^{\downdownarrows}\;\;\;\;
\mathcolor{red}{\overleftarrow{\mathbf{Cu^{2+}}}} \;\mathcolor{green}{_\uparrow O^{2-}} \;\;\;
\mathcolor{red}{\overrightarrow {\mathbf{Cu}}^{2+}}\;\;\mathcolor{green}{_\downarrow O^{2-}}\;\;\;
\mathcolor{red}{\overleftarrow {\mathbf{Cu}}^{2+}} \;\mathcolor{green}{_\uparrow O^{2-}}\;\;\;\;
\mathcolor{red}{\overrightarrow{\mathbf{Cu^{2+}}}}\;\;\;
\mathcolor{black}{_\downarrow} \mathbf{o}^{\upuparrows} \;\;\;\;
\mathcolor{red}{\overleftarrow{\mathbf{Cu^{2+}}}}\;\;\;\mathcolor{green}{_\downarrow O^{2-}}\;\;\;\mathcolor{red}{\overrightarrow {\mathbf{Cu}}^{2+}}\;\mathcolor{green}{_\downarrow O^{2-}}\;\;\;\mathcolor{red}{\overleftarrow {\mathbf{Cu^{2+}}}}\;\;\;\mathcolor{black}{_\uparrow} \mathbf{o}^{\downdownarrows}$   

    \noindent $\mathcolor{green}{\;\;\;\;\;\;\;\;\;\;\;\;O^{2-}} 
    \mathcolor{blue}{\;\;\;\;\;\;\;\;\;\;\;\;\;\;\; 
    O^{2-} \;\;\;\;\;\;\;\;\;\;\;\;\;\;\;\; 
    O^{2-}} \;\;\;\;\;\;\;\;\;\;\;\;\;\;\; 
    \mathcolor{green}{O^{2-} \;\;\;\;\;\;\;\;\;\;\;\;\;\;\;\; 
    O^{2-}} \;\;\;\;\;\;\;\;\;\;\;\;\;\;\;\;
    \mathcolor{green}{_\uparrow O^{2-}} \;\;\;\;\;\;\;\;\;\;\;\;\;\;\;\; \mathcolor{green}{O^{2-}}$

\noindent $\mathcolor{green}{_\uparrow O^{2-}}\;\;\;
\mathcolor{red}{\overrightarrow{\mathbf{Cu}}^{2+}} \;\mathcolor{blue}{_\downarrow O^{2-}} \;\;\;
\mathcolor{red}{\overleftarrow {C}u^{2+}}\;\;\mathcolor{blue}{_\uparrow O^{2-}}\;\;\;
\mathcolor{red}{\overrightarrow{C}u^{2+}}\;\;\mathcolor{blue}{_\downarrow O^{2-}}\;\;\;\;
\mathcolor{red}{\overleftarrow{\mathbf{Cu}}^{2+}}\;\;\mathcolor{green}{_\uparrow O^{2-}}\;\;
\mathcolor{red}{\overrightarrow{\mathbf{Cu}}^{2+}}\;\;\;\mathcolor{green}{ O^{2-}}\;\;
\mathcolor{red}{\mathbf{\Downarrow} \mathbf{Cu^{2+}}}\;\mathcolor{green}{ O^{2-}}\;\;\;
\mathcolor{red}{\overrightarrow {\mathbf{Cu}}^{2+}}\;\mathcolor{green}{_\downarrow O^{2-}}$ 

    \noindent $\mathcolor{green}{\;\;\;\;\;\;\;\;\;\;\;\;O^{2-}}
    \mathcolor{blue}{\;\;\;\;\;\;\;\;\;\;\;\;\;\;\; O^{2-} 
    \;\;\;\;\;\;\;\;\;\;\;\;\;\;\;\; O^{2-}} 
    \mathcolor{green}{\;\;\;\;\;\;\;\;\;\;\;\;\;\;\; O^{2-} 
    \;\;\;\;\;\;\;\;\;\;\;\;\;\;\;\; O^{2-}}
    \;\;\;\;\;\;\;\;\;\;\;\;\;\;\;\;\mathcolor{black}{_{\downarrow} \mathbf{o}^{\downdownarrows}}
    \mathcolor{green}{ \;\;\;\;\;\;\;\;\;\;\;\;\;\;\;\;\; O^{2-}}$

\noindent $ \mathcolor{green}{_\downarrow O^{2-}}\;\;\;
\mathcolor{red}{\overleftarrow{\mathbf{Cu}}^{2+}}\;\mathcolor{blue}{_\uparrow O^{2-}}\;\;\;
\mathcolor{red}{\overrightarrow {C}u^{2+}}\;\;\mathcolor{blue}{_\downarrow O^{2-}}\;\;\;
\mathcolor{red}{\overleftarrow {C}u^{2+}} \;\;\mathcolor{blue}{_\uparrow O^{2-}}\;\;\;\;
\mathcolor{red}{\overrightarrow{\mathbf{Cu}}^{2+}}\;\mathcolor{green}{_\downarrow O^{2-}}\;\;\;
\mathcolor{red}{\overleftarrow{\mathbf{Cu}}^{2+}}\;\;\;\mathcolor{green}{O^{2-}} \;\;
\mathcolor{red}{\mathbf{\Uparrow} \mathbf{Cu^{2+}}}\;\mathcolor{green}{ O^{2-}}\;\;\;
\mathcolor{red}{\overleftarrow {\mathbf{Cu}}^{2+}}\;\mathcolor{green}{_\uparrow O^{2-}}$ 

\noindent $\mathcolor{green}{\;\;\;\;\;\;\;\;\;\;\;\;O^{2-}}\;\;\;\;\;\;\;\;\;\;\;\;\;\;\; \mathcolor{blue}{O^{2-} \;\;\;\;\;\;\;\;\;\;\;\;\;\;\;\; O^{2-}} \;\;\;\;\;\;\;\;\;\;\;\;\;\;\; 
    \mathcolor{green}{O^{2-} \;\;\;\;\;\;\;\;\;\;\;\;\;\;\;\; O^{2-}} 
    \mathcolor{green}{ \;\;\;\;\;\;\;\;\;\;\;\;\;\;\;
    _\uparrow O^{2-}} \;\;\;\;\;\;\;\;\;\;\;\;\;\;\;\; \mathcolor{green}{O^{2-}}$

\noindent $\;\;\mathcolor{black}{_\uparrow} \mathbf{o}^{\upuparrows}\;\;\;
\mathcolor{red}{\overrightarrow{\mathbf{Cu^{2+}}}} \;\mathcolor{green}{_\downarrow O^{2-}}\;\;\;
\mathcolor{red}{\overleftarrow {Cu}^{2+}}\;\;\mathcolor{green}{_\uparrow O^{2-}}\;\;\;
\mathcolor{red}{\overrightarrow{Cu}^{2+}}\;\;\mathcolor{green}{_\downarrow O^{2-}}\;\;\;\;
\mathcolor{red}{\overleftarrow{\mathbf{Cu^{2+}}}}\;\;\mathcolor{black}{_\uparrow} \mathbf{o}^{\downdownarrows}\;\;\;\;\;
\mathcolor{red}{\overrightarrow{\mathbf{Cu^{2+}}}}\;\;\mathcolor{green}{_\downarrow O^{2-}}\;\;\;
\mathcolor{red}{\overleftarrow {Cu}^{2+}}\;\;\mathcolor{green}{_\uparrow O^{2-}}\;\;\;
\mathcolor{red}{\overrightarrow {\mathbf{Cu^{2+}}}}\;\;\mathcolor{black}{_\downarrow} \mathbf{o}^{\upuparrows}$   

\noindent $\mathcolor{green}{\;\;\;\;\;\;\;\;\;\;\;\;O^{2-}}\;\;\;\;\;\;\;\;\;\;\;\;\;\;\; \mathcolor{blue}{O^{2-} \;\;\;\;\;\;\;\;\;\;\;\;\;\;\;\; O^{2-}} \;\;\;\;\;\;\;\;\;\;\;\;\;\;\; 
    \mathcolor{green}{O^{2-} \;\;\;\;\;\;\;\;\;\;\;\;\;\;\;\; O^{2-}} 
    \mathcolor{blue}{ \;\;\;\;\;\;\;\;\;\;\;\;\;\;\;
    O^{2-}} \;\;\;\;\;\;\;\;\;\;\;\;\;\;\;\; \mathcolor{green}{O^{2-}}$

\noindent \footnotesize FIG. 20. $CuO_2$ plane of $La_{1.6-x}Nd_{0.4}Sr_xCuO_4$ at the quantum critical point $x^* = 0.23$. Primary and secondary superlattices interlace such that the $\mathbf{(4 \times 4)}$ and $\mathbf{(3 \times 3)}$ squares of the former (Fig. 19) are filled each with an $O^\upuparrows$ atom of the latter. The assembly contains four full $O^{\upuparrows/\downdownarrows}$ atoms of the primary superlattice and two of the secondary—--in agreement with Eq.  (16). Lines of demotted $\mathcolor{red}{\mathbf{Cu^{2+}}}$ ions (fifth $Cu$-$O$ row, first and forth $Cu$-$O$ column) traverse the assembly and thus extend throughout the crystal domain. \normalsize

\section{SHORT-RANGE CHARGE-ORDER DOME}

Besides the long-range charge order, coupled in lanthanum cuprates ({\it 214}) with long-range spin order (stripes),  Eqs. (11 - 13), a short-range charge-order pattern has been observed in hole-doped cuprates.\cite{70,72} This charge order is regarded a competitor to superconductivity. The profile of its onset temperature (upon cooling), $T_{co}(x)$, is a steep dome with maximum at doping $x_4 = 0.125$. Signal intensity vanishes at (or near) $x_5 = 0.08$ and $x=0.16$. What gives rise to this charge-order dome?

The top part of Fig. 21 shows for comparison a $4 \times 4$ $O^{\upuparrows}$ superlattice, familiar from
Fig. 13. In a repetitive pattern, the  $CuO_5O^{\upuparrows}$ octahedra that bracket each $O^{\upuparrows}$ atom, tilt toward each other (above the $CuO_2$ plane). Likewise, the in-between $CuO_6$ octahedra  tilt moder-

\bigskip

\noindent $\mathcolor{blue}{\;\;\;\;\;\;\;\;\;\;\;\; _\uparrow O^{2-} \;\;\;\;\;\;\;\;\;\;\;\;\;\;\; O^{2-} \;\;\;\;\;\;\;\;\;\;\;\;\;\;\; O^{2-} \;\;\;\;\;\;\;\;\;\;\;\;\;\;\;_\downarrow O^{2-} \;\;\;\;\;\;\;\;\;\;\;\;\;\;\;_\uparrow O^{2-} \;\;\;\;\;\;\;\;\;\;\;\;\;\;\;\; O^{2-} \;\;\;\;\;\;\;\;\;\;\;\;\;\;\;\; O^{2-}}$

\noindent $ \mathcolor{blue}{_\downarrow O^{2-}} \;\;\; \mathcolor{red}{\overleftarrow{Cu}^{2+}} \;\; \mathcolor{blue}{\mathbf{\uparrow} O^{2-}} \;\;\;\mathcolor{red}{\overrightarrow{Cu^{2+}}} \;\;\;\;\mathbf{\downarrow}\mathbf{o}^\mathbf{\upuparrows} \;\;\;\;\mathcolor{red}{\overleftarrow{Cu^{2+}}} \;\;\mathcolor{blue}{\mathbf{\uparrow} O^{2-}} \;\;\;\mathcolor{red}{\overrightarrow{Cu}^{2+}} \;\;\;\mathcolor{blue}{_\downarrow O^{2-}} \;\;\;\;\mathcolor{red}{\overleftarrow{Cu}^{2+}} \;\;\;\mathcolor{blue}{\mathbf{\uparrow} O^{2-}} \;\;\;\mathcolor{red}{\overrightarrow{Cu^{2+}}} \;\;\;\mathbf{\downarrow}\mathbf{o}^\mathbf{\downdownarrows} \;\;\;\;\mathcolor{red}{\overleftarrow{Cu^{2+}}} \;\;\mathcolor{blue}{\mathbf{\uparrow} O^{2-}}$  

\noindent $\mathcolor{blue}{\;\;\;\;\;\;\;\;\;\;\;\; _\downarrow O^{2-} \;\;\;\;\;\;\;\;\;\;\;\;\;\;\; O^{2-} \;\;\;\;\;\;\;\;\;\;\;\;\;\;\; O^{2-} \;\;\;\;\;\;\;\;\;\;\;\;\;\;\;_\uparrow O^{2-} \;\;\;\;\;\;\;\;\;\;\;\;\;\;\;_\downarrow O^{2-} \;\;\;\;\;\;\;\;\;\;\;\;\;\;\;\; O^{2-} \;\;\;\;\;\;\;\;\;\;\;\;\;\;\;\; O^{2-}}$

\noindent $ \mathcolor{blue}{_\uparrow O^{2-}} \; \mathcolor{red}{_{\nearrow}{Cu^{2+}}} \;\; \mathcolor{blue}{_\downarrow O^{2-}} \;\;\;\;\mathcolor{red}{\overleftarrow{Cu}^{2+}} \;\;\mathcolor{blue}{_\uparrow O^{2-}}\mathcolor{red}{ \;\;\; \mathcolor{red}{\overrightarrow{Cu}^{2+}}} \;\;\mathcolor{blue}{_\downarrow O^{2-}} \;\mathcolor{red}{_{\swarrow}  {Cu^{2+}}} \;\;\mathcolor{blue}{_\uparrow O^{2-}}\mathcolor{red}{\mathcolor{red}{ \;_{\nearrow}  {Cu^{2+}}}} \;\;\mathcolor{blue}{_\downarrow O^{2-}} \;\;\;\;\mathcolor{red}{\overleftarrow{Cu}^{2+}} \;\;\mathcolor{blue}{_\uparrow O^{2-}} \;\;\;\;\mathcolor{red}{\overrightarrow{Cu}^{2+}} \;\;\mathcolor{blue}{_\downarrow O^{2-}}$ 

\noindent $\mathcolor{blue}{\;\;\;\;\;\;\;\;\;\;\;\; _\uparrow O^{2-} \;\;\;\;\;\;\;\;\;\;\;\;\;\;_\downarrow O^{2-} \;\;\;\;\;\;\;\;\;\;\;\;\;\;_\uparrow O^{2-} \;\;\;\;\;\;\;\;\;\;\;\;\;\;\;_\downarrow O^{2-} \;\;\;\;\;\;\;\;\;\;\;\;\;\;\;_\uparrow O^{2-} \;\;\;\;\;\;\;\;\;\;\;\;\;\;\;_\downarrow O^{2-} \;\;\;\;\;\;\;\;\;\;\;\;\;\;\;_\uparrow O^{2-}}$

\noindent $ \mathcolor{blue}{_\downarrow O^{2-}} \; \mathcolor{red}{_{\swarrow}Cu^{2+}} \;\; \mathcolor{blue}{_\uparrow O^{2-}} \;\mathcolor{red}{_{\nearrow}  Cu^{2+}} \;\;\mathcolor{blue}{_\downarrow O^{2-}} \;\mathcolor{red}{_{\swarrow} {Cu^{2+}}} \;\;\mathcolor{blue}{_\uparrow O^{2-}} \;\mathcolor{red}{_{\nearrow}  {Cu^{2+}}} \;\;\mathcolor{blue}{_\downarrow O^{2-}} \;\mathcolor{red}{_{\swarrow}  {Cu^{2+}}} \;\;\mathcolor{blue}{_\uparrow O^{2-}} \;\mathcolor{red}{_{\nearrow}  {Cu^{2+}}} \;\;\mathcolor{blue}{_\downarrow O^{2-}} \;\mathcolor{red}{_{\swarrow} {Cu^{2+}}} \;\;\mathcolor{blue}{_\uparrow O^{2-}} $ 

\noindent $\mathcolor{blue}{\;\;\;\;\;\;\;\;\;\;\;\; _\downarrow O^{2-} \;\;\;\;\;\;\;\;\;\;\;\;\;\;_\uparrow O^{2-} \;\;\;\;\;\;\;\;\;\;\;\;\;\;_\downarrow O^{2-} \;\;\;\;\;\;\;\;\;\;\;\;\;\;\;_\uparrow O^{2-} \;\;\;\;\;\;\;\;\;\;\;\;\;\;\;_\downarrow O^{2-} \;\;\;\;\;\;\;\;\;\;\;\;\;\;\;_\uparrow O^{2-} \;\;\;\;\;\;\;\;\;\;\;\;\;\;\;_\downarrow O^{2-}}$

\noindent $ \mathcolor{blue}{_\uparrow O^{2-}} \; \mathcolor{red}{_{\nearrow}{Cu^{2+}}} \;\; \mathcolor{blue}{_\downarrow O^{2-}} \;\;\;\;\mathcolor{red}{\overleftarrow{Cu}^{2+}} \;\;\mathcolor{blue}{_\uparrow O^{2-}}\mathcolor{red}{ \;\;\; \mathcolor{red}{\overrightarrow{Cu}^{2+}}} \;\;\mathcolor{blue}{_\downarrow O^{2-}} \;\mathcolor{red}{_{\swarrow}  {Cu^{2+}}} \;\;\mathcolor{blue}{_\uparrow O^{2-}}\mathcolor{red}{\mathcolor{red}{ \;_{\nearrow}  {Cu^{2+}}}} \;\;\mathcolor{blue}{_\downarrow O^{2-}} \;\;\;\;\mathcolor{red}{\overleftarrow{Cu}^{2+}} \;\;\mathcolor{blue}{_\uparrow O^{2-}} \;\;\;\;\mathcolor{red}{\overrightarrow{Cu}^{2+}} \;\;\mathcolor{blue}{_\downarrow O^{2-}}$ 

\noindent $\mathcolor{blue}{\;\;\;\;\;\;\;\;\;\;\;\; _\uparrow O^{2-} \;\;\;\;\;\;\;\;\;\;\;\;\;\;\; O^{2-} \;\;\;\;\;\;\;\;\;\;\;\;\;\;\; O^{2-} \;\;\;\;\;\;\;\;\;\;\;\;\;\;\;_\downarrow O^{2-} \;\;\;\;\;\;\;\;\;\;\;\;\;\;\;_\uparrow O^{2-} \;\;\;\;\;\;\;\;\;\;\;\;\;\;\;\; O^{2-} \;\;\;\;\;\;\;\;\;\;\;\;\;\;\;\; O^{2-}}$

\noindent $ \mathcolor{blue}{_\downarrow O^{2-}} \;\;\; \mathcolor{red}{\overleftarrow{Cu}^{2+}} \;\; \mathcolor{blue}{\mathbf{\uparrow} O^{2-}} \;\;\;\mathcolor{red}{\overrightarrow{Cu^{2+}}} \;\;\;\;\mathbf{\downarrow}\mathbf{o}^\mathbf{\downdownarrows} \;\;\;\;\mathcolor{red}{\overleftarrow{Cu^{2+}}} \;\;\mathcolor{blue}{\mathbf{\uparrow} O^{2-}} \;\;\;\mathcolor{red}{\overrightarrow{Cu}^{2+}} \;\;\;\mathcolor{blue}{_\downarrow O^{2-}} \;\;\;\;\mathcolor{red}{\overleftarrow{Cu}^{2+}} \;\;\;\mathcolor{blue}{\mathbf{\uparrow} O^{2-}} \;\;\;\mathcolor{red}{\overrightarrow{Cu^{2+}}} \;\;\;\mathbf{\downarrow}\mathbf{o}^\mathbf{\upuparrows} \;\;\;\;\mathcolor{red}{\overleftarrow{Cu^{2+}}} \;\;\mathcolor{blue}{\mathbf{\uparrow} O^{2-}}$

\noindent $\mathcolor{blue}{\;\;\;\;\;\;\;\;\;\;\;\; _\downarrow O^{2-} \;\;\;\;\;\;\;\;\;\;\;\;\;\;\; O^{2-} \;\;\;\;\;\;\;\;\;\;\;\;\;\;\; O^{2-} \;\;\;\;\;\;\;\;\;\;\;\;\;\;\;_\uparrow O^{2-} \;\;\;\;\;\;\;\;\;\;\;\;\;\;\;_\downarrow O^{2-} \;\;\;\;\;\;\;\;\;\;\;\;\;\;\;\; O^{2-} \;\;\;\;\;\;\;\;\;\;\;\;\;\;\;\; O^{2-}}$

\medskip
------------------------------------------------------------------------------------------------------------------
\medskip



\noindent $\mathcolor{blue}{\;\;\;\;\;\;\;\;\;\;\;\; _\uparrow O^{2-} \;\;\;\;\;\;\;\;\;\;\;\;\;\;_\downarrow O^{2-} \;\;\;\;\;\;\;\;\;\;\;\;\;\;_\uparrow O^{2-}} \;\;\;\;\;\;\;\;\;\;\;\;\;\;\;\mathcolor{blue}{_\downarrow O^{2-} \;\;\;\;\;\;\;\;\;\;\;\;\;\;\;_\uparrow O^{2-}} \;\;\;\;\;\;\;\;\;\;\;\;\;\;\;\mathcolor{blue}{_\downarrow O^{2-} \;\;\;\;\;\;\;\;\;\;\;\;\;\;\;_\uparrow O^{2-}}$

\noindent $ \mathcolor{blue}{_\downarrow O^{2-}} \;\;\;\;
\mathcolor{red}{\overleftarrow{Cu^{2+}}}\; 
\;\;\mathcolor{black}{\mathbf{\uparrow} \mathbf{o^{\upuparrows}}} \;\;\;\;
\mathcolor{red}{\overrightarrow {Cu^{2+}}} \;\;\;
\mathcolor{blue}{_\downarrow O^{2-}} \;\;\;\;
\mathcolor{red}{\overleftarrow {Cu}^{2+}} \;\;
\mathcolor{blue}{_\uparrow O^{2-}} \;
\mathcolor{red}{_{\nearrow}Cu^{2+}}\;\; 
\mathcolor{blue}{_\downarrow O^{2-}} \;\;
\mathcolor{red}{\overleftarrow {Cu}^{2+}} \;\;
\mathcolor{blue}{_\uparrow O^{2-}} \;\;\;
\mathcolor{red}{\overrightarrow{Cu^{2+}}} \;\;\;
\mathcolor{black}{\;\;\mathbf{\mathbf{\downarrow} o^{\downdownarrows}}} \;\;\;\;\;
\mathcolor{red}{\overleftarrow{Cu^{2+}}}\;\;
\mathcolor{blue}{_\uparrow O^{2-}}$

\noindent $\mathcolor{blue}{\;\;\;\;\;\;\;\;\;\;\;\; _\downarrow O^{2-} \;\;\;\;\;\;\;\;\;\;\;\;\;\;_\uparrow O^{2-} \;\;\;\;\;\;\;\;\;\;\;\;\;\;_\downarrow O^{2-}} \;\;\;\;\;\;\;\;\;\;\;\;\;\;\;\mathcolor{blue}{_\uparrow O^{2-} \;\;\;\;\;\;\;\;\;\;\;\;\;\;\;_\downarrow O^{2-}} \;\;\;\;\;\;\;\;\;\;\;\;\;\;\;\mathcolor{blue}{_\uparrow O^{2-} \;\;\;\;\;\;\;\;\;\;\;\;\;\;\;_\downarrow O^{2-}}$

\noindent $\mathcolor{blue}{ _\uparrow O^{2-}} \; \mathcolor{red}{_{\nearrow}Cu^{2+}} \;\; \mathcolor{blue}{_\downarrow O^{2-}} \;\mathcolor{red}{_{\swarrow}  Cu^{2+}} \;\;\mathcolor{blue}{_\uparrow O^{2-}} \;\mathcolor{red}{_{\nearrow} {Cu^{2+}}} \;\;\mathcolor{blue}{_\downarrow O^{2-}} \;\mathcolor{red}{_{\swarrow}  Cu^{2+}} \;\;\mathcolor{blue}{_\uparrow O^{2-}} \;\mathcolor{red}{_{\nearrow}  Cu}^{2+} \;\;\mathcolor{blue}{_\downarrow O^{2-}} \;\mathcolor{red}{_{\swarrow}  {Cu^{2+}}} \;\;\mathcolor{blue}{_\uparrow O^{2-}} \;\mathcolor{red}{_{\nearrow} {Cu^{2+}}} \;\;\mathcolor{blue}{_\downarrow O^{2-}}$ 

\noindent $\mathcolor{blue}{\;\;\;\;\;\;\;\;\;\;\;\; _\uparrow O^{2-} \;\;\;\;\;\;\;\;\;\;\;\;\;\;_\downarrow O^{2-} \;\;\;\;\;\;\;\;\;\;\;\;\;\;_\uparrow O^{2-} \;\;\;\;\;\;\;\;\;\;\;\;\;\;\;_\downarrow O^{2-} \;\;\;\;\;\;\;\;\;\;\;\;\;\;\;_\uparrow O^{2-} \;\;\;\;\;\;\;\;\;\;\;\;\;\;\;_\downarrow O^{2-} \;\;\;\;\;\;\;\;\;\;\;\;\;\;\;_\uparrow O^{2-}}$

\noindent $\mathcolor{blue}{ _\downarrow O^{2-}} \; \mathcolor{red}{_{\swarrow}Cu^{2+}} \;\; \mathcolor{blue}{_\uparrow O^{2-}} \;\mathcolor{red}{_{\nearrow}  Cu^{2+}} \;\;\mathcolor{blue}{_\downarrow O^{2-}} \;\mathcolor{red}{_{\swarrow} {Cu^{2+}}} \;\;\mathcolor{blue}{_\uparrow O^{2-}} \;\mathcolor{red}{_{\nearrow}  {Cu^{2+}}} \;\;\mathcolor{blue}{_\downarrow O^{2-}} \;\mathcolor{red}{_{\swarrow}  {Cu^{2+}}} \;\;\mathcolor{blue}{_\uparrow O^{2-}} \;\mathcolor{red}{_{\nearrow}  {Cu^{2+}}} \;\;\mathcolor{blue}{_\downarrow O^{2-}} \;\mathcolor{red}{_{\swarrow} {Cu^{2+}}} \;\;\mathcolor{blue}{_\uparrow O^{2-}}$

\noindent $\mathcolor{blue}{\;\;\;\;\;\;\;\;\;\;\;\; _\downarrow O^{2-} \;\;\;\;\;\;\;\;\;\;\;\;\;\;_\uparrow O^{2-} \;\;\;\;\;\;\;\;\;\;\;\;\;\;_\downarrow O^{2-} \;\;\;\;\;\;\;\;\;\;\;\;\;\;\;_\uparrow O^{2-} \;\;\;\;\;\;\;\;\;\;\;\;\;\;\;_\downarrow O^{2-} \;\;\;\;\;\;\;\;\;\;\;\;\;\;\;_\uparrow O^{2-} \;\;\;\;\;\;\;\;\;\;\;\;\;\;\;_\downarrow O^{2-}}  $

\noindent $\mathcolor{blue}{ _\uparrow O^{2-}} \; \mathcolor{red}{_{\nearrow}Cu^{2+}} \;\; \mathcolor{blue}{_\downarrow O^{2-}} \;\mathcolor{red}{_{\swarrow}  Cu^{2+}} \;\;\mathcolor{blue}{_\uparrow O^{2-}} \;\mathcolor{red}{_{\nearrow} {Cu^{2+}}} \;\;\mathcolor{blue}{_\downarrow O^{2-}} \;\mathcolor{red}{_{\swarrow}  Cu^{2+}} \;\;\mathcolor{blue}{_\uparrow O^{2-}} \;\mathcolor{red}{_{\nearrow}  Cu}^{2+} \;\;\mathcolor{blue}{_\downarrow O^{2-}} \;\mathcolor{red}{_{\swarrow}  {Cu^{2+}}} \;\;\mathcolor{blue}{_\uparrow O^{2-}} \;\mathcolor{red}{_{\nearrow} {Cu^{2+}}} \;\;\mathcolor{blue}{_\downarrow O^{2-}}$ 

\noindent $\mathcolor{blue}{\;\;\;\;\;\;\;\;\;\;\;\; _\uparrow O^{2-} \;\;\;\;\;\;\;\;\;\;\;\;\;\;_\downarrow O^{2-} \;\;\;\;\;\;\;\;\;\;\;\;\;\;_\uparrow O^{2-}} \;\;\;\;\;\;\;\;\;\;\;\;\;\;\;\mathcolor{blue}{_\downarrow O^{2-} \;\;\;\;\;\;\;\;\;\;\;\;\;\;\;_\uparrow O^{2-}} \;\;\;\;\;\;\;\;\;\;\;\;\;\;\;\mathcolor{blue}{_\downarrow O^{2-} \;\;\;\;\;\;\;\;\;\;\;\;\;\;\;_\uparrow O^{2-}}$

\noindent $\mathcolor{blue}{ _\downarrow O^{2-}} \; \mathcolor{red}{_{\swarrow}Cu^{2+}} \;\; \mathcolor{blue}{_\uparrow O^{2-}} \;\mathcolor{red}{_{\nearrow}  Cu^{2+}} \;\;\mathcolor{blue}{_\downarrow O^{2-}} \;\mathcolor{red}{_{\swarrow} {Cu^{2+}}} \;\;\mathcolor{blue}{_\uparrow O^{2-}} \;\mathcolor{red}{_{\nearrow}  {Cu^{2+}}} \;\;\mathcolor{blue}{_\downarrow O^{2-}} \;\mathcolor{red}{_{\swarrow}  {Cu^{2+}}} \;\;\mathcolor{blue}{_\uparrow O^{2-}} \;\mathcolor{red}{_{\nearrow}  {Cu^{2+}}} \;\;\mathcolor{blue}{_\downarrow O^{2-}} \;\mathcolor{red}{_{\swarrow} {Cu^{2+}}} \;\;\mathcolor{blue}{_\uparrow O^{2-}}$

\noindent $\mathcolor{blue}{\;\;\;\;\;\;\;\;\;\;\;\; _\downarrow O^{2-} \;\;\;\;\;\;\;\;\;\;\;\;\;\;_\uparrow O^{2-} \;\;\;\;\;\;\;\;\;\;\;\;\;\;_\downarrow O^{2-} \;\;\;\;\;\;\;\;\;\;\;\;\;\;\;_\uparrow O^{2-} \;\;\;\;\;\;\;\;\;\;\;\;\;\;\;_\downarrow O^{2-} \;\;\;\;\;\;\;\;\;\;\;\;\;\;\;_\uparrow O^{2-} \;\;\;\;\;\;\;\;\;\;\;\;\;\;\;_\downarrow O^{2-}}$

\noindent $ \mathcolor{blue}{_\uparrow O^{2-}} \;\;\;\;
\mathcolor{red}{\overrightarrow{Cu^{2+}}}\; 
\;\;\mathcolor{black}{\mathbf{\downarrow} \mathbf{o^{\downdownarrows}}} \;\;\;\;
\mathcolor{red}{\overleftarrow {Cu^{2+}}} \;\;\;
\mathcolor{blue}{_\uparrow O^{2-}} \;\;\;\;
\mathcolor{red}{\overrightarrow {Cu}^{2+}} \;\;\;
\mathcolor{blue}{_\downarrow O^{2-}} \;
\mathcolor{red}{_{\swarrow}Cu^{2+}}\;\; 
\mathcolor{blue}{_\uparrow O^{2-}} \;
\mathcolor{red}{\overrightarrow {Cu}^{2+}} \;\;
\mathcolor{blue}{_\downarrow O^{2-}} \;\;\;
\mathcolor{red}{\overleftarrow{Cu^{2+}}} \;\;\;
\mathcolor{black}{\;\;\mathbf{\mathbf{\uparrow} o^{\upuparrows}}} \;\;\;\;\;
\mathcolor{red}{\overrightarrow{Cu^{2+}}}\;\;
\mathcolor{blue}{_\downarrow O^{2-}}$

\noindent $\mathcolor{blue}{\;\;\;\;\;\;\;\;\;\;\;\; _\uparrow O^{2-} \;\;\;\;\;\;\;\;\;\;\;\;\;\;_\downarrow O^{2-} \;\;\;\;\;\;\;\;\;\;\;\;\;\;_\uparrow O^{2-}} \;\;\;\;\;\;\;\;\;\;\;\;\;\;\;\mathcolor{blue}{_\downarrow O^{2-} \;\;\;\;\;\;\;\;\;\;\;\;\;\;\;_\uparrow O^{2-}} \;\;\;\;\;\;\;\;\;\;\;\;\;\;\;\mathcolor{blue}{_\downarrow O^{2-} \;\;\;\;\;\;\;\;\;\;\;\;\;\;\;_\uparrow O^{2-}}$

\noindent \footnotesize FIG. 21. $CuO_2$ plane with a $\mathbf{4 \times 4}$ $O^{\upuparrows}$ superlattice at doping $x_4 = 1/8 = 0.125$ of heavily and moderately tilted octahedra having maximal phase coherence (top). In contrast, a $\mathbf{5 \times 5}$ $O^{\upuparrows}$ superlattice at doping $x_5 = 0.08$ has minimal phase coherence (bottom). In order to prevent distraction from tilt phase relations, stressed $O^{2-}$ ions are {\it not} marked green and demotted $Cu^{2+}$ ions are {\it not} marked bold. \normalsize

\noindent ately toward one another. Phase coherence of the highly symmetric pattern gives rise to maximal constructive interference in diffraction experiments, manifesting the peak intensity at $x_4 = 1/8$.

In contrast, at doping $x_5 = 0.08$, pairs of $CuO_5O^{\upuparrows}$ octahedra that bracket subsequent $O^{\upuparrows}$ atoms tilt {\it oppositely} in the corresponding $5 \times 5$ $O^{\upuparrows}$ superlattice (lower part of Fig. 21). Between them, the neighboring $CuO_6$ octahedra tilt moderately in the same direction but opposite to the moderately tilted $CuO_6$ octahedra of the parallel row. The reverse pattern holds in the neighboring superlattice cell. Opposite phases at corresponding sites cause minimal constructive interference in diffraction experiments causing a minimal signal at 
$x_5$. 


The short-range charge-order dome is the result of phase coherence of heavy $CuO_5O^{\upuparrows}$ tilts and moderate $CuO_6$ octahedra tilts, peaking at doping $x_4 = 0.125$ and dropping to minima at $x_5 = 0.08$ and $x \approx 0.16$. The charge-order dome coincides with a {\it dent} in 
the ``superconducting dome,'' $T_c(x)$, indicating a competition of charge density with superconductivity. It also coincides with a local minimum (maximum) at $x \simeq 0.12$ ($x \simeq 0.08$) in the doping dependence of the electronic heat per temperature (i. e., entropy), $[C_{el}/T](x)$,\cite{5} due to order (disorder).

Because of the triplet-state $O^{\upuparrows}$ atoms in {\it 214} cuprates, an accompanying short-range {\it spin-order dome} can be expected for these compounds. A ``magnetically ordered volume fraction,'' $V_M(x)$, is detected with muon-spin rotation ($\mu$SR) experiments on LTT-phase $La_{2-x}Ba_xCuO_4$, peaking at $x_4 = 0.125$ with $V_M(x_4) = 0.93$ and falling off approximately dome-like at neighboring doping, $V_M(0.115) = 0.90$ and $V_M(0.135) = 0.86$.\cite{73,74,75} Assuming a parabolic doping dependence, $V_M(x) \propto (x-x_4)^2$, and average of neighbor values, $\overline{V_M}(0.115) =\overline{V_M}(0.135) = 0.88$, extrapolates to $V_M(x) = 0$ near $x_5 = 0.08$ and $x \approx 0.16$---a remarkable correlation with the phase coherence of the short-range charge-order dome.

With respect to doping $x$, a {\it local} maximum of phase coherence can be expected, by symmetry, at every doping $x_n = 2/n^2$ with \textit{even} index $n$, associated with a commensurate $O^{\upuparrows}$ superlattice of spacing $A_0 = na_0$. This would explain the prominence of $x_6$ and $x_{10}$, causing corresponding changes of crystal properties (see Appendix C, Table IV).

\section{CHARGE ORDER IN OXYGEN-DOPED/ENRICHED CUPRATES}

Some kind of charge order is also found in oxygen \emph{doped} $YBa_2Cu_3O_{6+y}$ {\it (123)}, as well as in oxygen {\it enriched} cuprates (OEC), like $LaCuO_{4+\delta}$, $HgBa_2CuO_{4+\delta}$, $Bi_2Sr_2CuO_{6+\delta}$, $Bi_2Sr_2CaCu_2O_{8+\delta}$, and $Bi_2Sr_2Ca_2Cu_3O_{10+\delta}$. With resonant elastic and inelastic X-ray scattering (REXS, RIXS) an incommensurability $q(p)$ is observed\cite{76,77,78,79,80,81,82,83,84,85,86,87,88} that {\it decreases} with increasing hole doping $p$---linearly for OEC and slightly quadratically for {\it 123} (Fig. 22)---in contrast to {\it increasing} $q_{co}(x)$ in the {\it 214} family (Fig. 11).  The RIXS signals appear beyond a rather large N\'{e}el doping, $p > p_N$ (qualitatively explained in Sec. IX) and fade out near $p = 0.17 \pm 0.03$. How can that be understood? 

\noindent \textbf{Assumption 11: While RIXS reveals charge order in \emph{214}-type lanthanum cuprates by $\mathbf{O^{\upuparrows}}$ superlattices, this method fails to detect charge order in oxygen-doped/enriched cuprates by $\mathbf{O^{\uparrow \downarrow}}$ superlattices.}

\noindent A tell-tale sign of the observed incommensurability is its back extrapolation to $q(0) = 0.354 \; r.l.u.$---the wave number of the diagonal  tilt pattern of $CuO_6$ octahedra in pristine $La_2CuO_4$ (Sec. X, Fig. 10). What is being observed with X-ray scattering in oxygen-doped/enriched cuprates is the receding host tilt pattern, gradually destroyed with advancing charge order. The charge order of oxygen-doped/enriched cuprates {\it itself} is not detected. 

\includegraphics[width=4.3in]{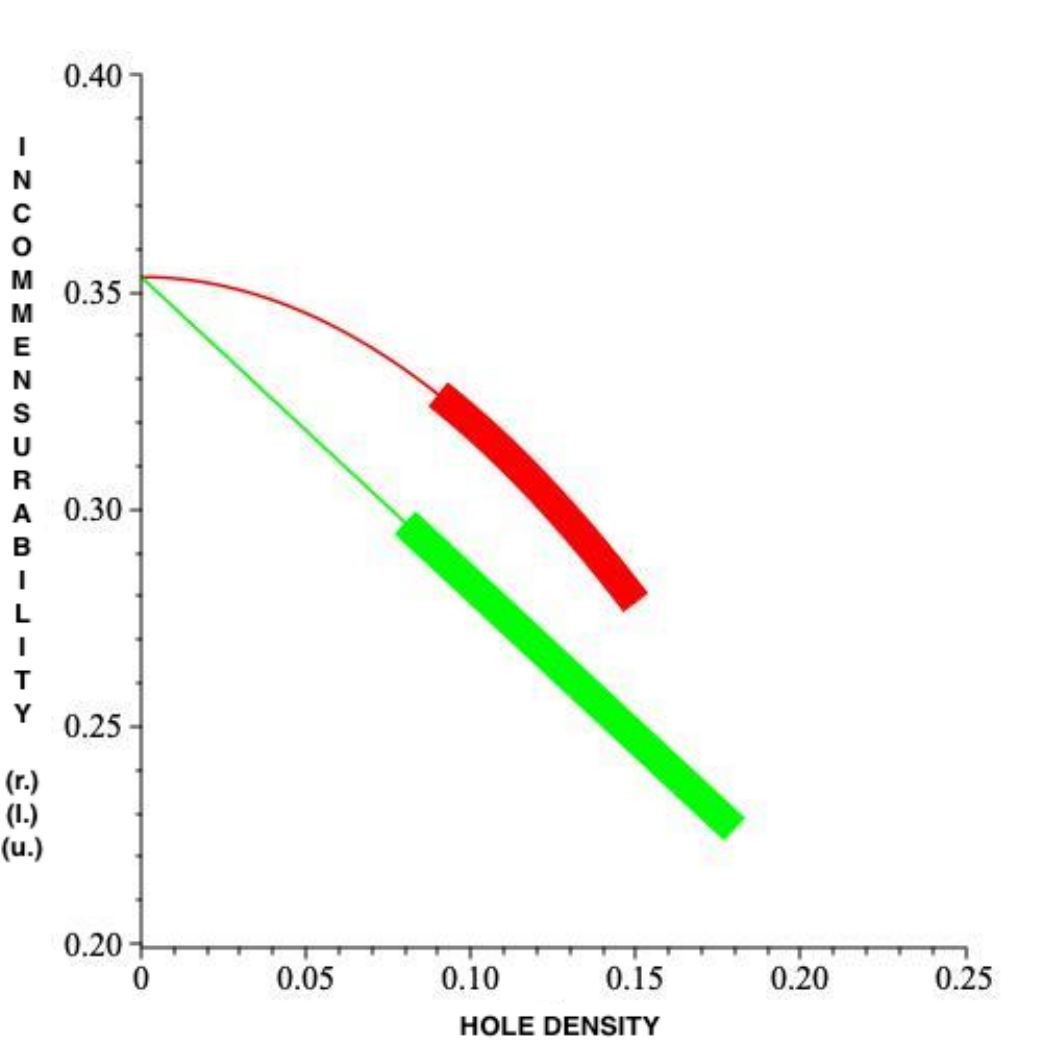}  

\noindent \footnotesize FIG. 22. Incommensurability $q_{CuO_6}(p)$, observed with RIXS, in oxygen-doped $YBa_2Cu_3O_{6+y}$ (red) and oxygen-enriched cuprates $HgBa_2CuO_{4 + \delta}$, $Bi_2Sr_2CuO_{6+\delta}$ and $Bi_2Sr_2CaCu_2O_{8+\delta}$ (green) in dependence on hole density $p$. Thick lines {\it summarize} observed data.\cite{76,77,78,79,80,81,82,83,84,85,86,87,88}  Thin lines show  back extrapolation, Eqs. (18,19).   \normalsize

\noindent In {\it 214} cuprates, detection of the host tilt pattern is overpowered by the stronger RIXS signal from the $O^{\upuparrows}$ superlattice but in
{\it 123}/OEC the host tilt pattern comes to the fore. (Only when the $CuO_6$ octahedra are sufficiently thinned out, highly sensitive RIXS can detect weak $O^{\uparrow \downarrow}$ charge order.\cite{98})

By replacing in Figs. 12-14 the triplet-state $\mathbf{O^{\upuparrows}}$ atoms by singlet-state $\mathbf{O^{\uparrow \downarrow}}$ atoms, one can illustrate the hidden charge order of {\it 123}/OEC. A common base member of heavily tilted, neighboring $CuO_5O^{\uparrow \downarrow}$ octahedra, each $O^{\uparrow \downarrow}$ atom eliminates the slight diagonal tilt of {\it two} $CuO_6$ octahedra (see Fig. 12). This dilutes the density of the host octahedra by twice the $O^{\uparrow \downarrow}$ density, $o(p) = p/2$, which in turn increases the effective average spacing of the remaining $CuO_6$ octahedra. The incommensurability of the diluted $CuO_6$ octahedra decreases correspondingly. If the charge-order incommensurabilty of the (tetragonal) OEC is (approximately) $q_{co}(p) = \sqrt{p}$, then the incommensurabilty of the host tilt in OEC becomes
\begin{equation}
     q_{CuO_6}^{OEC}(p) = \frac{\sqrt{2}}{2} [0.5  - q_{co}^{2}(p)] = \frac{\sqrt{2}}{2} [0.5  - 2o(p)] =   0.354 - 0.71\; p \;.
\end{equation}
A limit is reached at hole doping  $p_2 \equiv 2/2^2 = 0.5$, corresponding to a commensurate $O^{\uparrow \downarrow}$ superlattice of spacing $A_0 = 2a_0$,  when no more $CuO_6$ host octahedra exist, $q_{CuO_6}^{OEC}(p_2) = 0.354 - 0.71/2 = 0$.

$YBa_2Cu_3O_{6+y}$ has an orthorhombic crystal structure with $CuO$ chains (Fig. 6). If (in tetragonal approximation) its charge-order incommensurability is $q(p) = \gamma \; p$, then a square dependence of the  host-tilt incommensurability would result, 
\begin{equation}
     q_{CuO_6}^{Y123}(p) = \frac{\sqrt{2}}{2} [0.5  - q^2(p)] =   0.354 - \frac{\gamma^2 \sqrt{2}}{2}\; p^2  = 0.354 - 3.33 \; p^2\;.
\end{equation}

\section{QUANTUM CRITICAL POINT OF OXYGEN-ENRICHED CUPRATES}
In $YBa_2Cu_3O_{6+y}$, $ HgBa_2CuO_{4+\delta}$, and $Bi_2Sr_2CaCu_2O_{8+\delta}$
a quantum critical point $p^* = 0.19$ is observed in transport-based experiments\cite {89,90,91,92,93,94,95,96,97}---a value close to, but different from $x^* = 0.18$ of $La_{2-x}Sr_xCuO_4$. Above the superconducting transition temperature $T_c(p)$, and also somewhat below (with superconductivity suppressed by a magnetic field), the pseudogap temperature $T^*(p)$ of $YBa_2Cu_3O_{6+y}$ and $ HgBa_2CuO_{4+\delta}$ is a straight line. In some experiments the line aims directly at $p^*$. In other cases it aims at a value $p > p^*$, as in  $La_{2-x}Sr_xCuO_4$ where a discontinuous collapse occurs at $x^*$, accompanied by a saturation to constant incommensurability of $q_{co}(x) = 0.235$ for $x > \hat{x} = 0.125$, Eq. (11), due to interlacing $O^{\upuparrows}$ superlattices. The ambiguity is eliminated by experiments on $Bi_2Sr_2CaCu_2O_{8+\delta}$. With angular resolved photoemission spectroscopy (ARPES) and REXS a discontinuous change of incommensurability is observed from $q \simeq 0.3$ to $q \simeq 0.25$ at $x \approx 1/8$.\cite{81} Moreover, high-precision RIXS detects (beyond AFM, $p > p_N$) an incommensurabilty $q(p)$ that decreases as in Eq. (18),  but saturates in the doping range $0.13 < p < 0.16$ to a constant level of $q(p) = 0.25$, extending to doping $p = 0.22$.\cite{98} A periodic modulation of the local density of state by $4a_0$ has been observed with scanning tunneling microscopy (STM) in a slightly overdoped crystal, $p \simeq 0.17$.\cite{99} Charge order with $q = 0.25$ has also been observed in trilayer $Bi_2Sr_2Ca_2Cu_3O_{10+\delta}$.\cite{88}

The constant $q(p)$ value is indicative of a  $4 \times 4$ $O^{\uparrow \downarrow}$ superlattice, akin to the one in $La_{2-x}Sr_xCuO_4$ where it is spanned by $O^{\upuparrows}$ atoms. The intensity of the RIXS signal decreases steeply for $p \rightarrow 0.13$ but then rises again in the range $0.13 < p < 0.22$.\cite{98} This can be interpreted as a crossover of two mechanisms with increasing doping $p$: The first is the waning signal of $q_{CuO_6}^{OEC}(p)$, Eq. (18), from the receding $CuO_6$ host octahedra. When its intensity falls below that of the  $4 \times 4$ $O^{\uparrow \downarrow}$ superlattice with its tilted $CuO_5O^{\uparrow \downarrow}$ octahedra, the latter comes to the fore. Its rise in intensity may be caused by the filling of $(4 \times 4)$ $O^{\uparrow \downarrow}$ squares to $(4 \times 4 + 1)$ islands. The different  saturated $q_{co}(p)$ values at $p > \hat{p}$---that is, $q_{co} = 0.235$ for $La_{2-x}Sr_xCuO_4$ but $q_{co} \simeq 0.245$ for {\it 123}/OEC---may result from slightly different watershed values, $\hat{x} = 0.125$ and $\hat{p} \simeq 0.13$, respectively, together with somewhat heavier and lesser tilts of the $O^{\upuparrows}$ and $O^{\uparrow \downarrow}$ superlattices, accounted for by tilt corrections $\grave{t} = 0.015$ and $\grave{t} \simeq 0.010$, Eq. (11).

To summarize, the saturation of a primary $(4 \times 4)$ $O^{\upuparrows}$ or $O^{\uparrow \downarrow}$ superlattice at watershed doping $\hat{p} \simeq 1/8 = 0.125$ is common to both $La_{2-x}Sr_xCuO_4$ and the oxygen-doped/enriched cuprates. An interlaced secondary superlattice grows until the pseudogap closes at doping $x^* = 0.18$ for the former but at $p^* = 0.19$ for the latters. Why the difference? The relation between $p^*$,  watershed doping $\hat{p}$, and tilt correction $\grave{t}$ is given by $p^* = (3\hat{p} - \grave{t})/2$, Eq. (15). With the above values,  $p^* = 0.19$ is obtained for {\it 123}/OEC.

\section{CHARGE DENSITY FLUCTUATIONS}

In addition to static 
charge order (Secs. XI - XVIII), {\it dynamic} charge order---also called ``charge density fluctuations'' (CDF)---has been observed with RIXS in $La_{2-x}Sr_xCuO_4$\cite{100,101,102,103,104} as well as in $YBa_2Cu_3O_{6+y}$, $BiSr_2CaCu_2O_{8+\delta}$, and $HgBa_2CuO_{4+\delta}$.\cite{105,106,107} 

\noindent For $La_{2-x}Sr_xCuO_4$ the main findings are:\cite{103} 

(i) In the explored doping range, $0.11 \le x \le 0.22, $ a chopped-off charge-order dome is observed, extending over $0.11 < x < 0.16$; peaking at $x = 0.125$; having a constant incommensurability, $q_{co}(x) \simeq 0.24$ {\it r.l.u.} and a correlation length $\xi \approx 13 a_0$. It corresponds to a $4 \times 4$ lattice periodicity. The symmetry of the observed dome implies that its full doping range extends over $0.08 < x < 0.16$. This coincides with the short-range charge order covered in Sec. XVI. The incommensurabilty agrees (within error bars) with that of the saturated primary $O^{\upuparrows}$ superlattice, $q_{co}(\hat{x}) = 0.235$ {\it r.l.u.}, Eq. (11).

(ii) The charge order is tracked by CDF of the same incommensurability, $q_{cdf}^{co}(x) \simeq 0.24$ {\it r.l.u.}, but with shorter correlation length, $\xi \approx 5 a_0$. These CO-CDF extend to $x = 0.18$. Their doping range coincides with that of the saturated primary $O^{\upuparrows}$ superlattice, $0.125 = x_4 < x < x^* = 0.18$.

(iii) Another CDF signal is observed with incommensurability $q_{cdf}^\#(x) \approx 0.29$ {\it r.l.u.} and correlation length, $\xi \approx 5 a_0$. By the sensitivity of the RIXS experiment it disappears at $x = 0.22$. However, extrapolation suggests a flat dome, extending over $0.05 \preccurlyeq x \preccurlyeq 0.25$ (with the lower limit inferred from phonon softening).\\

Charge density fluctuations can be regarded as arising from temporal displacements, $\mathbf{s}(t)$, of double holes from $O^{\upuparrows}$ atoms on their superlattice site to neighboring $O^{2-}$ ions, Eq. (3)---most probably by $\mathbf{s} = (\pm a_0/2, \pm b_0/2)$, accompanied by an orientational change of the demotted $\mathcolor{red}{\mathbf{Cu^{2+}_2}}$ dimers between $[100]$ and $[010]$ directions, as illustrated in Fig. 26. This would explain the equality of their incommensurabilities, $q_{cdf}^{co}(x) = q_{co}(x)$, for $\hat{x} < x < x^*$, given by Eq. (12).

\noindent \textbf{Assumption 12: RIXS signals of CDF originate with fluctuating $\mathbf{[100]}$-$\mathbf{[010]}$ reorientations of $\mathcolor{red}{\mathbf{Cu^{2+}_2}}$ dimers.}

In the doping range $0.056 = x_6 < x <\hat{x} = 0.125$, $CuO_5O^{\upuparrows}$ octahedra are well separated (no overlap, Fig. 13) and tilt axially.
Figure 26 illustrates an elementary step of a charge density fluctuation in this situation: When a double hole hops from an $O^{\upuparrows}$ atom, say at a ``$b$-site'' (left), to the $O^{2-}$ at a nearest ``$a$-site'' (right)---turning it to an $O^{\upuparrows}$ atom, Eq. (3)---the flanking $\mathcolor{red}{\mathbf{Cu}}O_5O^{\upuparrows}$ octahedra change their tilt orientation from $[100]$  to $[010]$. The change gives rise to a RIXS signal. Such CDF-RIXS signals appear only along {\it axial} directions in the Brillouin zone, $(H,0)$ and $(0,K$), not diagonally, $(H,H)$.

\pagebreak

A different scenario holds for lower doping, $x < x_6$, where $CuO_5O^{\upuparrows}$ octahedra tilt diagonally. The $\mathcolor{red}{\mathbf{Cu}}O_5^{2-}O^{\upuparrows}$ octahedra flanking the $O^{\upuparrows}$ atom then have opposite diagonal tilts, $[110]$ and $[\underline{11}0]$, before {\it and} after the fluctuation (Fig. 27), causing \emph{no} RIXS signal in that doping range. 

In the previously considered range, $x_6 < x <\hat{x}$, the intensity of the CDF component of the RIXS signal is proportional to doping, $I_{cdf}^{co}(x) = \kappa (x - x_6)$ with a coefficient $\kappa$. Beyond watershed doping, $x > \hat{x}$, $4 \times 4$ $O^{\upuparrows/\downdownarrows}$ squares are being filled with a central $O^{\upuparrows}$ atom, leading to ``$4 \times 4 +1$ islands'' where {\it all} $Cu^{2+}$ ions are demotted: $\mathcolor{red}{\mathbf{Cu^{2+}}}$  (Fig. 14). 
Then a charge density fluctuation by a double hole's hop is no longer discernible
by a reorientation of  $\mathcolor{red}{\mathbf{Cu^{2+}_2}}$ dimers.
Moreover, the $\mathcolor{red}{\mathbf{Cu^{2+}_2}}$ dimers about the $O^{\upuparrows}$ atoms at the corners of the $4 \times 4 +1$ islands 

\medskip 

\noindent $\mathcolor{blue}{\;\;\;\;\;\;\;\;\;\;\;\; _\uparrow O^{2-}} \;\;\;\;\;\;\;\;\;\;\;\;\;\;\;\mathcolor{blue}{ O^{2-} \;\;\;\;\;\;\;\;\;\;\;\;\;\;_\uparrow O^{2-}} \;\;\;\;\;\;\;\;\;\;\;\;\;\; 
| \;\;\;\;\;\;\;\;\;\;\;\;\;\;\mathcolor{blue}{_\uparrow O^{2-} \;\;\;\;\;\;\;\;\;\;\;\;\;\;\;_\downarrow O^{2-}} \;\;\;\;\;\;\;\;\;\;\;\;\;\;\;\mathcolor{blue}{_\uparrow O^{2-}}$
 
\noindent $\mathcolor{blue}{_\downarrow O^{2-}} \;\;\;\;\mathcolor{red}{\overleftarrow{Cu}^{2+}} \;\;\mathcolor{green}{_\uparrow O^{2-}}\mathcolor{red}{ \;\;\; \mathcolor{red}{\overrightarrow{Cu}^{2+}}} \;\;\mathcolor{blue}{_\downarrow O^{2-}} \;\mathcolor{red}{_\swarrow  Cu^{2+}} \;\;\mathcolor{blue}{_\uparrow O^{2-}}\;\; 
| \;\;\;\mathcolor{blue}{_\downarrow O^{2-}} \;\mathcolor{red}{_\swarrow  Cu^{2+}} \;\;\mathcolor{blue}{_\uparrow O^{2-}} \;\;\mathcolor{red}{\mathbf{\uparrow} Cu^{2+}} \;\;\mathcolor{blue}{_\downarrow O^{2-}} \;\;\;\mathcolor{red}{_\swarrow Cu^{2+}} \;\;\mathcolor{blue}{_\uparrow O^{2-}}$ 

\noindent $\;\;\;\;\;\;\;\;\;\;\;\;
\mathcolor{green}{O^{2-} \;\;\;\;\;\;\;\;\;\;\;\;\;\;\;\; O^{2-}} \;\;\;\;\;\;\;\;\;\;\;\;\;\;\;\;\mathcolor{blue}{ O^{2-}} \;\;\;\;\;\;\;\;\;\;\;\;\;\; | \;\;\;\;\;\;\;\;\;\;\;\;\;\;\;\; \mathcolor{blue}{O^{2-}} \;\;\;\;\;\;\;\;\;\;\;\;\;\;\; \mathcolor{green}{_\uparrow O^{2-}} \;\;\;\;\;\;\;\;\;\;\;\;\;\;\;\; \mathcolor{blue}{_\downarrow O^{2-}}$

\noindent $ \mathcolor{green}{\mathbf{\uparrow} O^{2-}} \;\;\mathcolor{red}{\overrightarrow{\mathbf{Cu^{2+}}}} \;\;\;\;_\downarrow\mathbf{o}^\mathbf{\upuparrows} \;\;\;\;\mathcolor{red}{\overleftarrow{\mathbf{Cu^{2+}}}}\;\;\;\mathcolor{green}{\mathbf{\uparrow} O^{2-}} \;\;\mathcolor{red}{\overrightarrow{Cu}^{2+}} \;\;\mathcolor{blue}{_\downarrow O^{2-}} \;\; | \;\;\;\mathcolor{blue}{_\uparrow O^{2-}} \;\;\;\mathcolor{red}{\mathbf{\uparrow} Cu^{2+}} \;\;\;\mathcolor{green}{ O^{2-}} \;\mathcolor{red}{\mathbf{\Downarrow}\mathbf{Cu^{2+}}} \;\;\;\mathcolor{green}{ O^{2-}} \;\;\;\;\mathcolor{red}{\mathbf{\uparrow} Cu^{2+}} \;\;\mathcolor{blue}{_\downarrow O^{2-}}$  

\noindent $ \;\;\;\;\;\;\;\;\;\;\;\;\mathcolor{green}{O^{2-} \;\;\;\;\;\;\;\;\;\;\;\;\;\;\;\; O^{2-}} \;\;\;\;\;\;\;\;\;\;\;\;\;\;\;\;\mathcolor{blue}{ O^{2-}} \;\;\;\;\;\;\;\;\;\;\;\;\;\;
| \;\;\;\;\;\;\;\;\;\;\;\;\;\;\;\;\;\mathcolor{green}{O^{2-}}\;\;\;\;\;\;\;\;\;\;\;\;\;\;\; _\downarrow\mathbf{o}^{\upuparrows}\;\;\; \mathcolor{green}{ \;\;\;\;\;\;\;\;\;\;\;\;\;\;\;\; O^{2-}}  $

\noindent $\mathcolor{blue}{_\downarrow O^{2-}} \;\;\;\;\mathcolor{red}{\overleftarrow{Cu}^{2+}} \;\;\mathcolor{green}{_\uparrow O^{2-}}\mathcolor{red}{ \;\;\; \mathcolor{red}{\overrightarrow{Cu}^{2+}}} \;\;\mathcolor{blue}{_\downarrow O^{2-}} \;\mathcolor{red}{_{\swarrow}  {Cu^{2+}}} \;\;\mathcolor{blue}{_\uparrow O^{2-}}\;\;\; | \;\;\;\mathcolor{blue}{_\downarrow O^{2-}} \;\;\mathcolor{red}{ \mathbf{\downarrow} Cu^{2+}} \;\;\;\mathcolor{green}{ O^{2-}} \;\;\;\mathcolor{red}{\mathbf{\Uparrow} \mathbf{Cu}^{2+}} \;\;\;\mathcolor{green}{ O^{2-}} \;\;\;\;\mathcolor{red}{\mathbf{\downarrow} Cu^{2+}} \;\;\mathcolor{blue}{_\uparrow O^{2-}}$

\noindent $\mathcolor{blue}{\;\;\;\;\;\;\;\;\;\;\;\; _\uparrow O^{2-}} \;\;\;\;\;\;\;\;\;\;\;\;\;\;\;\mathcolor{blue}{ O^{2-} \;\;\;\;\;\;\;\;\;\;\;\;\;\;_\uparrow O^{2-}} 
\;\;\;\;\;\;\;\;\;\;\;\;\;\;\;  |  \;\;\;\;\;\;\;\;\;\;\;\;\;\;\;\mathcolor{blue}{_\uparrow O^{2-}} \;\;\;\;\;\;\;\;\;\;\;\;\;\;\;\mathcolor{green}{_\uparrow O^{2-}} \;\;\;\;\;\;\;\;\;\;\;\;\;\;\;\;\mathcolor{blue}{_\uparrow O^{2-}}$


\noindent \footnotesize FIG. 26. Elementary step of a {\it charge density fluctuation} in the  $CuO_2$ plane at doping $x_6 < x <\hat{x}$ with an $O^{\upuparrows}$ atom at a ``$b$-site'' (left) and displaced by $(a_0/2, -b_0/2)$ to a nearest ``$a$-site'' (right) with associated heavy tilts of $CuO_5O^{\upuparrows}$ octahedra (long overarrow, vertical double arrow) and medium tilts of  $CuO_6$ neighbors (short overarrow, vertical single arrow), demotted $\mathcolor{red}{\mathbf{Cu^{2+}}}$ ions, and stressed $\mathcolor{green}{O^{2-}}$ ions. The fluctuation changes the opposite axial tilt orientation of the $\mathcolor{red}{\mathbf{Cu^{2+}}}\mathcolor{green}{O^{2-}_5}O^{\upuparrows}$ octahedra from $[100]$ (left) to $[010]$ (right).
\normalsize

\bigskip
\noindent $\mathcolor{blue}{\;\;\;\;\;\;\;\;\;\;\;\; _\uparrow O^{2-}} 
\;\;\;\;\;\;\;\;\;\;\;\;\;\;\mathcolor{blue}{_\downarrow O^{2-}} 
\;\;\;\;\;\;\;\;\;\;\;\;\;\;\mathcolor{blue}{_\uparrow O^{2-}}
\;\;\;\;\;\;\;\;\;\;\;\;\;\;\;
 | 
\;\;\;\;\;\;\;\;\;\;\;\;\;\;\;\mathcolor{blue}{_\uparrow O^{2-}} 
\;\;\;\;\;\;\;\;\;\;\;\;\;\;\;\mathcolor{blue}{_\downarrow O^{2-}}
\;\;\;\;\;\;\;\;\;\;\;\;\;\;\;\mathcolor{blue}{_\uparrow O^{2-}}$

\noindent $ \mathcolor{blue}{_\downarrow O^{2-}} \; \mathcolor{red}{{\swarrow}Cu^{2+}} \; \mathcolor{green}{_\uparrow O^{2-}} \mathcolor{red}{\mathbf{\nearrow}  Cu^{2+}} \;\; \mathcolor{blue}{_\downarrow O^{2-}} \;\mathcolor{red}{{_\swarrow} {Cu^{2+}}} \;\;\mathcolor{blue}{_\uparrow O^{2-}} \;
\;\;\;
| 
\;\;\;\;\;\mathcolor{blue}{_\downarrow O^{2-}} \mathcolor{red}{_{\swarrow} {Cu^{2+}}} \;\;\mathcolor{blue}{_\uparrow O^{2-}} \;\mathcolor{red}{\mathbf{\nearrow} Cu^{2+}} \;\;\mathcolor{blue}{_\downarrow O^{2-}} \;\;
\mathcolor{red}{\mathbf{_\swarrow} Cu^{2+}}\;\;\;\mathcolor{blue}{_\uparrow O^{2-}} $ 

\noindent $\mathcolor{green}{\;\;\;\;\;\;\;\;\;\;\;\; _\downarrow O^{2-} \;\;\;\;\;\;\;\;\;\;\;\;\;\;_\uparrow O^{2-} \;\;\;\;\;\;\;\;\;\;\;\;\;\;\;\mathcolor{blue}{_\downarrow O^{2-}}} 
\;\;\;\;\;\;\;\;\;\;\;\;\;\;\;
 | \;\;\;\;\;\;\;\;\;\;\;\;\;\;\;\;\mathcolor{blue}{_\downarrow O^{2-}} \;\;\;\;\;\;\;\;\;\;\;\;\;\;\; \mathcolor{green}{_\uparrow O^{2-}} \;\;\;\;\;\;\;\;\;\;\;\;\;\;\;\mathcolor{blue}{_\downarrow O^{2-}}$

\noindent $\mathcolor{green}{ _\uparrow O^{2-}}  
\mathcolor{red}{\mathbf{\Nearrow}\mathbf{Cu^{2+}}} \; 
_\uparrow\mathbf{o}^{\upuparrows} \;
\mathcolor{red}{\mathbf{\Swarrow}  \mathbf{Cu^{2+}}} \;
\mathcolor{green}{_\downarrow O^{2-}}\; 
\mathcolor{red}{\mathbf{\nearrow} Cu^{2+}} \;\;
\mathcolor{blue}{_\downarrow O^{2-}} \;\;\;\;
| 
\;\;\;\;\;\mathcolor{blue}{_\uparrow O^{2-}}  \; \mathcolor{red}{\mathbf{\nearrow}Cu^{2+}}\; \mathcolor{green}{_\downarrow O^{2-}}  \mathcolor{red}{\mathbf{\Swarrow}\mathbf{Cu^{2+}}} \;\; \mathcolor{green}{_\uparrow O^{2-}} \;  \mathcolor{red}{\mathbf{\nearrow} Cu^{2+}}\;\; \mathcolor{blue}{\mathbf{_\downarrow} O^{2-}}$ 

\noindent $\mathcolor{green}{\;\;\;\;\;\;\;\;\;\;\;\; 
_\uparrow O^{2-}} \;\;\;\;\;\;\;\;\;\;\;\;\;\;
\mathcolor{green}{_\downarrow O^{2-}} \;\;\;\;\;\;\;\;\;\;\;\;\;\;\;
\mathcolor{blue}{_\uparrow O^{2-}} \;\;\;\;\;\;\;\;\;\;\;\;\;\;\;\;
 | \;\;\;\;\;\;\;\;\;\;\;\;\;\;\;\;\;
\mathcolor{green}{_\uparrow O^{2-}} \;\;\;\;\;\;\;\;\;\;\;\;\;\;\;
_\uparrow\mathbf{o}^{\upuparrows} \;\;\;\;\;\;\;\;\;\;\;\;\;\;\; \mathcolor{green}{_\uparrow O^{2-}}$

\noindent$\mathcolor{blue}{_\downarrow O^{2-}}\;\mathcolor{red}{\mathbf{\swarrow} Cu^{2+}}\;\; 
\mathcolor{green}{_\uparrow O^{2-}} \;\mathcolor{red}{{\nearrow}  Cu^{2+}} \;\;
\mathcolor{blue}{_\downarrow O^{2-}} \;\mathcolor{red}{_{\swarrow} {Cu^{2+}}} \;\;\mathcolor{blue}{_\uparrow O^{2-}} \;\;\;\;
| 
\;\;\;\;\mathcolor{blue}{_\downarrow O^{2-}}\;\;\mathcolor{red}{_{\swarrow} {Cu^{2+}}}\;\;
\mathcolor{green}{_\uparrow O^{2-}} \;\mathcolor{red}{\mathbf{\Nearrow} \mathbf{Cu^{2+}}} \;\;\mathcolor{green}{_\downarrow O^{2-}} \mathcolor{red}{\mathbf{\swarrow} Cu^{2+}} \;\;\mathcolor{blue}{_\uparrow O^{2-}}$

\noindent $\mathcolor{blue}{\;\;\;\;\;\;\;\;\;\;\;\;\;\; _\downarrow O^{2-} \;\;\;\;\;\;\;\;\;\;\;\;\;\;_\uparrow O^{2-}} \;\;\;\;\;\;\;\;\;\;\;\;\;\;\mathcolor{blue}{_\downarrow O^{2-}} \;\;\;\;\;\;\;\;\;\;\;\;\;\;\;
\;
| \;\;\;\;\;\;\;\;\;\;\;\;\;\;\;\;\;
\mathcolor{blue}{_\downarrow O^{2-}} \;\;\;\;\;\;\;\;\;\;\;\;\;\;\;\mathcolor{green}{_\uparrow O^{2-}} \;\;\;\;\;\;\;\;\;\;\;\;\;\;\;\mathcolor{blue}{_\downarrow O^{2-}}$



\noindent \footnotesize FIG. 27. Same as Fig. 26 but at lower doping, $p < p_6 = 0.056$.
The fluctuation does {\it not} alter the opposite diagonal tilt orientation, $[110]$ and $[\underline{11}0]$,  of the $\mathcolor{red}{\mathbf{Cu^{2+}}}\mathcolor{green}{O^{2-}_5}O^{\upuparrows}$ octahedra in the figure's left and right part. \normalsize

\noindent no longer contribute to  the CO-CDF signal, causing a {\it decrease} of intensity, $I_{cdf}^{co}(x) = I_{cdf}^{co}(\hat{x}) - \kappa (x - \hat{x})$, and disappearance, $I_{cdf}^{co}(x^*) \approx 0$, at doping $x^*$ when all $4 \times 4 +1$ islands become contiguous at their corners (Fig. 16). Likewise, no reorientation of $\mathcolor{red}{\mathbf{Cu^{2+}_2}}$ dimers is discernible with additional $4 \times 4 +1$ islands at doping $x > x^*$.

An explanation of the other, CDF$^\#$ signal of low flat dome and with incommensurability $q_{cdf}^*(x) \approx 0.29$, will be given below.

In oxygen-doped/enriched cuprates ($123/OEC$), the situation is quite different from $La_{2-x}Sr_xCuO_4$, as CO-CDF are affected by both the extent of $CuO_6$ host octahedra and the tilt of the hidden $CuO_5O^{\uparrow \downarrow}$ octahedra.
Like the charge-order incommensurability $q_{co}(p)$ in $123/OEC$ [Fig. 22, Eqs. (18,19)], $q_{cdf}^{co}(p)$ {\it decreases} with doping (Table III), however, with a doping {\it delay}, $q_{cdf}^{co}(p) \simeq q_{CuO_6}(p - p_6)$. The decrease of $q_{cdf}^{co}(p)$ is a strong indication of the receding pattern of $CuO_6$ host octahedra. The doping delay by $p_6$ is indicative of a change of tilt orientation of the $CuO_5O^{\uparrow \downarrow}$ octahedra. In $La_{2-x}Sr_xCuO_4$ a change of stripe orientation at doping $x_6$ was origionally discovered with neutron scattering,\cite{48} not with RIXS. (No connection with $CuO_5O^{\upuparrows}$ octahedra tilts was known at the time.) Due to its hidden nature, no change of stripe orientation has been reported for the $O^{\uparrow \downarrow}$ superlattice in $Y123/OEC$. The $p_6$-delay of $q_{cdf}(p)$ supports indirectly the same change of tilt orientation, that is, at $p_6$, in {\it both} $La_{2-x}Sr_xCuO_4$ and $123/OEC$.
 
As in $La_{2-x}Sr_xCuO_4$, no reorientation of $\mathcolor{red}{\mathbf{Cu^{2+}_2}}$ dimers is discernible for doping $p <p_6$ of $123/OEC$ due to diagonal $CuO_5O^{\uparrow \downarrow}$ octahedra tilts. Thus, as neighbor-hops of double holes maintain opposite diagonal $\mathcolor{red}{\mathbf{Cu}}O_5O^{\uparrow \downarrow}$ tilts, no fluctuation of the host $CuO_6$ octahedra is

 \medskip 

    \begin{table}[ht]

\begin{tabular}{|p{3.7cm}|p{1.6cm}|p{1.4cm}|p{1.4cm}|p{3.2cm}|p{1cm}|}
 \hline  \hline
\;\;\;Compound\;\; &$\;\;\;\;\;\;p$&$\;\;q_{cdf}^{expt}$&$\;\;q_{cdf}^{calc}$& \;\;\;\;\;\; Formula& \; Eq. \\ \hline \hline 
$\;\;\;\;YBa_2Cu_3O_{6+y}$&$\;\;\;\;0.06 $&$\;\;\;0.35$&$\;\;\;0.35$&$\;\leftarrow \;\;q_{CuO_6}^{123}(p - p_6)$&$\;\;\; 19$\\ \hline 
$\;\;\;\;YBa_2Cu_3O_{6+y}$&$\;\;\;\;0.11 $&$\;\;\;0.33$&$\;\;\;0.34$&$\;\leftarrow \;\;q_{CuO_6}^{123}(p - p_6)$&$\;\;\; 19$\\ \hline 
$\;\;\;\;YBa_2Cu_3O_{6+y}$&$\;\;\;\;0.185 $&$\;\;\;0.30$&$\;\;\;0.30$&$\;\leftarrow \;\;q_{CuO_6}^{123}(p - p_6)$&$\;\;\; 19$\\ \hline 
$\;\;\;\;YBa_2Cu_3O_{6+y}$&$\;\;\;\;0.22 $&$\;\;\;0.27$&$\;\;\;0.26$&$\;\leftarrow \;\;q_{CuO_6}^{123}(p - p_6)$&$\;\;\; 19$\\ \hline \hline
$\;\;\;\;HgBa_2CuO_{4+\delta}$&$\;\;\;\;0.09 $&$\;\;\;0.28$&$\;\;\;0.29$&$\;\leftarrow \;\;q_{CuO_6}^{OEC}(p - p_6)$&$\;\;\; 18$\\ \hline \hline 
$\;\;\;\;BiSr_2CaCu_2O_{8+\delta}$&$\;\;\;\;0.19 $&$\;\;\;0.25$&$\;\;\;0.26$&$\;\leftarrow \;\;q_{CuO_6}^{OEC}(p - p_6)$&$\;\;\; 18$\\ \hline \hline 
\end{tabular}
\label{table:3}
\footnotesize \caption{Doping dependence of the incommensurability of charge-density fluctuations, $q_{cdf}(p)$: Experimental values\cite{105,106,107} and calculations with $p_6 \equiv 2/6^2 \simeq 0.056$.  } \normalsize
\end{table}

\noindent  observed by RIXS in that doping range. As far as {\it fluctuations} are concerned, the crystal ap- pears  pristine, $q_{cdf}^{co}(p) \simeq 0.35$, for $p < p_6$. At larger doping, axial $[100]$-$[010]$ tilt fluctuations of $\mathcolor{red}{\mathbf{Cu}}O_5O^{\uparrow \downarrow}$ octahedra dilute the pattern of $CuO_6$ octahedra, resulting in  $q_{cdf}^{co}(p) \simeq q_{CuO_6}(p - p_6)$.

Low-intensity charge-density fluctuations of a different kind (here denoted as CDF$^\#$) have been observed in both $La_{2-x}Sr_xCuO_4$ and oxygen-doped/enriched cuprates.\cite{103,105,106} In contrast to the CO-CDF, which have an incommensurability $q_{cdf}^{co}(p) = q_{co}(p)$, the CDF$^\#$ extend over a wide doping range, $0.05 \preccurlyeq p \preccurlyeq 0.25$ with a flat intensity dome, $I_{cdf}^\#(p)$, that is, with only weak doping dependence. The CDF$^\#$ pervade all temperatures, up to $T \approx 300$ K but are essentially temperature independent, that is, {\it not} thermally activated.

\textbf{Assumption 13: The low-intensity charge-density fluctuations of} (essentially) \textbf{doping and temperature independence are slow quantum fluctuations, originating with spontaneous demotting/remotting pairs of $\mathbf{Cu \; 3d}$-$\mathbf{O \; 2p}$ holes in the host lattice}. 


The incommensurability of the CDF$^\#$  implies an underlying order of fluctuating $Cu  3d$-$O  2p$ hole pairs,
$ \;  Cu^{2+} \;\; O^{2-} \;\; Cu^{2+} \Longleftrightarrow \;\; \mathcolor{red}{\mathbf{Cu^{2+}}}\ \; O^{2-} \;\; \mathcolor{red}{\mathbf{Cu^{2+}}},$ 
of density (per unit cell) $p^\# = 2(q_{cdf}^\#)^2$. With $q_{cdf}^\#(p) \approx 0.30$ {\it r.l.u.}, this gives $p^\#  \approx 0.18$---a value that coincides with the doping $p^* = 0.185 \pm 0.005$ that marks the closing of the pseudogap. Altough $p^*$ is  {\it no true} critical point (no large quantum fluctuations, no diverging correlation length), there seems to be a connection of CDF$^\#$ with $p^*$. The CDF$^\#$ represent a slight instability of the half-filled state over the entire doping range, $p > p_N$, and may assist the transition to the strange-metal phase.

\section{FERMI SURFACE RECONSTRUCTION, QUANTUM OSCILLATIONS}

When subjected to a strong magnetic field, a reconstruction of the Fermi surface of $YBa_2Cu_3O_{6+y}$ and oxygen-enriched cuprates has been suspected for a long time and was confirmed with the discovery of quantum oscillations in $YBa_2Cu_3O_{6+y}$ and $HgBa_2CuO_{4+\delta}$.\cite{108,109} Quantum oscillations, detected in de Haas-van Alphen and Shubnikov-de Haas experiments, are a standard tool to determine size and shape of (closed) Fermi surfaces of metals. In said cuprates they occur in the underdoped range, $0.98 \ge p \le 0.126$ and $p \approx 0.09$, respectively, that is, in the pseudogap regime where Fermi arcs are found. The existence of quantum oscillations in underdoped cuprates implies a {\it closed} Fermi surface. In the present cases, its size is only $2\pm 0.2$\% and 3\% of the Brillouin zone, respectively; hence called Fermi ``pocket.'' Implied by Hall effect and Seebeck effect experiments, its nature is considered electron-like (``electron pocket'').\cite{110,111,112,113} How can a Fermi pocket be constructed out of Fermi arcs?

Fermi-surface reconstruction is commonly attributed to charge order. In the oxygen-doped/enriched cuprates the influence of charge order is {\it indirect} by (directly) reducing the momentum dependence $q_{CuO_6}(p)$ of the diagonal tilt of host lattice $CuO_6$ octahedra, Eqs. (18, 19). It has been noticed that vectors of such length, and along the \textit{a-} and $b-$direction, connect Fermi-arc tips of neighboring BZ quadrants of $YBa_2Cu_3O_{6+y}$ and $HgBa_2CuO_{4+\delta}$ for doping $p_{QO}$ where quantum oscillations occur.\cite{114,115,116,117} Figure 28 shows a Fermi arc in the first quadrant of the BZ. With the $q_b$ axis serving as a mirror axis, the Fermi arc in the second quadrant is mirror symmetric to the first. When the second Fermi arc is shifted to the right by $q = q_{CuO_6}(p_{QO})$, the top Fermi-arc tips $\hat{Q}^\Uparrow$ of both arcs coincide. With a vertical line through $\hat{Q}^\Uparrow$ serving as a new mirror axis, half of the Fermi pocket is constructed. Similar shifts of the Fermi arcs from the forth and third BZ quadrant complete the closure of the  (diamont shaped) Fermi pocket. It is common practice to shift energy bands of quantum states with energy $\epsilon(\mathbf{q})$  from higher BZs by a reciprocal lattice constant, say $q_a = 1/a_0$, back into the first BZ to obtain a {\it reduced} BZ. However, what justifies the shift of Fermi arcs by $q_{CuO_6}(p_{QO})$ to construct a Fermi pocket?

The justication is given by the magnetic aspect. Along with the pattern of diagonally tilted $CuO_6$ octahedra of {\it non}-doped/enriched cuprates goes the AFM of the $^\uparrow Cu^{2+}$ spins that spans the {\it magnetic} BZ. Its borders connect the antinodal points of the BZ, $\mathbf{q} = (\pm 0.5, \pm0.5)$ (dashed red line in Fig. 28). The perpendicular distance of the magnetic BZ border from the origin of the BZ is $q_{CuO_6}(0) \simeq 0.35 \; r.l.u.$ (dashed red arrow in Fig. 28). Oxygen doping/enrichment eliminates some host octaheadra, $CuO_6 \rightarrow CuO_5O^{\uparrow\downarrow}$, causing a reduction of AFM. The receded magnetic BZ border is depicted by a dashed purple line in Fig. 28, a distance $q$ from the origin of the BZ (diagonal purple arrow), here $q = q_{CuO_6}(p) = 0.29$ for $p=0.09$, Eq. 18. Horizontal and vertical purple arrows of length $p/2$ indicate half of the shift of Fermi arcs from neighboring BZ quadrants to construct a Fermi pocket. 

Quantum oscillations from a reconstructed Fermi surface have {\it not} been detected in oxygen-enriched $Bi_2Sr_2CuO_{6+\delta}$ and $Bi_2Sr_2CaCu_2O_{8+\delta}$ due to large residual impurity scattering. The electron mean free path in the bismuth cuprates is about seven times shorter than in $YBa_2Cu_3O_{6+y}$ and about half as long as in $HgBa_2CuO_{4+\delta}$ under comparable conditions.\cite{118}

Why don't the {\it 214} lanthanum cuprates show quantum oscillations? The maximum charge-order incommensurabilty of $La_{2-x}Sr_xCuO_4$ is $q_{co} = 0.235$  {\it r.l.u.} and $q_{co} = 0.266$ {\it r.l.u.} for $La_{1.6-x}Nd_{0.4}Sr_xCuO_4$---short and, respectively, shy of $q \approx 0.30$ {\it r.l.u.} (Table II and Fig. 9), necessary to connect Fermi-arc tips of the Brillouin-zone quadrants.

\includegraphics[width=3.9in]{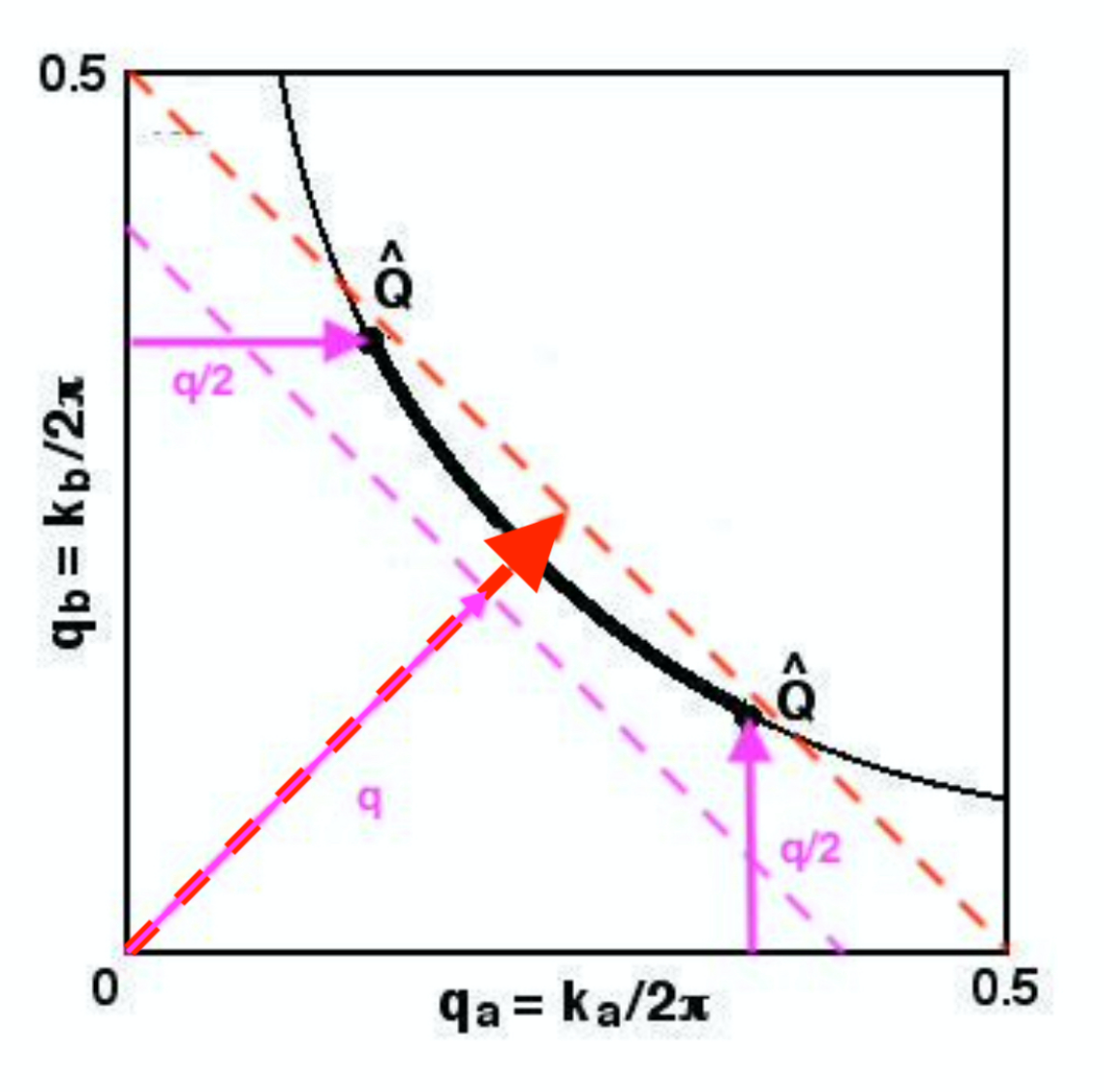}  

\noindent\footnotesize  FIG. 28. Schematic drawing of the first quadrant of the lattice Brillouin zone (BZ) with Fermi arc of oxygen-enriched cuprate (OEC) at doping $p=0.09$ (thick black line) and underlining Fermi surface (thin black line). The dashed red line $(0, 0.5)\mathcolor{red}{- - -}(0.5, 0)$ shows the boundary of the {\it magnetic} BZ of a {\it non-doped} cuprated, with distance $q_{CuO_6}(0) = \sqrt{2}/4 \simeq 0.35$ from the origin of the BZ (dashed red arrow). The purple dashed line shows the boundary of the  {\it shrunk magnetic} BZ of an OEC with distance $q_{CuO_6}(p) \simeq 0.29 \; r.l.u.$ for $p = 0.09$, Eq. (18), from the BZ origin (purple diagonal arrow). The horizontal and vertical purple arrow of length $q/2$ connect to Fermi-arc tips in the second and forth BZ quadrant, respectively. \normalsize 

\section{FERMI ARCS REVISITED}{}   

The present approach, with its emphasis on ion configurations, falls short of revealing the nature of the Fermi arcs in momentum space. 
Although they may be related to the distribution of \emph{trapped} demotted $Cu 3d$-$O  2p$ holes in the $CuO_2$ plane, no quantitave relationship with the arcs' geometry could be discerned.




\section{SUMMARY}
Pristine $La_2CuO_4$---like the related multi-lanthanide cuprates $La_{2-z}Ln_zCuO_4$ ($Ln = Nd, Eu; z = 0.4, 0.2$)---is an AFM Mott insulator with half-filled $Cu3d$-$O2p$ quantum states. Doping those {\it 124} compounds with $Sr$---or, in the first case, with $Ba$---introduces additional holes into the $CuO_2$ planes. The ``doped'' holes reside pairwise at oxygen atoms,  $O^{2-} + 2e^+ \rightarrow O$. The $O$ atoms' lack of net charge {\it demottes} $Cu3d$-$O2p$ holes, hosted about the flanking $Cu^{2+}$ ions, by delocalizing them at their hosts. The demotted  $Cu3d$-$O2p$ holes are trapped between (motted) $Cu^{2+}$ ions along $Cu$-$O$ bonds, preventing them from moving freely through the crystal.

In the {\it 214} family, the oxygen atoms are in a triplet state ($S=1$), with the double hole's spins in parallel, $O^{\upuparrows}$. On the other hand, when cuprates are oxygen-doped/enriched, as in $YBa_2Cu_3O_{6+y}$ and $LaCuO_{4+\delta}$, $HgBa_2CuO_{4+\delta}$, $Bi_2Sr_2CuO_{6+\delta}$, $Bi_2Sr_2CaCu_2O_{8+\delta}$, oxygen atoms likewise occupy anion sites in the $CuO_2$ planes but with antiparallel spins of their double hole, $O^{\uparrow \downarrow}$ ($S=0$). The presence and absence of a net magnetic moment of the $O^{\upuparrows}$ and $O^{\uparrow \downarrow}$ atoms, respectively, affects the collapse of 3D AFM (small or larger N\'{e}el doping $p_N$) as well as the presence or absence of magnetic order accompanying charge order (stripes).

$CuO_6$ octahedra of $La_2CuO_4$ have a slight diagonal tilt---alternately oriented along the $Cu$-$O$ bonds. When doped with $Sr$, each common oxygen {\it atom} at the base of neighboring $CuO_5O^{\upuparrows}$ octahedra {\it tends} to tilt them axially. At low doping, $x < x_6 \simeq 0.056$, the tendency is overwhelmed by the diagonal tilt of the host $CuO_6$ octahedra, resulting in a heavy {\it diagonal} tilt of the $CuO_5O^{\upuparrows}$ octahedra.
However, enforced by stronger magnetic $^\uparrow Cu^{2+}$-$O^{\upuparrows}$ and $O^{\upuparrows}$-$O^{\downdownarrows}$ interaction at higher doping, $x > x_6$, the axial tilt tendency gains the upper hand, resulting in a heavy {\it axial} tilt of the $CuO_5O^{\upuparrows}$ octahedra. The $CuO_6$ octahedra next to them are moderately tilted in a transition to the slight tilt of the host $CuO_6$ octahedra. 

We call $O^{2-}$ base ions of heavily tilted $CuO_5O^{\upuparrows}$ octahedra, and also $O^{2-}$ ions between moderately tilted $CuO_6$ octahedra, {\it ``stressed'',} marked green, $\mathcolor{green}{O^{2-}}$, in configurations figures. Being in close range of a demottifying  $O^{\upuparrows}$ atom, stressed $O^{2-}$ ions are convenient in determining whether a  $Cu3d$-$O2p$ hole is demotted or not: If a copper ion is adjacent to at least {\it three} stressed $O^{2-}$ ions, then the corresponding  $Cu3d$-$O2p$ hole is {\it ``demotted''} (marked as $\mathcolor{red}{\mathbf{Cu^{2+}}}$), that is, delocalized at that copper ion but trapped by (motted) $Cu^{2+}$ ions along $Cu$-$O$ bonds.

The distribution of demotted  $Cu3d$-$O2p$ holes at $\mathcolor{red}{\mathbf{Cu^{2+}}}$ ions in the $CuO_2$ plane is reflected by Fermi arcs in the Brillouin zone, extending from nodal points $\mathbf{\dot{Q}} = (\pm 0.21, \pm 0.21)$ toward anti-nodal points $\mathbf{X} = (\pm 0.5, 0)$ and $\mathbf{{Y}} = (0, \pm 0.5)$ at the border of the Brillouin zone.
A {\it partial} energy gap---conventionally called ``pseudogap''---extends from the tips of the Fermi arcs, $\mathbf{\hat{Q}}$, to the border of the Brillouin zone, with maximum value $\Delta^*(p)$ at points $\mathbf{X}$ and $\mathbf{Y}$. At temperature $T \ge T^*(p) = \Delta^*(p)/k_B$, thermal energy {\it demottes}  the remaining  $Cu3d$-$O2p$ holes, setting {\it all} half-filled  $Cu3d$-$O2p$ holes free to hop along the $Cu$-$O$ bonds in leaps of one-lattice constant length. This is the {\it strange metal} phase, characterized by a resistivity $\rho \propto T$. The latter is caused by a ``Plankian'' scattering rate $1/\tau = k_B T/\hbar$. The pseudogap temperature, $T^*(p) = T^*_0 - sp$, decreases with increased doping, as less (motted)  $Cu3d$-$O2p$ holes need to be demotted.

By contrast, charge transport in the {\it pseudogap} phase, $T < T^*$, is carried by zig-zag hops of double holes from $O^{\upuparrows}$ or $O^{\uparrow \downarrow}$ atoms to neighboring $O^{2-}$ ions. If their scattering rate is also Planckian, then the {\it two} hops, necessary for advancement by one lattice constant, invoke dual scattering with a net rate $(1/\tau)^2$. This could explain the observed resistivity $\rho \propto T^2$.

Because of their charge deviation from the host lattice's $O^{2-}$ ions, the $O^{\upuparrows}$ atoms in the $CuO_2$ planes of {\it 214} systems, as well as the $O^{\uparrow \downarrow}$ atoms in the $CuO_2$ planes of oxygen-doped/enriched cuprates, form {\it charge orders,} described (in quasi-tetragonal approximation) by an incommensurability $q_{co}(p)$ that depends on the hole density $p$. The predominant method to detect charge order is by resonant inelastic X-ray spectroscopy (RIXS), although historically preceded for {\it 214} cuprates with neutron scattering.\cite{29} 
In the {\it 214} family the RIXS signal is attributed to the heavily tilted $CuO_5O^{\upuparrows}$ octahedra.
The corresponding $O^{\upuparrows}$ atoms form an {\it incommensurate superlattice}. Its incommensurability, $q_{co}(x) \propto \sqrt{x - \grave{t}}$, in reciprocal lattice units ({\it r.l.u.}), {\it increases} with increasing hole density $p = x$, up to a ``watershed'' doping, $x < \hat{x}$. (The quantity $\grave{t}$ is a tilt correction.) The  orientation of the charge order changes at doping $x_6  \simeq 0.056$ with the change of tilt of  $CuO_5O^{\upuparrows}$ octahedra from diagonal to axial. 

At the ``watershed'' doping $\hat{x}$, the incommensurability saturates to a constant value, $q_{co}(x) \propto \sqrt{\hat{x} - \grave{t}}$. For $La_{2-x}Sr_xCuO_4$, the watershed doping is $\hat{x} = x_4 = 0.125$. In this case the $O^{\upuparrows}$ superlattice, of spacing $A_0 = 4 a_0$, is commensurate with the crystal lattice of spacing $a_0$. With larger doping, $x > \hat{x}$, a {\it secondary} superlattice of $O^{\upuparrows}$ atoms commences whereby the additional $O^{\upuparrows}$ atoms occupy the center of the $4 \times 4$ $O^{\upuparrows}$ quartets of the primary superlattice to form ``$4 \times 4 + 1$'' islands. Such islands fill up with demotted $\mathcolor{red}{\mathbf{Cu^{2+}}}$ ions. However, their demotted  $Cu3d$-$O2p$ holes are trapped by surrounding (motted) $Cu^{2+}$ ions. 

At doping $x^*$ the ``$4 \times 4 + 1$'' islands become contiguous at their corners. Continuous lines of $\mathcolor{red}{\mathbf{Cu^{2+}}}$ ions now extend along $Cu$-$O$ bonds. The wave vectors of the Fermi-arc tips $\mathbf{\dot{Q}}$ coincide with the $\mathbf{X}$ and $\mathbf{Y}$ points of the Brillouin zone. Closing the pseudogap, the Fermi arcs join to a complete (2D) Fermi surface. With a sudden collapse of the pseudogap-temperature line $T^*(x)$ at doping $x^*$, the demotted  $Cu3d$-$O2p$ holes are free to move throughout crystal domains without thermal demotting, $T^*(x^*) = 0$. By the present, ion-configuration-based approach, that is all there is to $x^*$. Lacking evidence for large-scale quantum fluctuations or divergence of correlation length, $x^*$ is not a {\it true} ``quantum critical point'' but merely marks the doping where the strange-metal phase is reached without thermal demotting.

The different values of watershed doping, $\hat{x} = 0.125$ of $La_{2-x}Sr_xCuO_4$ but $\hat{x} = 0.16$ of $La_{1.6-x}Nd_{0.4}Sr_xCuO_4$---and by $x^* = (3\hat{x} - \grave{t})/2$, of the corresponding quantum critical points, $x^* = 0.18$ but $x^* = 0.23$---are a consequence of different low-temperature orthorhombic (LTO) and tetragonal (LTT) phases, respectively.

At  doping $x_n \equiv 2/n^2$, the high symmetry of {\it commensurate} {\it 214} $O^{\upuparrows}$ superlattices of spacing $A_0 = n  a_0$, gives rise to significant changes of crystal properties. With $O^{\upuparrows}$ atoms at ideal lattice site, commmensurate superlattices are easy to illustrate---incommensurable ones are not.
However, a generalization to non-integer $\tilde{n} = \sqrt{2/x} $ indicates how an incommensurate superlattice reconfigurates to a {\it composite} commensurate superlattice.

A dent in both the superconducting dome, $T_c(x)$, and the electronic specific heat, $[C_{el}/T](x)$, is caused by a dome of {\it short-range} charge order,
peaking near $x = 0.12$ and terminating at $x=0.08$ and $x=0.16$. The dome is stabilized and destabilized by structural phase coherence and phase decoherence, respectively, of $CuO_5O^{\upuparrows}$ octahedra tilts.

For oxygen-enriched cuprates the incommensurability, observed with RIXS, {\it decreases} with increasing hole density, $q_{CuO_6}(p) = 0.35 - 0.71 p$, and similarly for oxygen-doped $YBa_2Cu_3O_{6+y}$. The signal is attributed to a pattern of diluted diagonally tilted $CuO_6$ host octahedra, reduced by heavily tilted $CuO_5O^{\uparrow\downarrow}$ octahedra (not detected with RIXS). Indirect evidence suggests that the $O^{\uparrow\downarrow}$ superlattice saturates at a watershed doping $\hat{p} = 1/8$, as in $La_{2-x}Sr_xCuO_4$, succeeded by an interlacing secondary superlattice, as well as by a discontinuous collapse of the pseudogap temperature $T^*(p)$ at $p^* = 0.19$.

Besides static charge order, due to $O^{\upuparrows}$ or $O^{\uparrow \downarrow}$ superlattices, charge-density {\it fluctuations} (CDF) are detected with RIXS. Their doping-dependent relation to charge order results from spontaneous back-and-forth hopping of double holes from $O^{\upuparrows}$ or $O^{\uparrow \downarrow}$ atoms to neighboring $O^{2-}$ ions with consequent reorientation of demotted ${\mathcolor{red}{\mathbf{Cu^{2+}_2}}}$ dimers. 
Two types of CDF are observed. One type tracks charge order of equal incommensurability, $q_{cdf}^{co}(p) = q_{co}(p)$. The other type, with $q_{cdf}^{\#} \approx 0.30$, is a low-intensity quantum fluctuation.

Subjected to a strong magnetic field, the Fermi surface of $YBa_2Cu_3O_{6+y}$ and $HgBa_2CuO_{4+\delta}$ reconstructs to a small diamont-shaped Fermi pocket, detected by quantum oscillations. The Fermi pocket results from bidirectional shifts of Fermi arcs by $q_{CuO_6}(p) \simeq 0.30$ for $p \simeq 0.10$, related to a reduced magnetic Brillouin zone. No quantum oscillations occur in {\it 214} cuprates as their incommensurabilities, $q_{co}(x) \le 0.27$, fail to connect Fermi-arc tips of Brillouin-zone quadrants.

Having addressed all questions posed in the introduction---except the nature of Fermi arcs---this concludes the summary.

\appendix

\section{CONTRACTION OF PLANAR LATTICE CONSTANTS}
The observed planar lattice constants of $La_{2-x}Sr_xCuO_4$, $a_0(x)$ and $b_0(x)$, are shown in Fig. 29, in their dependence on $Sr$ doping $x$ (solid lines).\cite{2} The range $0 < x < 0.21$ corresponds to the low-temperature orthogonal (LTO) phase with $b_0(x) > a_0(x)$ above and below. The region $x > 0.21$ corresponds to the low-temperature tetragonal (LTT) phase where $a_0(x)=b_0(x)$. 
{\it Quasi-tetragonal} lattice constants (dashed) are simulated, for $0<x<0.21$,  by
\begin{equation}
    \tilde{a}_0(x)= \tilde{b}_0(x)= \sqrt{2}\; [2r(Cu^{2+}) + (1-x)\; 2r(\overlinesegment{O^{2-}_{ab}}) + x \;2r(\tilde{O})] \;.
\end{equation}

\includegraphics[width=2.25in]{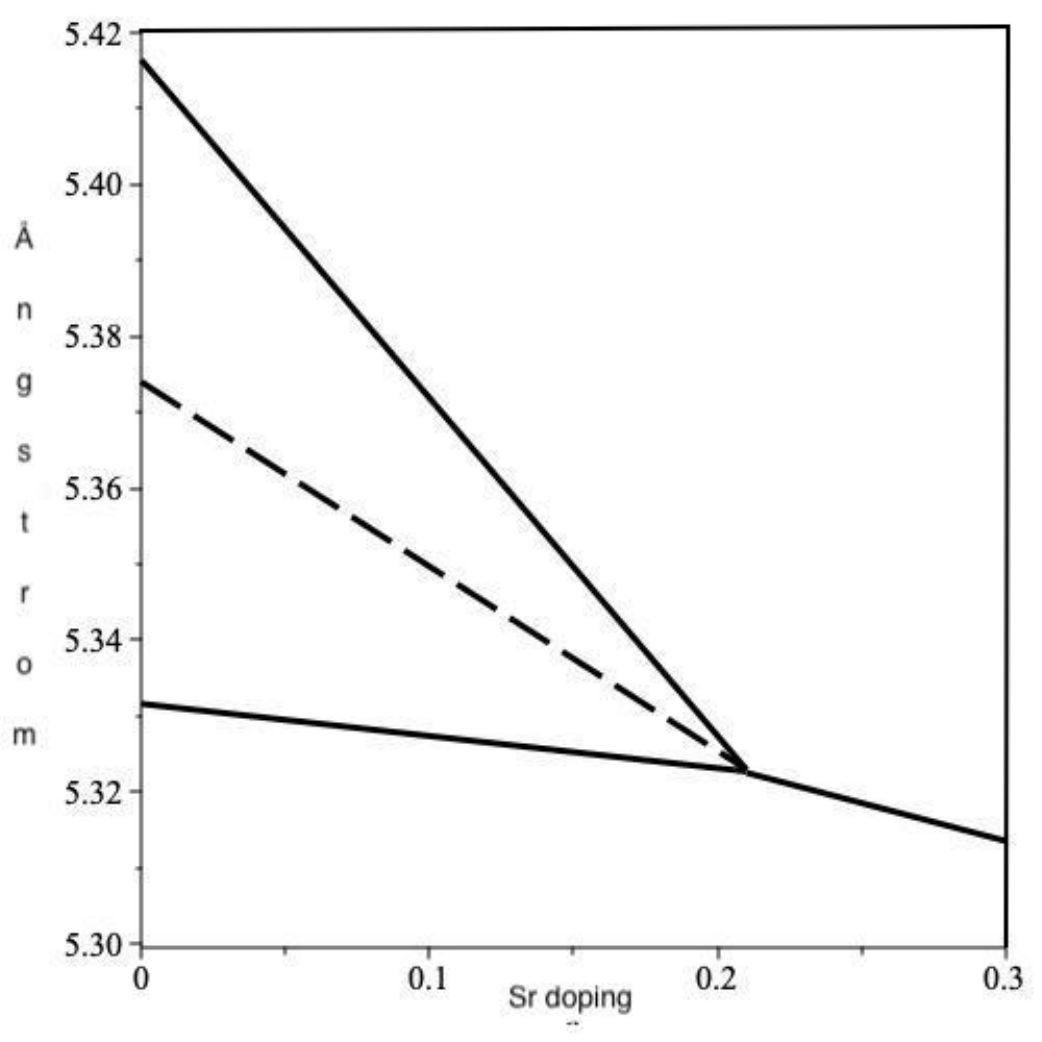}

\noindent \footnotesize FIG. 29. Doping dependence of observed lattice constants $a_0(x)$ (bottom branch) and $b_0(x)$ (top branch) of $La_{2-x}Sr_xCuO_4$ in the LTO (left) and LTT (right) phase. The dashed line shows a simulation of quasi-tetragonal lattice constants $\tilde{a}_0(x) = \tilde{b}_0(x)$, Eq. (A1). \normalsize

Here, $r(Cu^{2+})$ is the radius of the copper ion, the quantity  $2r(\overlinesegment{O^{2-}_{ab}}) = [2r(\overlinesegment{O^{2-}_a}) + 2r(\overlinesegment{O^{2-}_b}]/2$ is the average spacing between neighboring copper ions in the $a$- and $b$-direction, and $r(\tilde{O})$ is the {\it effective} radius of a relaxed oxygen atom (see Table I in Sec. II). (The $\sqrt{2}$ in Eq. (A1) is due to orthorhombic axes notation.) The different slope of the (dashed) $\tilde{a}_0(x)$  line at $x<0.21$ and the (solid) $a_0(x)$ line at $x>0.21$ in Fig. 29 originates with the different tilt of the $CuO_6$ octahedra---diagonal in the LTO phase but axial in the LTT phase (alternating along $a$ and $b$ in subsequent $CuO_2$ planes). 

\section{SUPERCONDUCTING DOME}

The doping dependence of the superconducting transition temperature on hole doping, $T_c(p)$, called the ``superconducting dome,'' has a parabolic profile once the the dent near $p=1/8$, as observed in $La_{2-x}Sr_xCuO_4$, is ignored. The superconducting dome has been modeled\cite{14} with good accuracy as
\begin{equation}
\frac{T_c(p)}{\wideparen{T}_c} =  1 - \frac{(\wideparen{p}-p)^2}{(\wideparen{p}-\acute{p})^2} 
= 1 - 82.6 \; (0.16 - p)^2 \;,
\end{equation}
with maximal transition temperature at optimal doping, $\wideparen{T}_c = T_c(\wideparen{p})$, and doping onset of superconductivity, $T_c(\acute{p}) = 0$ (see Fig. 30). The onset and optimal doping, $\acute{p} \simeq 0.05$ and 
$\wideparen{p} \simeq 0.16$, were obtained from $La_{2-x}Sr_xCuO_4$ where $x=p$. This gives the numerical values on the RHS of Eq. (B1). 
Inverting the formula furnishes the underdoped {\it (ud)} and overdoped {\it (od)} hole density in the $CuO_2$ plane as
\begin{equation}
p_{ud}^{od}(T_c) = 0.16 \pm \sqrt{\frac{\hat{T}_c-T_c}{82.6\;\hat{T}_c}} \;.
\end{equation} 
\includegraphics[width=3.8in]{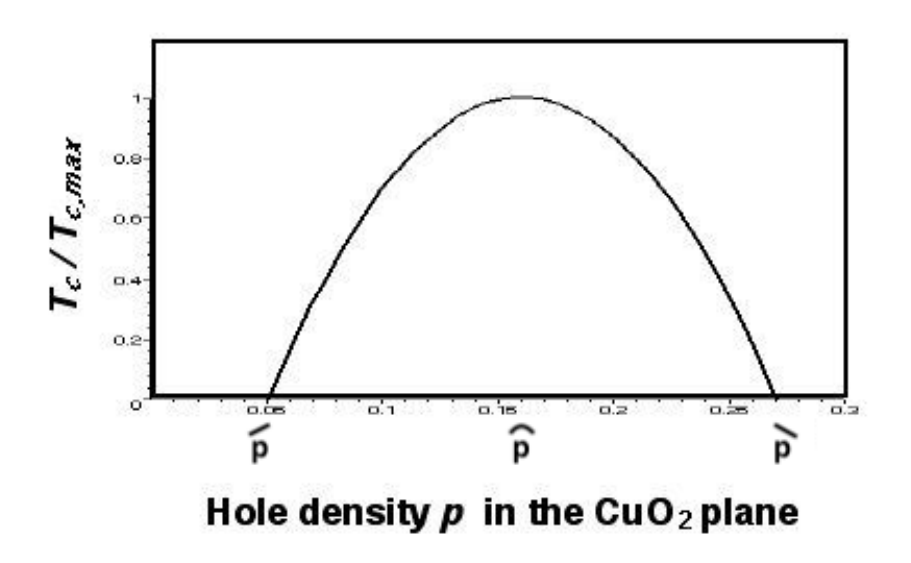}

\noindent \footnotesize FIG. 30.  Universal dome of superconducting cuprates, $\tau = T_c/\wideparen{T}_c$ {\it vs} hole density $p$,  Eq. (B1).  \normalsize

\section{COMMENSURATE SUPERLATTICES}
Superlattices of double-hole carrying $O$ atoms in $La_{2-x}Sr_xCuO_4$ are {\it commensurate} with the host lattice at doping $x_n = 2/n^2$, where the superlattice spacing is $A_0 = n a_0$ and $B_0 = n b_0$. 

\bigskip

\begin{table}[ht]

\begin{tabular}{|p{1cm}|p{1.8cm}|p{12cm}|}
 \hline  \hline
$\;\;\;\;n\;\;$ &$\;\;\;\;\;x_n$&$\;\;\;$Effect\\
 \hline  \hline
$\;\;\;10$&$\;\;\;\;0.02 $&$\;\;\;$Collapse of 3D-antiferromagnetism at N\'{e}el doping $x_N$\\ \hline \hline
$\;\;\;\;\;6$&$\;\;\;\;0.056 $&\;\;\;Change of $CuO_6$ octahedra tilt from diagonal to axial\\ \hline
$\;\;\;\;\;6$&$\;\;\;\;0.056 $&$\;\;\;$Doping onset of superconductivity \\ \hline \hline
$\;\;\;\;\;5$&$\;\;\;\;0.08 $&$\;\;\;$Least phase coherence of short-range charge order; Fig. 21 bottom\\ \hline \hline
$\;\;\;\;\;4$&$\;\;\;\;0.125 $&$\;\;\;$Maximal stripe intensity; Fig. 21 top\\ \hline
$\;\;\;\;\;4$&$\;\;\;\;0.125 $&$\;\;\;$Leveling off of incommensurability at watershed $\hat{x}$; Figs. 11, 13 \\ \hline 
$\;\;\;\;\;4$&$\;\;\;\;0.125 $&$\;\;\;$Dent in superconductive dome\\ \hline \hline
\end{tabular}
\label{table:4}
\footnotesize \caption{Commensurate superlattices of $O^{\upuparrows}$ atoms in $La_{2-x}Sr_xCuO_4$ at doping $x_n = 2/n^2$.} \normalsize
\end{table}

A generalization of the  spacing of an $n \times n$ commensurate $O$ superlattice to the {\it incommensurate} case gives $\tilde {n} = \sqrt{2/x}$. Multiplying $\tilde{n}$ with an integer $\nu$ to equal a sum of integers, $\nu \tilde{n} =n_1 + n_2 + ...$, shows how a reconfiguration to a
{\it composite} $(n_1 + n_2 + ...) \times (n_1 + n_2 + ...)$ {\it commensurate} superlattice can be achieved. Table V gives examples.

\begin{table}[ht]

\begin{tabular}{|p{1.8cm}|p{1.1cm}|p{0.6cm}|p{1.3cm}|p{4.3cm}|p{5.7cm}|} \hline  \hline
$\;\;\;\;\;\;\;\;x $&\;\;\;\;$\tilde{n}$\;\;\;\;\;\;\;\;&\;\;$\nu$&$\;\;\;\;\nu\tilde{n}$&$\;\;\;$Crystal$\;\;\;\;\;\;\;\;$&$\;\;\;$Composite\\  \hline  \hline
$\;\;\hat{x} = 0.16 $&$\;\;3.54$&$\;\;2$&$\;\;\;7.07$&$\;\;\;La_{1.6-x}Nd_{0.4}Sr_xCuO_4$&$\;\;(4 + 3) \times (4 + 3)\;\;\;\;$Fig. 19\\ \hline 
$\;\;\hat{x} = 0.14$&$ \;\;3.79$&$\;\;4$&$\;\;15.12$&$\;\;\;La_{2-x}Ba_xCuO_4$&$\;\;(4 + 4 + 4 + 3) \times (4 + 4 + 4 + 4 + 3)$\\ \hline \hline
\end{tabular}
\label{table:5}
\footnotesize \caption{Composite commensurate superlattices of $O$ atoms in superconducting cuprates.} \normalsize
\end{table}

\bigskip \bigskip 

\centerline{ \textbf{ACKNOWLEDGMENT}}

\centerline{ I thank Preston Jones for help with graphics.}


\begin{thebibliography}{118}
\bibitem{60} The data from Refs. 57 and 58 were used with a $Ba$ concentration re-calibrated according to Ref. 59.
\end{thebibliography}
\end{document}